\documentclass[prd,aps,twocolumn,floatfix]{revtex4}
\pdfoutput=1
\usepackage{amssymb,graphicx}
\usepackage{epsfig}
\usepackage[usenames]{color}
\newcommand{\Alfven}{ Alfv\'{e}n }
\newcommand{\nn}{\mbox{} \nonumber \\ \mbox{} }

\begin{document}

\title{Intense Electromagnetic Outbursts from Collapsing Hypermassive
Neutron Stars}

\author{
Luis Lehner$^{1,2,3}$,
Carlos Palenzuela$^{4}$,\\
Steven L. Liebling$^{5}$,
Christopher Thompson$^{4}$,
Chad Hanna$^{2}$}
\affiliation{
$^{1}$Department of Physics, University of Guelph, Guelph, Ontario N1G 2W1, Canada, \\
$^{2}$Perimeter Institute for Theoretical Physics,Waterloo, Ontario N2L 2Y5, Canada,\\
$^{3}$CIFAR, Cosmology \& Gravity Program, Canada, \\
$^{4}$Canadian Institute for Theoretical Astrophysics, Toronto, Ontario M5S 3H8,
 Canada, \\
$^{5}$Department of Physics, Long Island University, New York 11548, USA}

\begin{abstract}
We study the gravitational collapse of a magnetized neutron star using
a novel numerical approach able to capture both the dynamics of the star and
the behavior of the surrounding plasma.
In this approach, a fully general relativistic magnetohydrodynamics 
implementation models the collapse of the star and provides
appropriate boundary conditions to a force-free model which describes the stellar
exterior. We validate this strategy by comparing with known results for
the rotating monopole and aligned rotator solutions and then apply it to study
both rotating and non-rotating stellar collapse scenarios, and contrast the behavior with
what is obtained when employing the electrovacuum approximation outside the star.
The non-rotating electrovacuum collapse is shown to agree qualitatively
with a Newtonian model of the electromagnetic field outside
a collapsing star.  We illustrate and discuss a fundamental difference between the force-free
and electrovacuum solutions, involving the appearance of large zones of
electric-dominated field in the vacuum case.  This provides a clear demonstration
of how dissipative singularities appear generically in the non-linear time-evolution
of force-free fluids.
In both the rotating and non-rotating cases, our simulations indicate that the
collapse induces a strong electromagnetic transient, which leaves behind
an uncharged, unmagnetized Kerr black hole.  In the case of sub-millisecond rotation,
the magnetic field experiences strong winding and the transient carries much
more energy.  This result has important implications for models
of gamma-ray bursts.  Even when the neutron star is surrounded by
an accretion torus (as in binary merger and collapsar scenarios), a magnetosphere
may emerge through a dynamo process operating in a surface shear layer.  When this
rapidly rotating magnetar collapses to a black hole, the electromagnetic energy released
can compete with the later output in a Blandford-Znajek jet. 
Much less electromagnetic energy is released by 
a massive magnetar that is (initially) gravitationally stable: its rotational energy 
is dissipated mainly by internal torques.  A distinct plasmoid structure is seen in
our non-rotating simulations, which will generate a radio transient with subluminal
expansion, and greater synchrotron efficiency than is expected in shock models. 
Closely related phenomena appear to be at work in the giant flares of Galactic magnetars.
\end{abstract}

\maketitle

\section{Introduction}
Understanding the gravitational collapse of a massive neutron star is of
central importance for its connection to some of the most energetic
astrophysical phenomena. Such an event may take place within a core-collapse
supernova \cite{woosley:1993}, or in the late stage of a binary neutron star 
merger \cite{eichler:1989,narayan:1992}, and is widely
believed to power some types of gamma-ray bursts.  In recent years it has been
realized that newly formed stellar mass black holes may be prodigious sources
of electromagnetic energy, in addition to driving strong kinetic outflows~\cite{woosley:1993,1994MNRAS.270..480T,meszaros:1997,mckinney:2006}.  We
are faced with the exciting possibility of probing the most extreme forms of
gravitational collapse using coordinated measurements of electromagnetic
transients and gravitational waves.  Refining models that make observational
predictions for joint gravitational and electromagnetic radiation is critical
in order to establish efficient observation campaigns for both traditional
astronomers and gravitational-wave astronomers. It is also a key step
toward predicting the delay between peak electromagnetic and gravitational
wave emission, the electromagnetic emission pattern (i.e. the beaming angle)
and the electromagnetic and gravitational wave spectrum.

Our focus here is on the evolution of the neutron star's magnetic field
during its collapse to a black hole.  We employ fully self-consistent relativistic
calculations that follow the dense stellar material as well as the
strong electromagnetic and gravitational fields.  A particularly intense
electromagnetic transient is generated if the initial magnetic field is very
strong ($\sim 10^{15}-10^{16}$ G), that is, if the star is a magnetar.
Rapid rotation will enhance the energy of the transient, to a degree
that can only be derived by a full time-evolution of the electromagnetic
and gravitational fields.  Rapid rotation also provides a context
for generating strong magnetic fields:  when the neutron star is 
accreting, the shear layer at its surface is
a promising site for dynamo action.  A distinct magnetosphere will
emerge and hold off the accretion flow
if even $\sim 10^{-4}-10^{-3}$ of the energy that is dissipated in the
shear layer is converted to a poloidal magnetic field.  Understanding the
evolution of the magnetic field around an {\it isolated} star is therefore
potentially of key relevance for the collapsar and binary merger scenarios
of gamma-ray bursts.

Recent success in studying the behavior of plasmas around magnetized, spinning,
stable neutron stars within flat spacetime has been presented
in Refs.~\cite{2006ApJ...643.1139C,Spitkovsky:2006np,McKinney:2006sc}.
To study the collapse problem, one needs general relativity -- to account for the role of
spacetime curvature; and general relativistic magnetohydrodynamics (GRMHD) -- to
determine the internal evolution of the star.
The electromagnetic (EM) phenomena outside the star can be approached in a variety of ways:  through
a full GRMHD calculation, which generally is very expensive given the low matter density; or
more approximately using vacuum EM and force-free equations.  In this paper, we compare
all three approaches (restricting to ideal MHD in the case of GRMHD).

Within general relativity, studies paying attention to neutron star collapse have been
presented in the context of isolated stellar collapse (e.g.~\cite{Stephens:2006cn,Stephens:2007kb,O'Connor:2010tk,Reisswig:2010cd,Giacomazzo:2011cv})
and binary neutron star mergers (e.g.~\cite{Anderson:2007kz,Baiotti:2008ra,Liu:2008xy,Anderson:2008zp,Hotokezaka:2011dh,Shibata:2011fj})
with different degrees of realism.  The role of electromagnetic fields are typically
examined within general relativistic {\em ideal} MHD and studies have been presented
for single~\cite{Duez:2005cj,Shibata:2005mz,Duez:2006qe,Kiuchi:2008ss,Liebling:2010bn,Liebling:2010qv} and
binary star systems~\cite{Anderson:2008zp,Liu:2008xy,Giacomazzo:2009mp}. Such simulations have illustrated
important details of the extreme dynamical behavior induced by the system which could trigger tremendously
energetic phenomena.

Newtonian and general relativistic simulations
have shed light on how a binary neutron star merger can amplify pulsar-strength magnetic fields
by several orders of magnitude~\cite{price:2006,Anderson:2008zp,2010A&A...515A..30O,Giacomazzo:2009mp}.
The resulting hypermassive star is generically unstable to black hole formation,
which opens up the
tantalizing possibility that the increasing rotation rate and magnetic field strength
would drive an intense electromagnetic outflow.
To study such a scenario, our goal here is to combine the GRMHD approach with a suitable
description of the magnetically-dominated region outside a collapsing star, 
by applying the force-free approximation.  
Our approach is related to that of Ref.~\cite{Baumgarte:2002b}, which matched a
simplified, analytic, relativistic solution for the interior of a collapsing star (dust-ball)
with a numerical solution of the coupled Einstein-Maxwell
equations in its exterior.  
Boundary conditions at the star's surface are provided by the analytic, ideal MHD solution. 

We go beyond this simplified scenario in three significant ways. First,
our stars are evolved through collapse consistently by evolving the ideal GRMHD equations, 
and we can therefore consider in principle any kind of compact star. 
In particular, we have studied both rigidly
rotating and non-rotating stars, whereas only the non-rotating case was considered 
in~\cite{Baumgarte:2002b}. 
Second, the magnetosphere is described within the force-free
approximation, which, as argued in Ref.~\cite{Goldreich:1969sb},
is a much more realistic model that, for instance, has been instrumental in understanding pulsar 
spin down~\cite{Spitkovsky:2006np}.
Third and last, our matching of the exterior solution with the
interior one is dynamical. Thus the force-free solution can adapt to time-dependent fields
sourced by the star.

Our approach thus allows us to examine many interesting scenarios and we apply it here to
study the behavior of collapsing, magnetized, compact stars (either rotating or not), 
the behavior of surrounding plasma (as described within the force-free approximation) 
and possible electromagnetic radiation induced 
by the system. We compare our results with recent estimates in Ref.~\cite{Lyutikov:2011tq}
for the non-rotating case which predict that the collapse process is smooth and that the 
magnetic field remains anchored to the star as a black hole forms, leaving a final black hole
with a split-monopole field configuration.
Our work, which follows the dynamics of the system, indicates that in both cases the stars radiate significant electromagnetic energy in which reconnection plays a crucial role, 
and that this radiation ceases shortly after the formation of a black hole. With force-free, such a
black hole loses its electromagnetic hair within a dynamical time scale.

We describe our hybrid approximation and the evolution equations in Sec.~\ref{sec:approach},
followed by a summary of the numerical techniques in
Sec.~\ref{sec:numerical_techniques}. Our choice of initial data is described
in Sec.~\ref{sec:id}, and several tests of the hybrid approach are presented
in Sec.~\ref{sec:testing}. The new results for collapsing, magnetized stars are explained 
in detail in Sec.~\ref{sec:physics}, while the astrophysical consequences are 
described in Secs.~\ref{sec:astrophysicsI} and ~\ref{sec:magnetar}.
We conclude in Sec.~\ref{sec:concluding_remarks} with some final comments.

\section{Approach}
\label{sec:approach}
In the presence of matter and 
electromagnetic fields,
the Einstein equations must be suitably coupled to both the Maxwell and hydrodynamics equations.
This coupling is achieved by considering the stress energy tensor
\begin{equation}
T_{ab} = T_{ab}^{\rm fluid} + T_{ab}^{\rm em} \, ,
\end{equation}
with contributions from matter and electromagnetic energy given respectively by
\begin{eqnarray}\label{stressenergy}
T_{ab}^{\rm fluid} &=& \left[ \rho_o\left(1+\epsilon\right) + P \right] u_a u_b + P g_{ab}  \, , \\
T_{ab}^{\rm em} &=& {F_a}^c F_{bc} - \frac{1}{4} g_{ab} F^{cd} F_{cd} \, .
\end{eqnarray}
A perfect fluid with pressure $P$, { energy density} $\rho_o$, internal energy $\epsilon$,
and four-velocity $u^a$ describes the matter state,
and the Faraday tensor $F_{ab}$ describes the electromagnetic field.
The fluid and electromagnetic components are directly coupled through 
Ohm's law, which closes the system of equations
by defining the electric current 4-vector $J_a$ as a function of the other fields.
A general relativistic expression can be obtained by considering a
multifluid system of charged species~\cite{2008MNRAS.385..883K}, leading to a fully
non-linear propagation equation for the spatial component of the current.
However, it usually suffices to consider a simplified version 
accounting for an algebraic relation between the current and the
fields~\cite{1978PhRvD..18.1809B}
\begin{equation}\label{ohm_law}
J^a + \left(u_b J^b\right) u^a = \sigma^{ab} F_{bc} u^c\,,
\end{equation}
where $\sigma^{ab}$ is the conductivity of the fluid. The well-known scalar Ohm's
law is recovered for $\sigma^{ab} = g^{ab}\,\sigma$.
The equation of motion for the fluid and electromagnetic field are obtained from
{\em the conservation laws}
\begin{eqnarray}
\nabla_a T^{ab} = 0 \, \, ; \, \, \nabla_a (\rho_o u^a) = 0\, \, ; \\
\nabla_a F^{ab} = J^b \, \, ; \, \,  \nabla_a {}^*F^{ab} = 0\, \, ;
\end{eqnarray}
which, together with the Einstein equations $G_{ab} = 8\pi T_{ab}$, complete the system
of equations governing the dynamics.

Once the appropriate form of Ohm's law and the conductivity have been specified,
using for instance the algebraic relation of Eq.~(\ref{ohm_law}), the resulting equations
typically involve vastly different scales, rendering the implementation of these equations
quite costly from a computational point of 
view \footnote{See however the possibility of doing so with recently developed techniques
~\cite{2007MNRAS.382..995K,Palenzuela:2008sf,2011ApJ...735..113T} }. 
Fortunately, for specific regimes certain useful approximations 
capture the {\em most relevant} physics while bypassing the most strenuous difficulties. 
For our current purposes,  involving magnetized neutron stars, the following relevant
approximations can be defined and employed in different regimes:
\begin{itemize}
\item The {\em ideal-MHD} equations are obtained by requiring that the current remains
finite in the limit of infinite conductivity,
$\sigma \rightarrow \infty$. This condition also implies the vanishing of the electric field measured by
an observer co-moving with the fluid ($F_{ab}u^b=0$). This approximation is appropriate for the
highly conducting matter expected in neutron stars. However, 
the numerical evolution of the ideal MHD equations typically fails 
in low density regions where the inertia of the electromagnetic field is a few orders
of magnitude larger than that of the fluid unless sufficient resolution is available.
Such resolution requirements increase as the ratio of the electromagnetic to the fluid's inertia 
(or magnetic to fluid's pressure) grows. Consequently, the approach becomes costlier in regions with decreasing physical relevance
with respect to bulk matter motion.  
Such a situation arises in particular in ``vacuum'' regions due to the standard practice
of maintaining  a density floor (a so called {\em atmosphere}) in regions of low density
to exploit advanced numerical techniques for relativistic hydrodynamics. The atmosphere's density
is much smaller than that inside the star, so this approach does not affect the relevant matter physics.
However outside the star, the fluid inertia (or pressure) is typically much smaller than
that of the electromagnetic field in magnetized cases and one generally encounters  
a large number of numerical difficulties.
Different numerical strategies are often introduced to avoid
them but such measures can limit one's ability to extract appropriate physics in these regions. 
\item The {\em force-free} Maxwell equations are obtained by assuming the fluid's inertia is
much smaller than that of the electromagnetic fields. As a result $F_{ab} J^{b}=0$ which
in turn implies $F_{ab} {}^*F^{ab}=0$ ($\rightarrow E.B=0$)~\cite{Goldreich:1969sb,1977MNRAS.179..433B}.
This assumption therefore allows one to ignore the explicit time-evolution of the fluid 
as its dynamics are implicitly prescribed by the charge and current distribution. This approximation is well suited to
the dynamics of the low density, magnetically dominated plasma surrounding a compact
object, but it cannot account for the physics in dense regions.
\item The {\em vacuum} Maxwell equations are trivially recovered assuming no coupling
with matter ($\sigma = 0 $). This is a natural approximation in vacuum regions,
far away from the compact objects of interest.
\end{itemize}

As mentioned, computational costs are presently major obstacles to employ the general equations. 
It is thus highly desirable to define a scheme able to model all the relevant regimes in highly
dynamical systems simultaneously since none of these three approaches captures all the
expected behavior.
In particular we
have in mind interacting binary neutron stars, black hole-neutron star binaries as well as the collapse of 
a magnetized star considered here. All three systems are believed to play an important role in understanding gamma ray bursts and
other energetic events driven by compact stellar-mass objects.

In what follows, we describe and apply a {\em hybrid} approach which, while approximate, can
account for the dynamical interaction of both gravitating and electromagnetically driven fluids.
Such an approach allows one to study the magnetosphere's behavior, in particular energetics and field
topology providing important clues for understanding relevant systems.
To do so, we take into account that the electromagnetic inertia of a region with high conductivity (i.e. inside the
star) would be orders of magnitude larger than that of the plasma region. Thus, we can ignore the back-reaction
of the electromagnetic field in the plasma region onto that inside the star (i.e., a ``passive'' magnetosphere).
We exploit this observation to define
our approach, which employs both the ideal and force-free approximations suitably matched around the stellar surface.
The matching procedure is such that the star's electromagnetic field provides the boundary conditions for the surrounding region
treated with the force-free approximation, but the behavior of the magnetospheric plasma on the high density stellar interior is ignored.

Notice that this approach, in a sense, can be regarded as a natural extension
of that adopted for studying pulsar-spin down by magnetosphere 
interactions~\cite{Spitkovsky:2006np,2006ApJ...643.1139C}
to general relativistic dynamical contexts. 
In these works, only the behavior of the magnetospheric plasma is studied and the star's influence is accounted for
by boundary conditions. We here  account for the possible
stellar dynamics and consequent influence on the boundary conditions defined for the force-free approximation.

In practical terms, we regard our system as described by two sets of electromagnetic fields: (i) the
ideal MHD fields $\{E^i,B^i\}$ and (ii) the force-free fields $\{\tilde E^i, \tilde B^i\}$, governed by their
respective equations. Both sets of fields are defined with respect to observers orthogonal to the spacelike hypersurfaces
employed to foliate the spacetime. Thus identifying the appropriate fields for matching is direct.
The ideal MHD equations are evolved over the entire computational domain, as is customary.
The matching region is determined by fluid density being some value above the vacuum region (though alternatives based on
other physical quantities are obviously possible). 
In particular, the force-free fields are evolved only in regions  where $\rho_o < \rho^{\rm match}$.
Dynamic boundary conditions are applied to the force-free fields on the surface at which $\rho_o = \rho^{\rm match}$ using the ideal MHD
fields
\begin{eqnarray}
\tilde E_{\rm bc}^i &=& -\epsilon^{i}_{jk} v^j B^k \, ; \label{EfromB} \\
\tilde B_{\rm bc}^i &=&  B^i \, .
\label{eq:bbc}
\end{eqnarray}
These conditions are applied in the spirit of penalty techniques~\cite{Carpenter_Gottlieb_Abarbanel_1993} in
which the equations are modified with driving terms at boundary points in order to enforce the desired boundary
condition.
In particular, we extend the penalty technique across a number of grid points to effect both
the boundary conditions and to ``turn-off'' the evolution of the force-free fields within the higher density regions.
Notice that this extension is not mathematically rigorous, but its usage
will be justified by examining the solution in different test applications and comparing with the expected
behavior. In practice we define a smooth kernel $F(x^i,x^i_{\rm match})$  defined as
\begin{equation}
F(\rho_o \,,\rho^{\rm match})= \frac{2}{1 + e^{2\,K\,(\rho_o - \rho^{\rm match})}}
\label{eq:kernel}
\end{equation}
where typically we adopt $K \approx 0.001/\rho_{\rm atmos}$ and 
$\rho^{\rm match} \approx 200-2000\, \rho_{\rm atmos}$, being $\rho_{\rm atmos}$ 
the value for the density of the atmosphere. In the collapsing cases studied, this value is rescaled in time by 
the ratio of maximum density to the initial maximum density (see appendix). 
The values found in Eqs.~(\ref{EfromB}-\ref{eq:bbc})
along with the kernel Eq.~(\ref{eq:kernel}) make their appearance in
the equations determining $\{ \tilde E^i, \tilde B^i \}$ 
(together with ``constraint cleaning'' fields $\{ \tilde \Psi, \tilde \phi \}$,
introduced to ensure constraints are well behaved through the evolution ~\cite{Palenzuela:2010xn}) which are
\begin{eqnarray}
  \partial_t \tilde E^{i} &=& F \left ( {\cal L}_{\beta} \, \tilde E^{i}   +
  \epsilon^{ijk}\nabla_j (\,\alpha \tilde B_k\,)
   - \alpha\,\gamma^{ij} \nabla_j\,\tilde \Psi \right . \nonumber \\
  & +&  \left . \alpha\, trK \, \tilde E^i - 4 \pi \alpha J^{\,i} \right ) +
   \lambda (1 - F) (E_{bc}^i - \tilde E^i)  \, ,
\label{maxwellext_3+1_eq1a} 
\\
  \partial_t \tilde B^{i} &=& F \left( {\cal L}_{\beta} \, \tilde B^{i} -
  \epsilon^{ijk} \nabla_j (\,\alpha \tilde E_k\,)
   - \alpha\, \gamma^{ij} \nabla_j\, \tilde \phi \right.  \nonumber \\
   & +& \left. \alpha\, trK \, \tilde B^i \right ) 
   + \lambda (1 - F) (B_{\rm bc}^i - \tilde B^i) \, ,
\label{maxwellext_3+1_eq1c} 
\\
  \partial_t \tilde \Psi &=& F \left( {\cal L}_{\beta}\,\tilde \Psi - \alpha\, \nabla_i \tilde E^i 
   4 \pi \alpha\, q -\alpha \sigma_2\, \tilde \Psi \right )  \nonumber 
\\ &  -& \lambda (1-F)  \tilde \Psi \, ,
\label{maxwellext_3+1_eq1b} \\
  \partial_t \tilde \phi &=& F \left( {\cal L}_{\beta}\, \tilde \phi - \alpha\, \nabla_i \tilde B^i 
   -\alpha \sigma_2\, \phi \right )  \nonumber 
\\ &  -& \lambda (1-F)  \tilde \phi
\label{maxwellext_3+1_eq1d}\,.
\end{eqnarray}
The final term on the right-hand side
of these equations is a penalty factor, 
which is introduced to impose interface conditions and ensure
that a discrete energy norm is bounded.   Details of this 
``penalty technique'' 
are presented in~\cite{Carpenter_Gottlieb_Abarbanel_1993} and examples of applications in general relativity
can be found in~\cite{multipatch}. The above equations, together with the Einstein and GRMHD equations
are implemented as described in~\cite{Liebling:2010bn,Palenzuela:2009yr,Palenzuela:2010nf} to which we refer
the reader for further details.

\section{Numerical Techniques}
\label{sec:numerical_techniques}
We adopt finite difference techniques on a regular Cartesian grid to solve
the system. To ensure sufficient resolution in an efficient
manner we employ adaptive mesh refinement (AMR) via the HAD  computational infrastructure
that provides distributed, Berger-Oliger
style AMR~\cite{had_webpage,Liebling} with full sub-cycling
in time, together with an improved treatment of artificial boundaries~\cite{Lehner:2005vc}.
The refinement regions are determined using truncation error estimation provided by a shadow
hierarchy~\cite{Pretoriusphd} which adapts dynamically to ensure the estimated error is bounded within a
pre-specified tolerance.
The spatial discretization of the geometry and force-free fields is performed
using a fourth order accurate scheme satisfying the summation by parts rule,
and High Resolution Shock Capturing methods 
based on the HLLE flux formulae with PPM reconstruction
are used to discretize the fluid variables~\cite{Anderson:2006ay,Anderson:2007kz}.
The time-evolution is performed through the method of lines using
a third order accurate Runge-Kutta integration scheme, which helps
to ensure stability of the numerical implementation~\cite{Anderson:2008zp}. We adopt a Courant
parameter of $\lambda = 0.2$ so that $\Delta t_l = 0.2 \Delta x_l$ on each refinement level $l$. On each
level, one therefore ensures that the
Courant-Friedrichs-Levy~(CFL) condition dictated by the principal part of
the equations is satisfied.

To extract physical information, we monitor several quantities: (i) the
matter variables and spacetime behavior, (ii) the electromagnetic field configuration and fluxes,
and (iii) the
electromagnetic Newman-Penrose (complex) radiative scalar ($\Phi_2$).
This scalar is computed by contracting the Maxwell tensor
with a suitably defined null tetrad
\begin{eqnarray}
  \Phi_2 = F_{ab} n^a \bar m^b ,
\end{eqnarray}
and it accounts for the energy carried off by outgoing waves to infinity.
The luminosity of the electromagnetic waves is
\begin{eqnarray}
  L_{\rm em} &=& \frac{{dE}^{\rm em}}{dt} = \int F_{\rm em} d\Omega \nonumber \\
         &=&
           \lim_{r \rightarrow \infty}  \int r^2 |\Phi_2|^2 d\Omega ~.
\label{FEM} 
\end{eqnarray}

Additionally we monitor the ratio of particular components
of the Maxwell tensor
\begin{equation}\label{omegaF2}
  \Omega_F = \frac{F_{tr}}{F_{r\phi}} = \frac{F_{t\theta}}{F_{\theta \phi}} ~~,
\end{equation}
which, in the stationary, axisymmetric case, can be interpreted as the rotation frequency of the
electromagnetic field~\cite{1977MNRAS.179..433B}.

\section{Initial Data}
\label{sec:id}
Initial data for the hybrid equations involve the intrinsic metric ($g_{ij}$) and extrinsic 
curvature ($K_{ij}$) on a given hypersurface, as well as the magnetized fluid configuration in terms of
its primitive variables $(\rho,\epsilon,v^i,B^i)$.  The initial data for the geometry and
the fluid of rigidly rotating neutron stars are provided by the LORENE package 
{\it Magstar}~\cite{lorene}, which adopts a polytropic equation of state
$P=K \rho^{\Gamma}$ with $\Gamma=2$ rescaled to $K=100$.
Because the fluid inertia of a neutron star is many orders of magnitude larger than
its electromagnetic one, the magnetic field will have a negligible effect on both the geometry
and the fluid structure, and so it can be specified freely. Unless noted otherwise, in our simulations
we have chosen a dipolar structure for the initial magnetic field. The electric fields are set by assuming
the ideal MHD condition Eq.~(\ref{EfromB}), with zero fluid velocity in the exterior region.
Additionally, we require initial data for the force-free fields
$\{ \tilde E^i, \tilde B^i \}$. Inside the star, they are defined to be exactly the same as
their ideal MHD counterparts. Outside the star, the magnetic field is well defined by the
dipolar solution, while the electric field is computed by assuming that the magnetosphere rotates
rigidly with the star up to $R_e = 2 R_s$ ($R_s$ is the stellar radius), and imposing
again the ideal MHD condition for the electric field. This configuration provides consistent data 
for the problem; however such data will not necessarily conform to the physical situation considered
and so an unphysical early transient will be generated.

\section{Testing the approach}
\label{sec:testing}
We first establish that the adopted procedure is indeed able to capture correctly the dynamics of relevant systems.
Since the ideal MHD approximation is self-consistently evolved throughout the computational domain, our
tests must address the behavior of the force-free fields.
To do so, we first examine the convergence of our implementation and then illustrate that the approach provides 
the expected behavior by comparing with certain recently studied cases.

\subsection{Non-rotating, magnetized star with a dipole magnetic field}
Adopting a stable and non-rotating stellar solution with mass $M=1.63 M_{\odot}$
and equatorial radius $R_{\rm eq}=8.62~{\rm km}$,
from which we remove all initial pressure,
we examine convergence of the
force-free fields as the star collapses. This scheme contains ``standard'' sources of error, such as:
(i)~truncation error of our finite difference approximation of Maxwell's equations,
(ii)~errors associated with our application of the force-free conditions,
(iii)~the various numerical errors (truncation and constraint violations)
associated with our GRMHD implementation as well as (iv)~errors associated with
our dynamic boundary condition that matches the force-free fields to the 
corresponding MHD fields inside the star. A detailed analysis
of these errors is delicate and involved; however we illustrate
that the fields obtained by this approach converge to a unique solution
with increasing grid resolution. 

\begin{figure}
\begin{center}
\epsfig{file=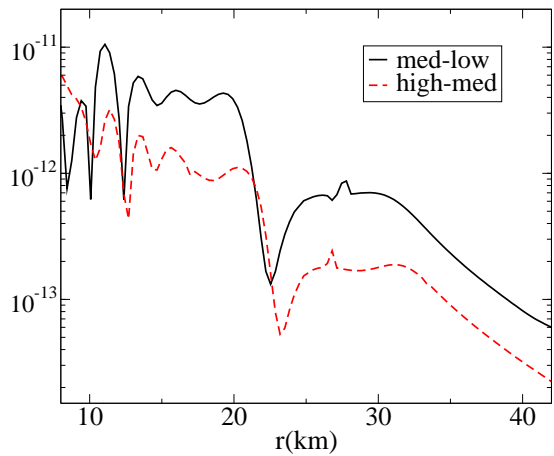,height=2.4in}
\caption{{\it Non-rotating, collapsing  solution}.
   Results from FMR solutions at three different resolutions for the same
nonrotating collapsing star. Shown are the two differences between the
resolutions. The difference in $\tilde B_z$ along $y=0$ in the equatorial plane
at 
$t=0.06 {\rm ms}$ is first interpolated onto a uniform mesh for each solution. 
The difference between the {\em medium} and {\em high} resolutions is smaller
than that between the {\em low} and {\em medium} resolutions, and this
decrease as one increases the resolution indicates that the scheme is
convergent. (Notice convergence in the central region is affected by initial data errors
induced by depleting the pressure to induce the collapse.) }
\label{fig:nonrotatingconvergence}
\end{center}
\end{figure}

We therefore evolve the collapse of a non-rotating star
at three different resolutions with fixed mesh refinement~(FMR).  We
subtract a high resolution run (with coarse level grid of $257^3$ points)
and a medium resolution run (of dimension $193^3$) and do similarly for
the medium and low resolutions (of dimension $129^3$). These two
differences are plotted at $t=0.06 {\rm ms}$ (i.e., when the radius of the star
starts to shrink) in Fig.~\ref{fig:nonrotatingconvergence} which
shows that the difference decreases with increasing resolution, consistent
with convergence.

\subsection{Magnetic monopole}
Next we consider a stationary force-free solution representing a rotating neutron star with a monopole
magnetic field~\cite{1973ApJ...180..207M}, and compare to our results.
To minimize effects due to oscillations of the star (induced by perturbations to the star
induced by the discretization),
the geometry and MHD fields are kept to their initial values and are not evolved. 
The initial data correspond to a rigidly rotating neutron star near the mass shedding limit,
with a mass $M=1.84 M_{\odot}$ and an equatorial radius $R_{\rm eq}=12~{\rm km}$,
rotating with a period of $T=0.886~{\rm ms}$.
The magnetic field is given by $B^r = \alpha B_0 (R_s/r)^2$,
regularized conveniently near the origin. As described earlier, the initial electric field
satisfies the ideal MHD condition in a rigidly co-rotating magnetosphere which extends
initially up to $R_e \sim 2\,R_s$. The evolution is performed in a cubic domain 
of length $L=136~{\rm km} \approx 14\, R_s$ with only two FMR grids and the star is placed at the origin. 
The maximum resolution is
$\Delta x = 0.72~{\rm km}$, so that roughly 30 grid points cover 
the star. This resolution is sufficient for this test as both the geometry
and matter variables are kept fixed and only the force-free equations are evolved
until reaching a  quasi-stationary configuration.

\begin{figure}
\begin{center}
\epsfig{file=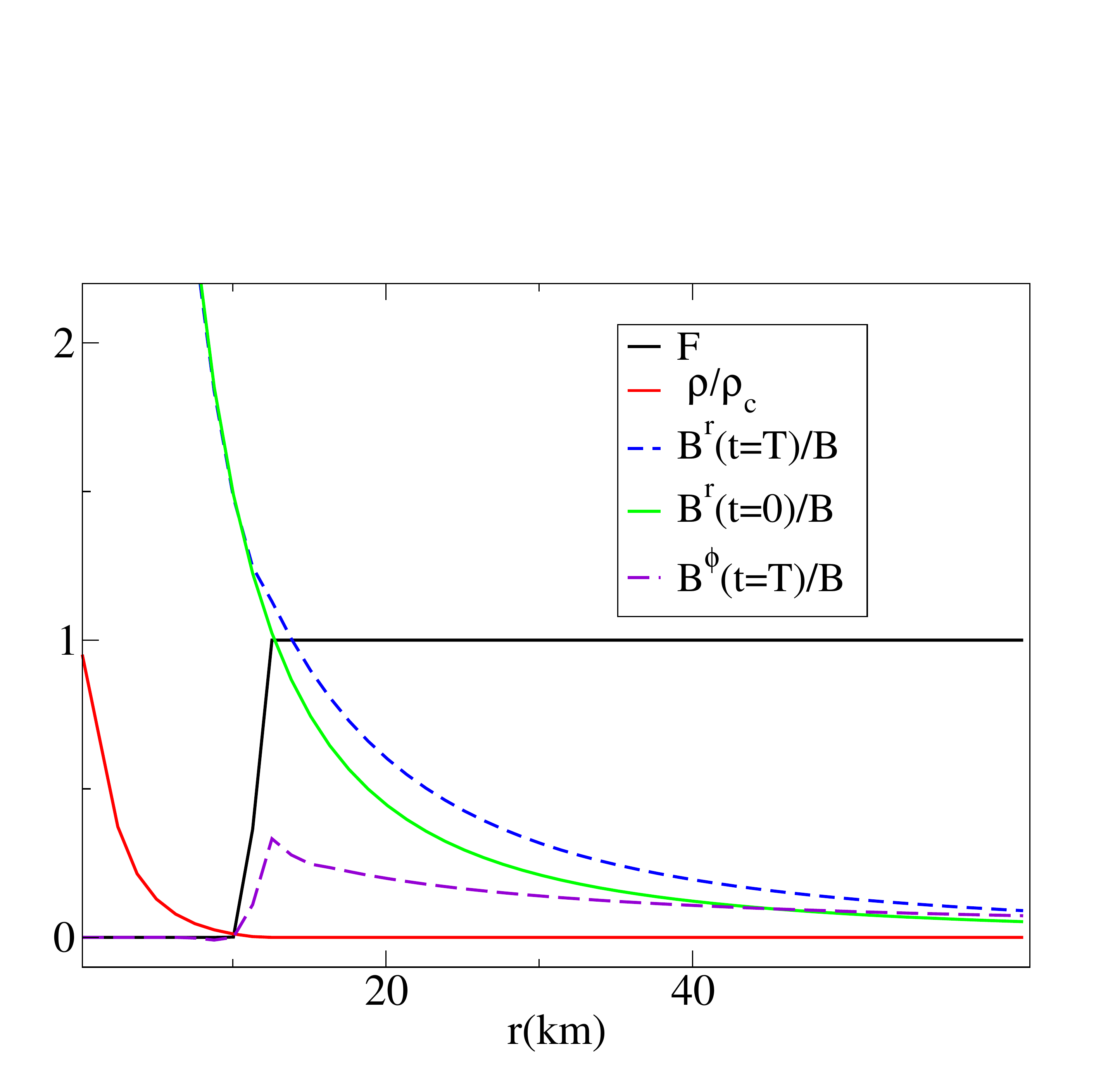,height=2.4in}
\caption{{\it Monopole solution}. Specific components of the magnetic field 
displayed along the $x=z=0$ line at the initial and final times, together with
the normalized density $\rho/\rho_c$ and the $F$ function. Notice that the
radial component has a smooth transition across the surface of the star, while
there appears a toroidal component in the magnetosphere.} 
\label{fig:monopole_bfield}
\end{center}
\end{figure}

\begin{figure}
\begin{center}
\epsfig{file=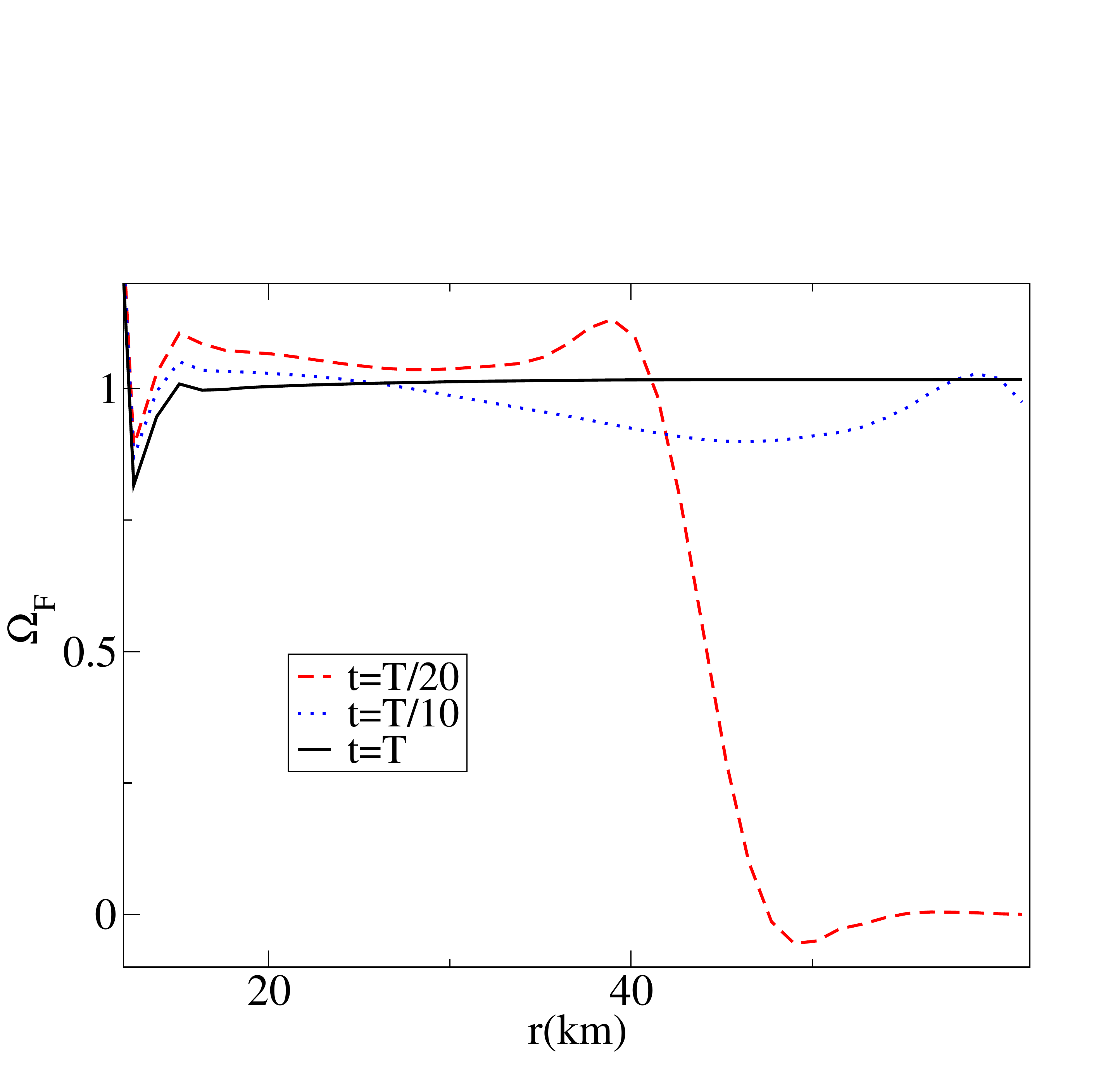,height=2.4in}
\caption{{\it Monopole solution}. The rotation of the magnetic field lines
(normalized with respect to its initial value) 
at different times inside and outside the domain, separated by the 
kernel function $F$ (continuous line). As time progresses, the rotation frequency
approaches the constant value expected for the monopole solution.} \label{fig:monopole_omega}
\end{center}
\end{figure}

The behaviour of different fields across the surface of the star after the solution has
settled are displayed in Fig.~\ref{fig:monopole_bfield},
which plots radial profiles of: the kernel function $F$,
the density $\rho_o$, and the non-trivial components of the magnetic field.
As illustrated in the figure, the radial magnetic field only changes outside the
star, where there also appears a toroidal component indicating the rotation of the
magnetic fields in the magnetosphere. 
Overall the electromagnetic fields are seen to relax
rather quickly to a state approaching the expected one. Another measure of
the obtained solution is given by the (normalized) rotational frequency of the magnetic
field $\Omega_F$ in Fig.~\ref{fig:monopole_omega}. As time progresses, this value
approaches $1$ as expected in a radially smooth way.

\subsection{Aligned rotator}
A particularly challenging test of our approach
is the aligned rotator solution~(see~\cite{Spitkovsky:2006np}). The aligned rotator
is a numerical solution of the force-free equations outside a rotating surface representing a star
that demonstrates closed field lines within the light cylinder (LC). This problem 
has been studied by a number of 
authors~\cite{Spitkovsky:2006np,2006ApJ...643.1139C,2006MNRAS.367...19K,2006MNRAS.368.1717B,2006MNRAS.368L..30M} 
working in flat space with a computational domain consisting only of the stellar exterior. In these
works the star's influence is accounted for through suitable boundary conditions derived from the
expected electromagnetic field at its surface.
These efforts were specifically motivated to obtain the solution in the magnetosphere of isolated
pulsars and, in particular, to understand a possible spin-down mechanism that works even when the star's dipolar
field is aligned with its angular momentum. 

As already mentioned, our approach differs in that we solve for the stellar interior, we work within curved space
and our computational grid does not conform to the star's geometry \footnote{This could be achieved 
by using multipatch methods~\cite{multipatch,Zink:2007xn} however.}.
Because of these differences and also because our initial data for the force-free fields is only nonzero in the immediate
neighborhood of the star, achieving the expected late-stage stationary solution
dynamically constitutes a demanding test.  We note that the time and length
scales required for such a test with respect to a realistic star are too computationally 
demanding if we are to capture the main physical aspects of the solution (i.e. field topology,
location of the light-cylinder, development of a current sheet, etc).
Instead of a realistic star, we adopt  a rapidly rotating
star so that the
light cylinder is brought closer to the star --where there is enough resolution to resolve
it-- and render the dynamical time scales shorter and easier to follow numerically.
To achieve such rapid rotation, we resort to an unstable star but prevent 
the instability from disrupting the star by artificially freezing both fluid and geometry. 
Consequently we only evolve the force-free equations and compare
the obtained configuration with the expected solution.

\begin{figure}
\begin{center}
\epsfig{file=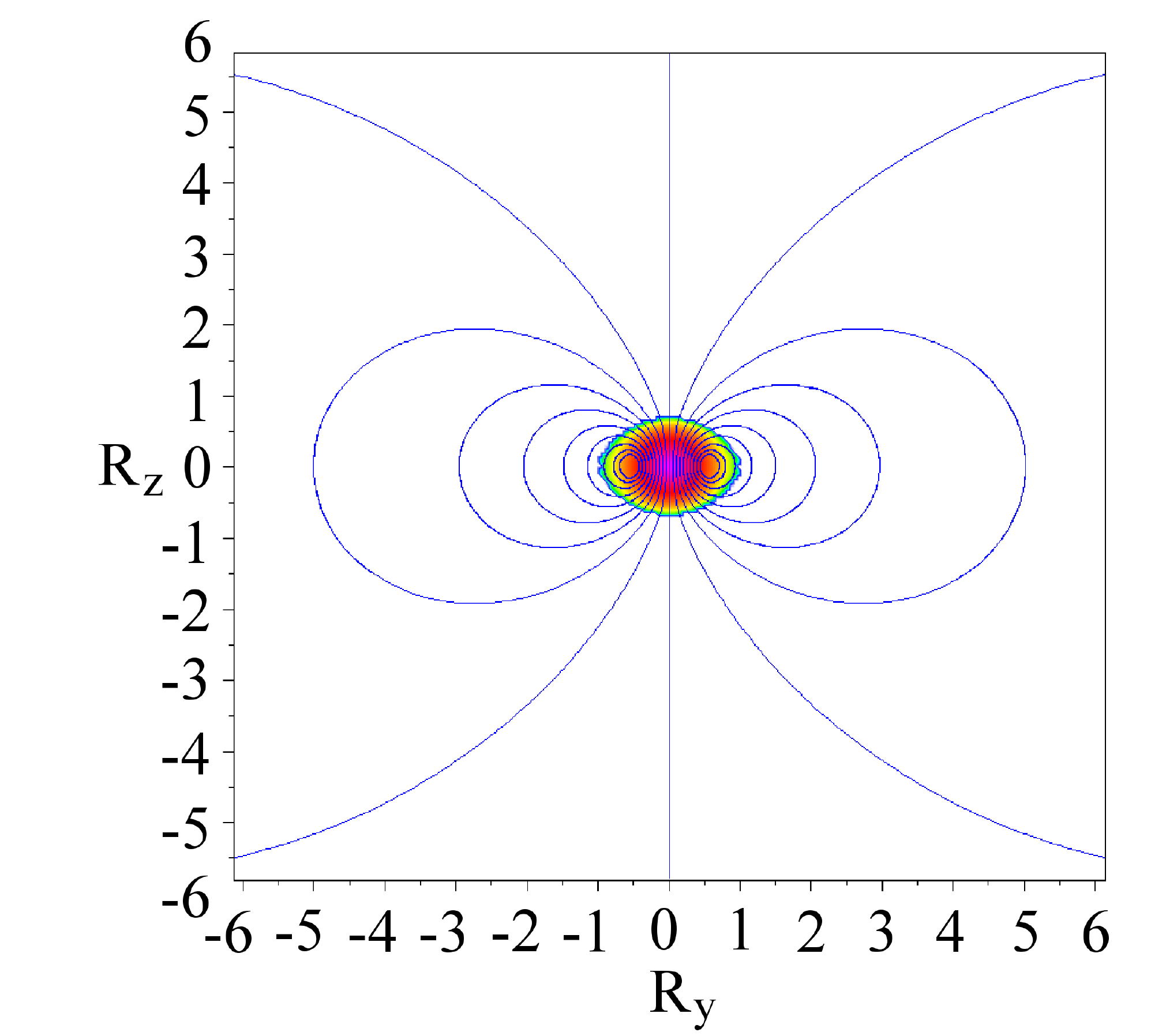,height=5.cm,width=4.8cm}
\epsfig{file=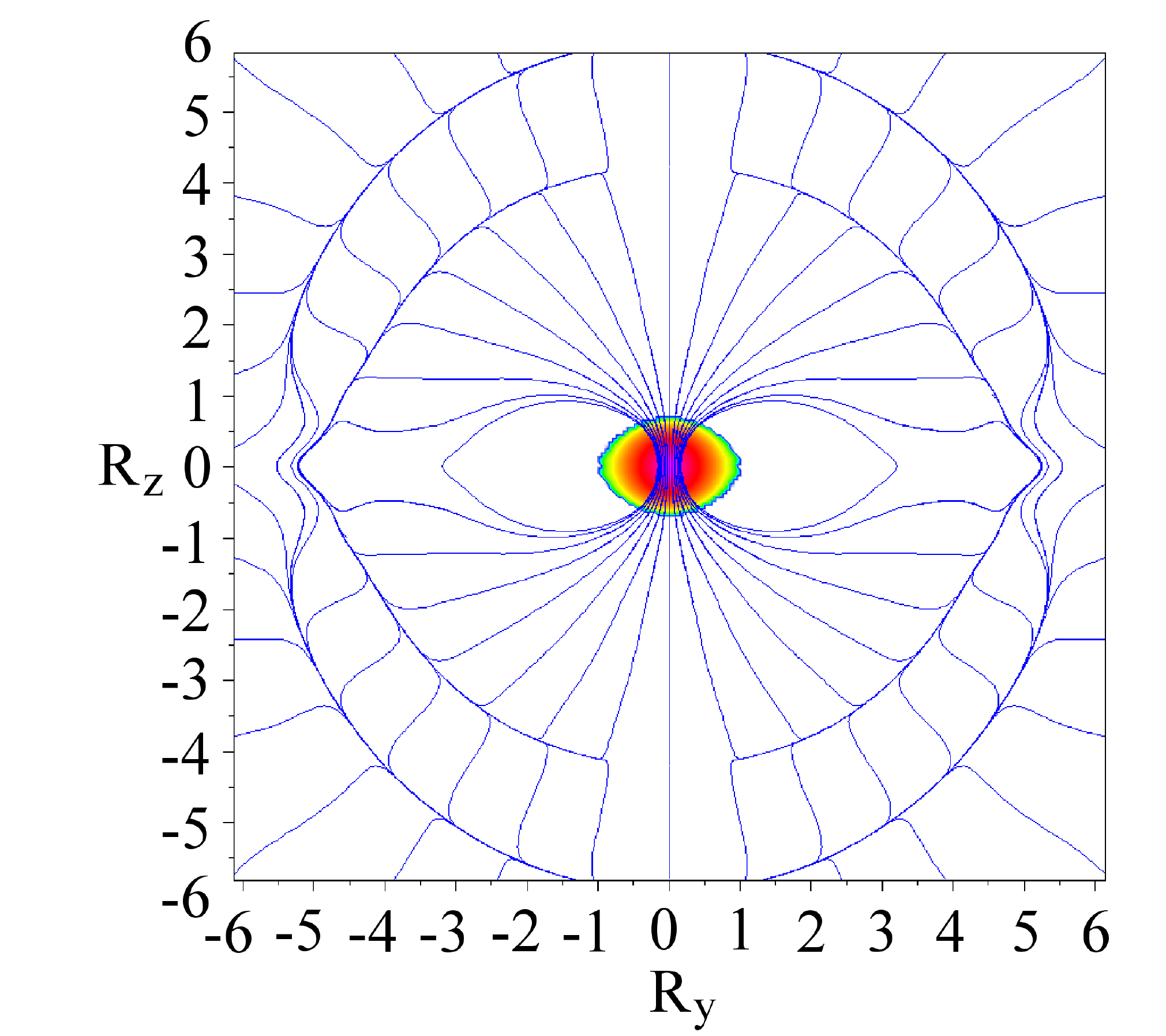,height=5.cm,width=4.8cm}
\epsfig{file=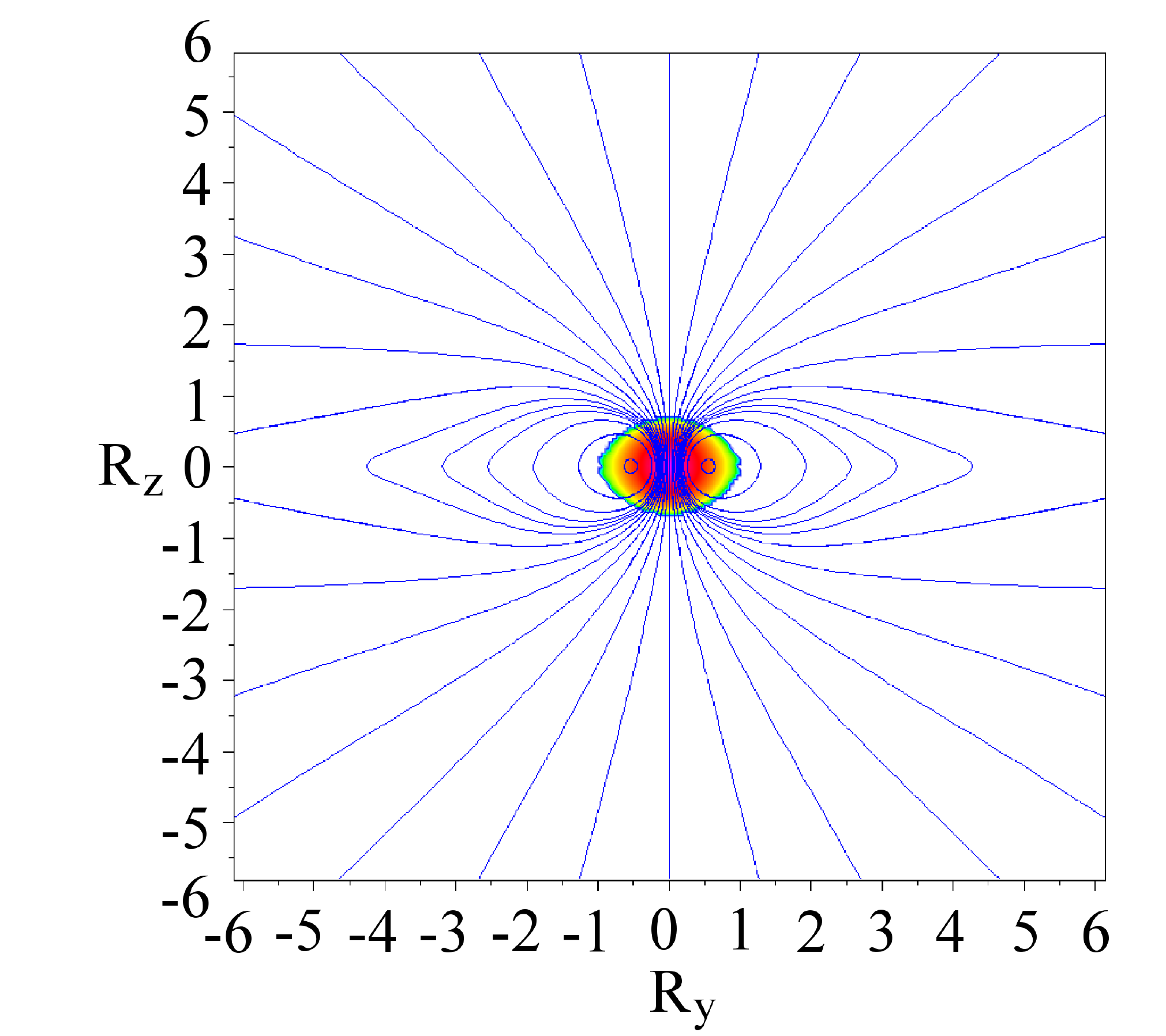,height=5.cm,width=4.8cm}
\caption{{\it Aligned rotator}. The fluid density and the magnetic field lines
on the $x=0$ plane at the four times $t =(0,1/3,2/3) T$. Even before
a complete rotational period, the solution exhibits the known properties of the aligned rotator
solution. The light cylinder is located roughly at the expected position
$R_{\rm LC} \approx 3.6\, R_s$ (large tick marks indicate one stellar radius $R_s$).
The intermediate plot illustrates the transient structure resulting as the LC forms and
the initial data, which only extends to $2 R_s$, relaxes to fill the computational domain.
These plots do not show the entire computational domain.
}
\label{fig:pulsar_bfield}
\end{center}
\end{figure}

We adopt a rotating star with a mass $M=1.7 M_{\odot}$,
equatorial/polar radius of $R_{\rm eq}=8.5/6.0~{\rm km}$ and rotational period 
$T=0.64~{\rm ms}$. The light cylinder for this star is located at
$R_{\rm LC} = c/\Omega = 31~{\rm km} \approx 3.6\, R_{s}$. The electromagnetic field is
initially set similar to that in the previous monopole test, but with a poloidal magnetic field given 
by a dipole outside the star. The evolution is performed in a cubic domain of
$L=184~{\rm km} \approx 22\, R_s$ with four FMR levels with resolutions
$\Delta x = (0.25, 0.5, 1.0, 2.0)~{\rm km}$. The FMR hierarchy consists of centered cubes
with side lengths $L= (2.8 , 5.6, 11.2, 22) R_{s}$, so that there are roughly 70 points across
the star in the equatorial plane. 

We evolve this star until
the solution settles to a quasi-stationary solution.
Fig.~\ref{fig:pulsar_bfield} illustrates the magnetic field
topology at three representative times.
In the last frame, we show the configuration to which the field settles, illustrating
features predicted by known solutions for the aligned
rotator. In particular, one can observe a region of closed field lines bounded at
the expected radius, denoting the light cylinder.

\section{Collapse of a magnetized star}
\label{sec:physics}

Having tested our hybrid approximation to EM field evolution, 
we now focus on the collapse of a magnetized star. The star may or may not be rotating,
the rotating case being of greater astrophysical relevance.  In particular,
we are interested in studying how the magnetosphere responds, and the distribution of escaping 
electromagnetic radiation.

Here it is essential to capture both the dynamics of the collapsing star and the surrounding magnetoplasma, since the two are tightly coupled.
It is worth recalling that the introduction of rotation 
introduces a fundamental difference in the radiative
output of stationary (non-collapsing) stars as calculated in the electrovacuum and force-free limits:  the
spindown torque of an aligned rotator vanishes in electrovacuum but does not in the force-free case. Furthermore,
this torque is strong and its magnitude varies only modestly with inclination angle~\cite{Spitkovsky:2006np}.

Previous studies of the magnetosphere of a collapsing star are fairly limited.
Some pioneering work in \cite{Baumgarte:2002vu} focused on the electrovacuum behavior
of a non-rotating star.   While this study included the effects of strong gravity,
by neglecting plasma or rotation it was not able to capture the winding of the magnetic
field during the collapse.   The analytic study by \cite{Lyutikov:2011tq} estimated
the EM output during the collapse by applying the formula for the spindown luminosity
of an equilibrium rotator, $L_{\rm sd} \propto B_s^2 R_s^6 \Omega^4$, where $R_s$ is
the stellar radius, $B_s$ the surface magnetic field, and $\Omega$ the spin frequency.  However, when the collapse time
is shorter than the rotation period (as must be the case if the star is to avoid a rotational hang-up),
the outer magnetosphere near the light cylinder is not able to follow the change in the surface 
magnetic field, and this equilibrium formula does not apply.  Instead, twisting of the
closed magnetic field lines is expected at a radius
 $r_{\rm max} \sim c t_{\rm col}/3$~\footnote{This is obtained by balancing the 
spin-up time $t_{\rm col}/2$ with the time for an Alfv\'en wave to propagate from 
the star out to $r_{\rm max}$ along a dipole field line.},
where $t_{\rm col} \equiv R_s/|\dot R_s|$ is the collapse time.  The corresponding 
toroidal magnetic field is $B_\phi \sim (\Omega r_{\rm max}/c) B(r_{\rm max})$, where
$B(r_{\rm max}) \sim (3 R_s/ct_{\rm col})^3 B_s$ in a dipole geometry.  The power injected
into the magnetosphere is
\begin{equation}\label{eq:lwind}
L_{\rm sd} \sim {1\over 6} B_s^2 R_s^2 c \left({\Omega R_s\over c}\right)^2 
                \left({3 R_s\over ct_{\rm col}}\right)^2
\end{equation}
which is larger than the equilibrium spindown power by a factor $\sim (\Omega t_{\rm col}/3)^{-2}$.  

To determine the dependence of $L_{\rm sd}$ on $R_s$, equation (\ref{eq:lwind})
must be combined with the conservation of
magnetic flux, $B_s R_s^2 = $ constant; and the appropriate scaling between
$\Omega$ and $R_s$ ($\Omega \propto R_s^{-2}$ for self-similar collapse).  The
collapse does not, in general, follow a simple power-law relation between
$t_{\rm col}$ and $R_s$, but in the special case of pressureless collapse from a
large radius ($t_{\rm col} \propto R_s^{3/2}$) one obtains $L_{\rm sd} 
\propto R_s^{-4}$.   It should be re-emphasized that equation (\ref{eq:lwind}) 
represents energy stored in the magnetosphere, and so the maximum amount of energy
that can only escape to infinity after the collapse is completed.

During the last, relativistic stages of the collapse, additional physical
effects arise. Spacetime curvature has a mixed effect on  magnetic 
field winding in the magnetosphere.
On the one hand, the rotation frequency is reduced with respect to the Newtonian
value;  on the other hand, torsional Alfv\'en waves in the force-free
magnetosphere slow down significantly as the horizon
approaches the surface of the star.

The net result, as we show, is that the magnetic field becomes strongly wound up if the
star is rotating close to breakup before the collapse, and develops $a/M \agt 0.5$ after
the collapse.  To clarify the influence of plasma and rotation, and to make useful
comparisons, we have made a parallel set of runs in the electrovacuum approximation.  These are
obtained straightforwardly by setting $J^i=0$ and not
enforcing the conditions $\{E.B =0, |E|<|B|\}$ in Eqs.~(\ref{maxwellext_3+1_eq1a}-\ref{maxwellext_3+1_eq1d}).

\subsection{Non-rotating stellar collapse}
\label{subsec:nonrotating}
The collapse of a magnetized, but non-rotating, neutron star was 
studied in~\cite{Baumgarte:2002b}, where a magnetic field frozen into
the stellar surface was matched to an exterior, vacuum solution of Maxwell's 
equations.   This calculation followed the transition to a black hole,
and the ringdown of the EM field threading the horizon.

In reality, the medium outside the star is an excellent electrical conductor.
As is argued in \cite{Goldreich:1969sb,1977MNRAS.179..433B}, a combination of large voltages and
strong gravitational fields will trigger runaway pair creation near compact objects
\cite{Goldreich:1969sb,1977MNRAS.179..433B}.  In 
a strongly dynamic situation, the number of particles
generated can be enhanced even further, e.g. by
a Kolmogorov-like transfer of energy from large-scale waves to internal 
plasma heat~\cite{Thompson:1998ss}. The remnant of a binary merger or a rapidly rotating stellar core
collapse is also a strong neutrino source and, therefore, its magnetosphere will be filled with
a much denser baryonic plasma compared with pulsar magnetospheres~\cite{1986ApJ...309..141D,
metzger:2007}.  But in most cases, the plasma around compact objects is magnetically dominated and
its dynamics is nearly force free.

\begin{figure}
\begin{center}
\epsfig{file=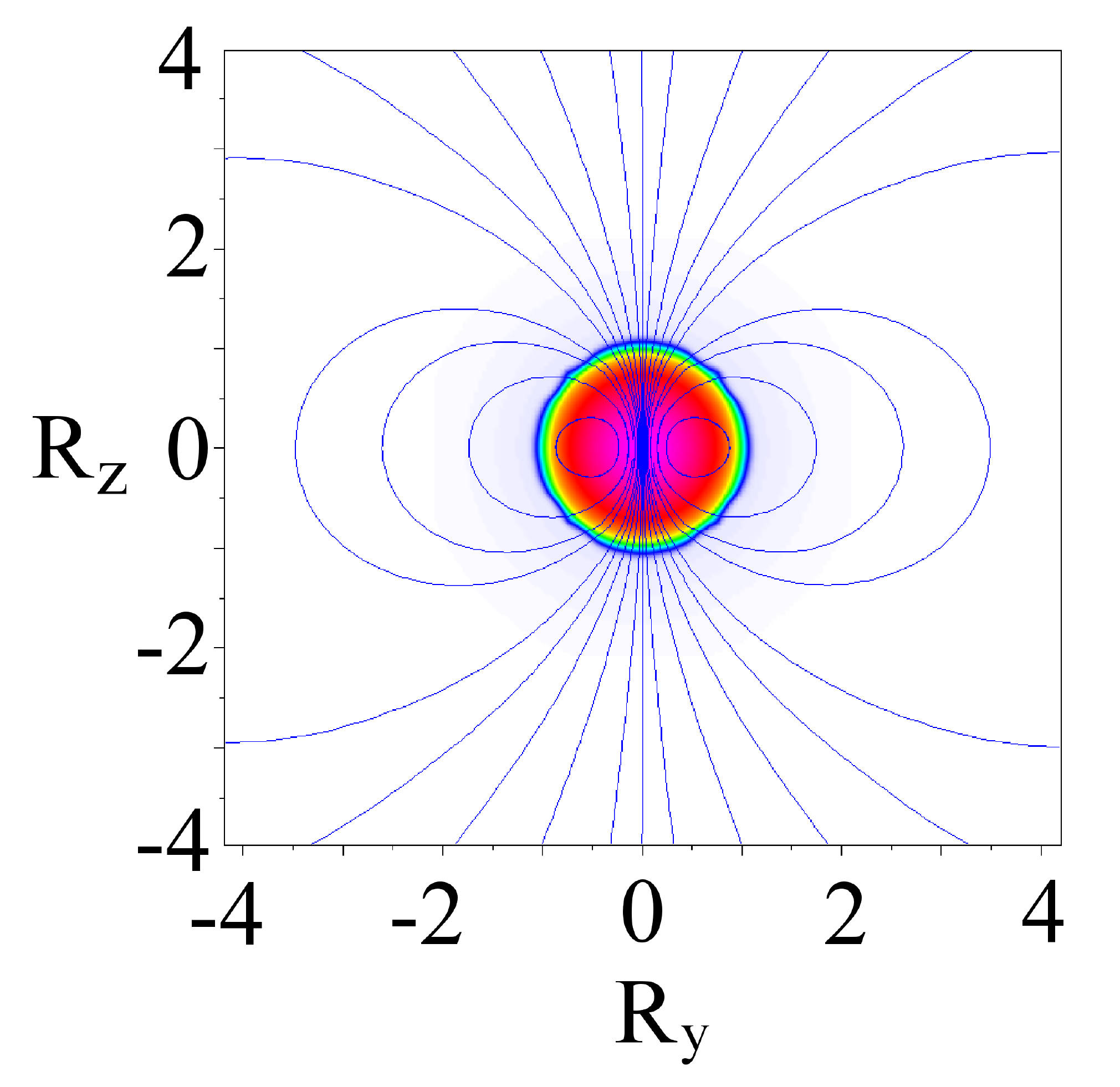,height=4.0cm}
\epsfig{file=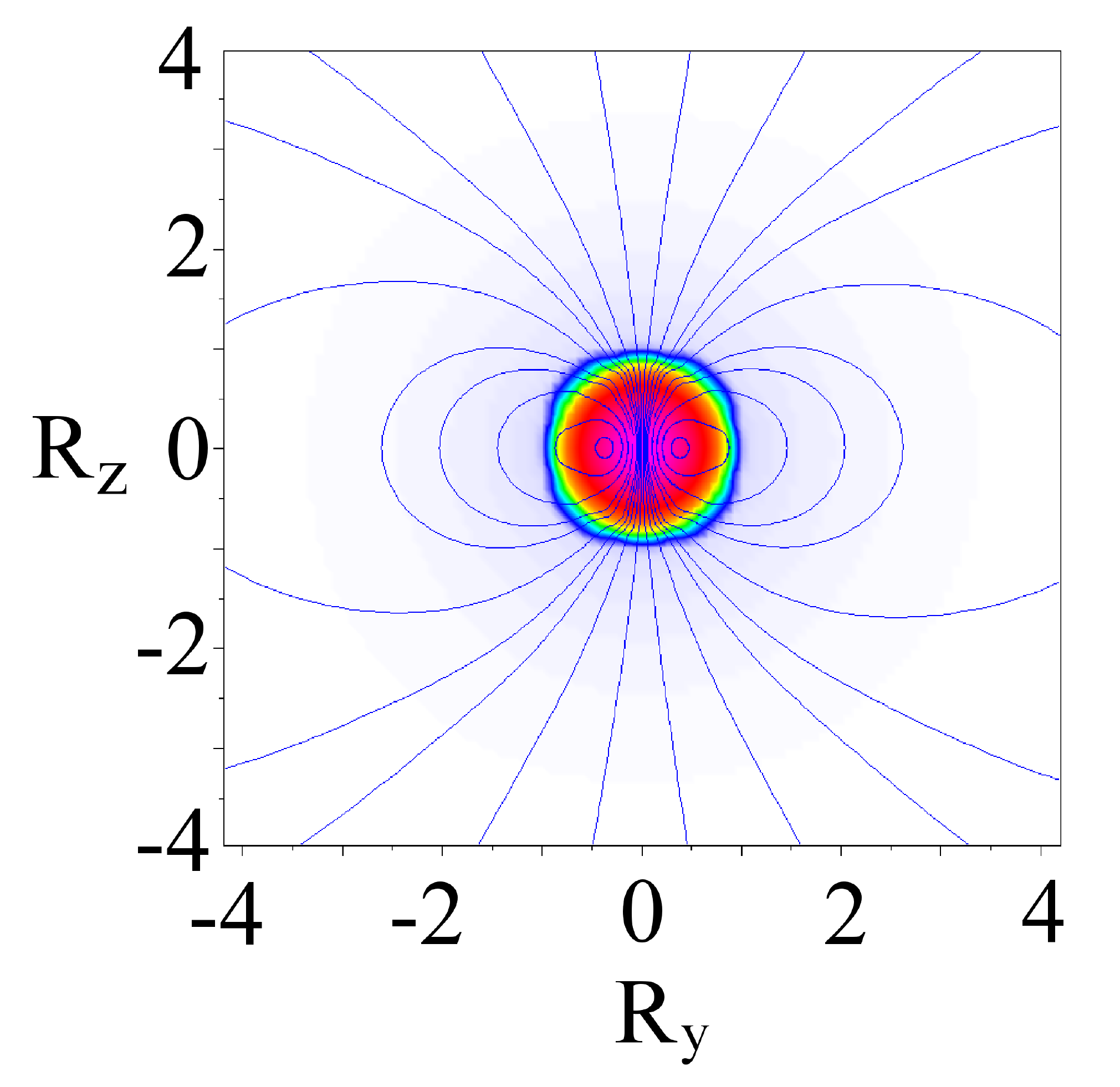,height=4.0cm} \\
\epsfig{file=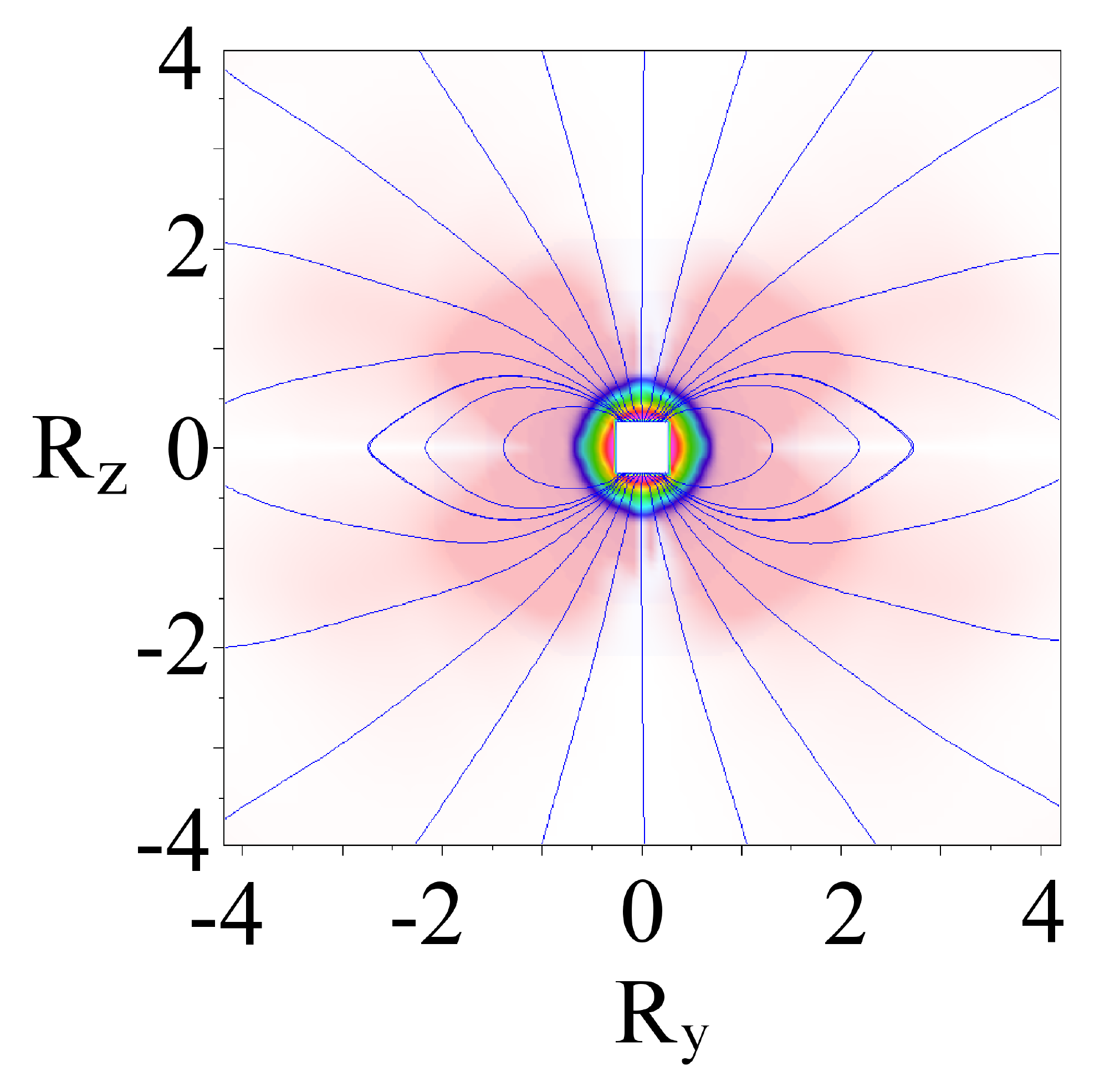,height=4.0cm}
\epsfig{file=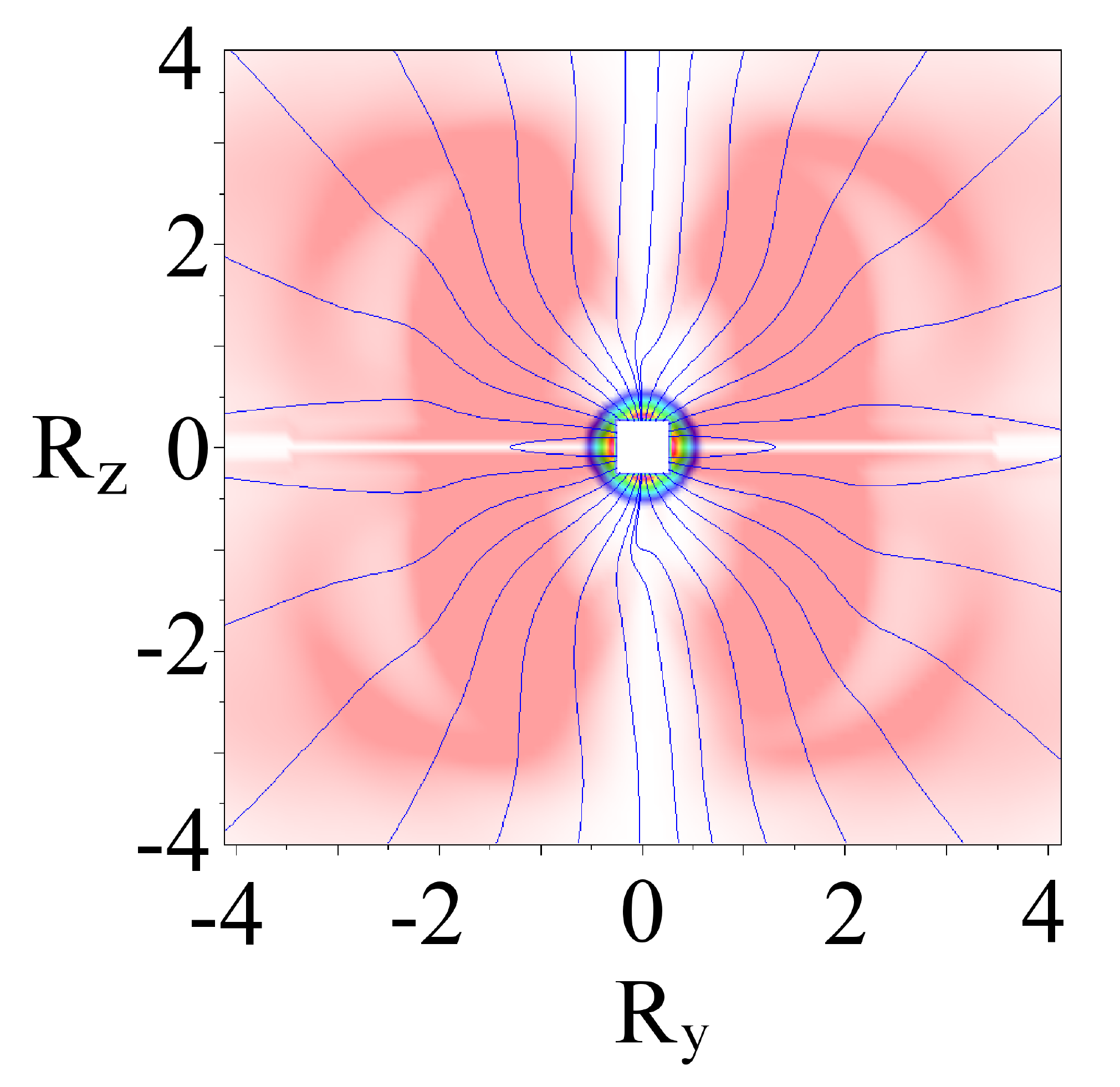,height=4.0cm} \\
\epsfig{file=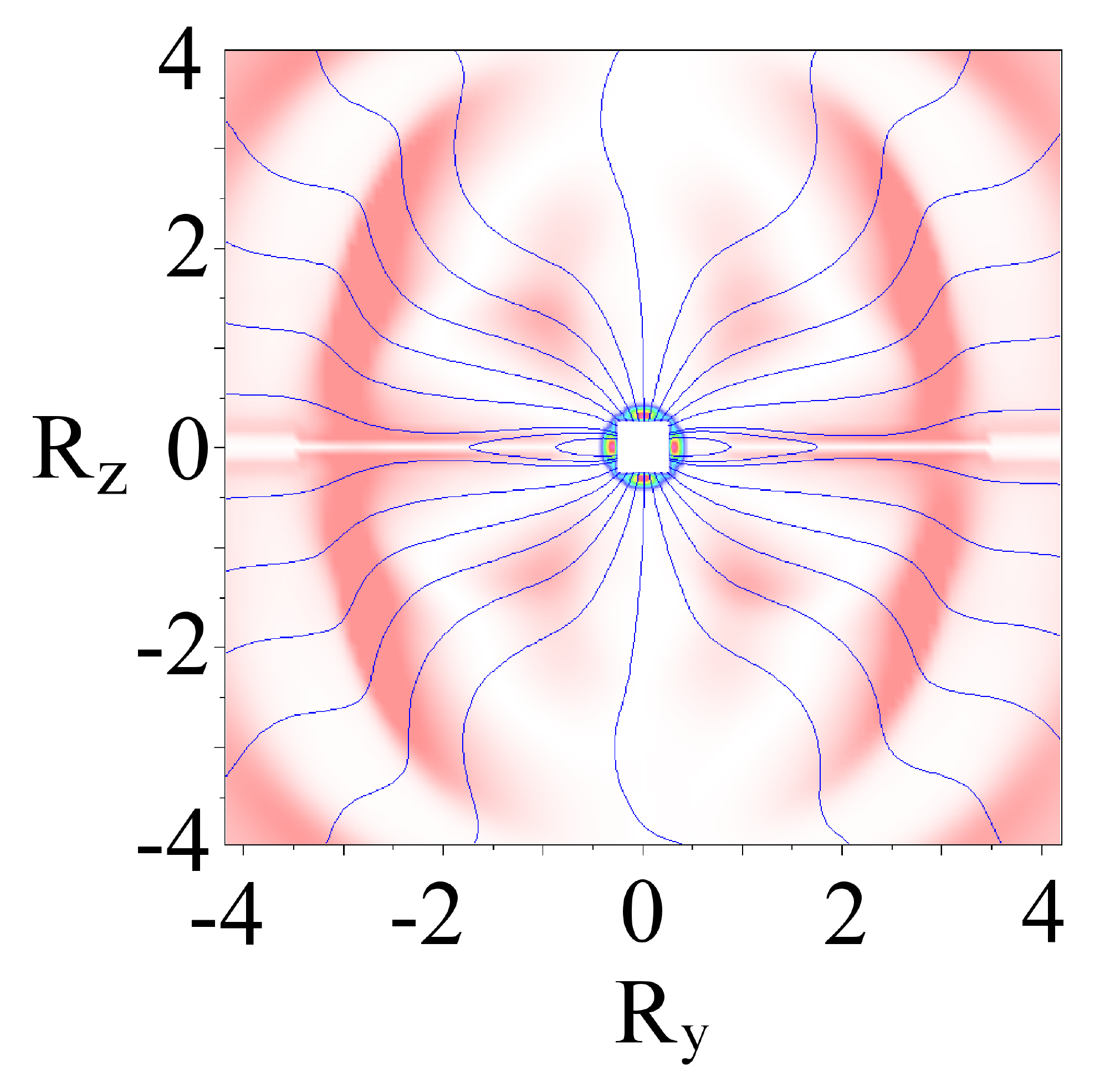,height=4.0cm}
\epsfig{file=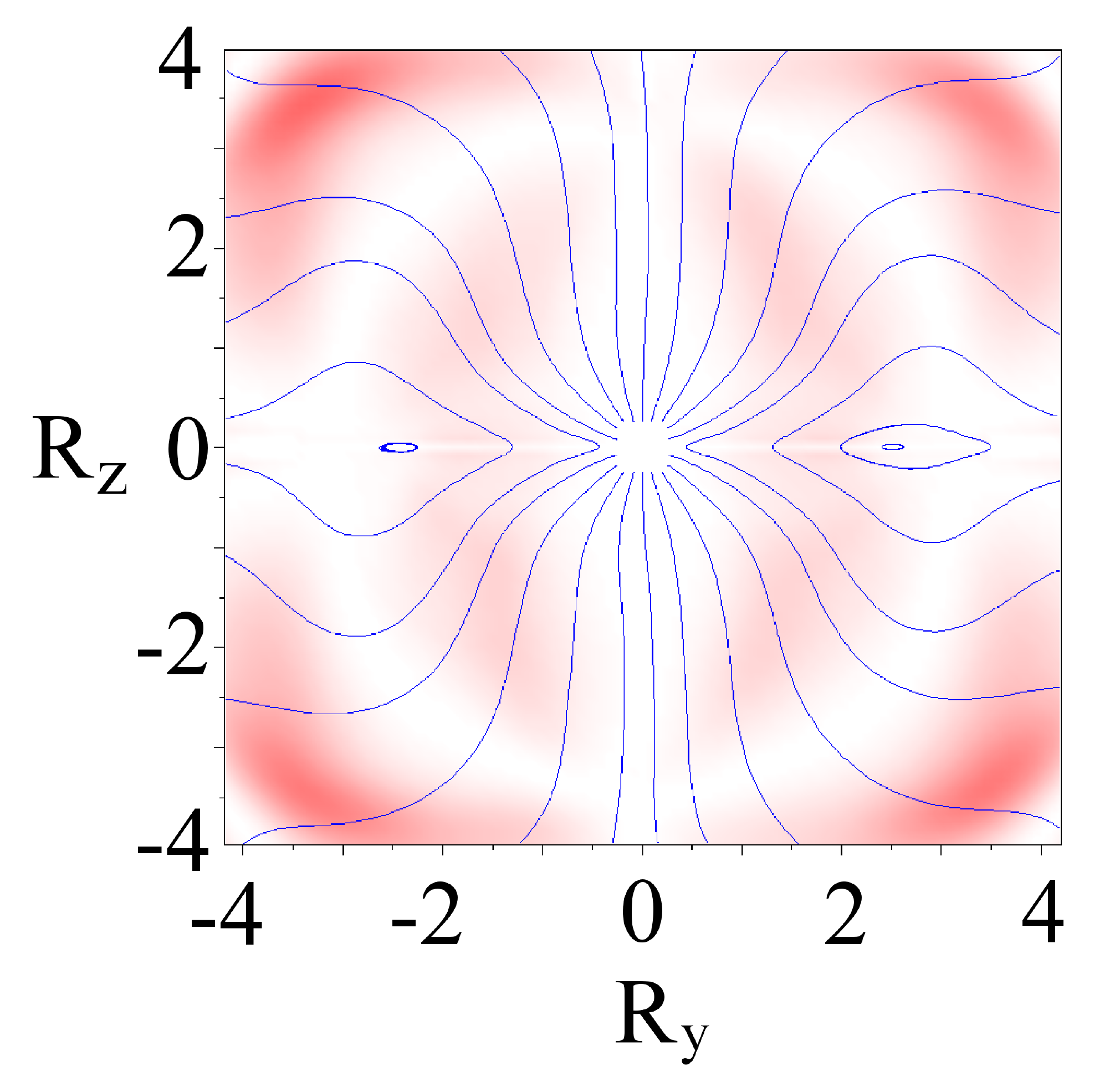,height=4.0cm}
\caption{{\it Non-rotating, unstable star (force-free)}. The fluid density (colors in
the central region), the
magnetic field lines (blue) and the EM radiation flux density (red) at  times
$t=(-1,\,-0.2,\,0.02,\,0.05,\,0.1,\,0.2)\,{\rm ms}$
(from left to right, top to bottom).
Here $t=0$ denotes the time that the horizon appears.
The magnetic field lines are dragged by the star during the collapse, producing
\Alfven waves in the magnetosphere that carry a small fraction of
the magnetospheric energy and stretch the magnetic field lines near the equatorial
plane. Most of the EM energy falls into the black hole (the solid white, central rectangle
denotes the excision region for the singularity).
} \label{fig:collapse_norot1}
\end{center}
\end{figure}
Therefore, we expect significant differences in the time evolution between
the electrovacuum exterior assumed by~\cite{Baumgarte:2002b} and a 
force-free magnetosphere.  To anchor the external force-free (or
electrovacuum) solution, we adopt a marginally unstable, non-rotating star 
with a mass $M_s= 1.63 M_{\odot}$ and a radius $R_s=8.62~{\rm km}$.
The numerical domain extends up to $L=16\,R_s$ and contains three centered FMR
grids of sequentially half sizes (and hence twice better resolved). 
The highest resolution grid has $\Delta x = 0.18~{\rm km}$.

The star collapses to a black hole in $\sim 1\, {\rm ms}$~ \cite{Liebling:2010bn}.
We set $t=0$ at the onset of an apparent horizon and display
the magnetic field and the density at various times in Fig.~\ref{fig:collapse_norot1}.
This force-free solution has several salient features.
There is an early transient in which the dipole magnetic field relaxes 
to a solution consistent with the physical configuration. 
Subsequently, as the star collapses,
the magnetic field lines are gradually stretched  along the equatorial
plane. After $~1.0\,{\rm ms}$ an apparent horizon appears in the interior of the star, which
grows as it swallows all the remaining fluid in $\sim 0.15\,{\rm ms}$.
As the outer layers of the star are accreted by the black hole,
the stretched magnetic field lines near the equatorial plane reconnect and form
closed loops that carry away electromagnetic energy and magnetic flux 
(note the field loops in the last frame of Fig.~\ref{fig:collapse_norot1}).

\begin{figure}
\begin{center}
\epsfig{file=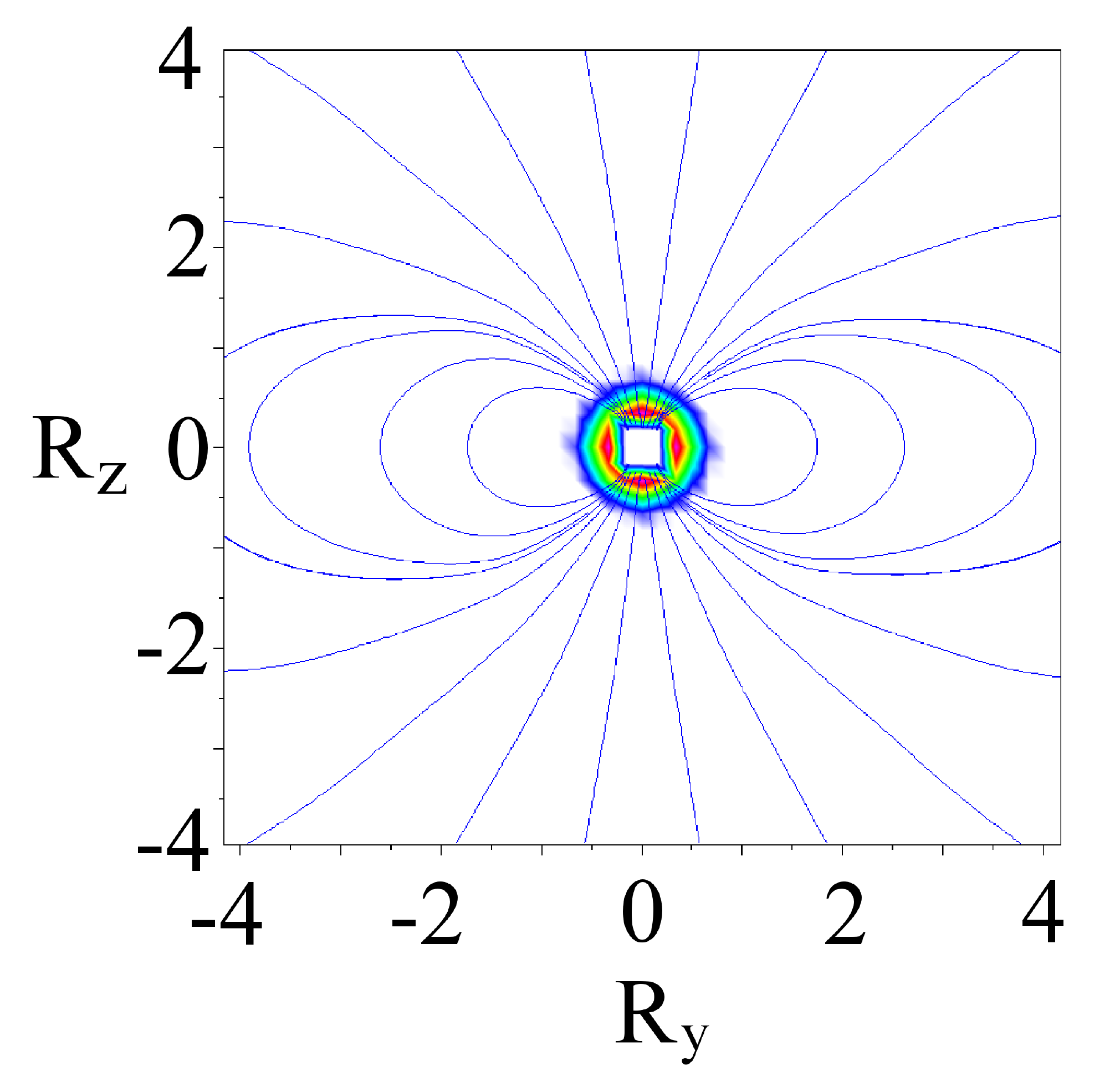,height=4.0cm} 
\epsfig{file=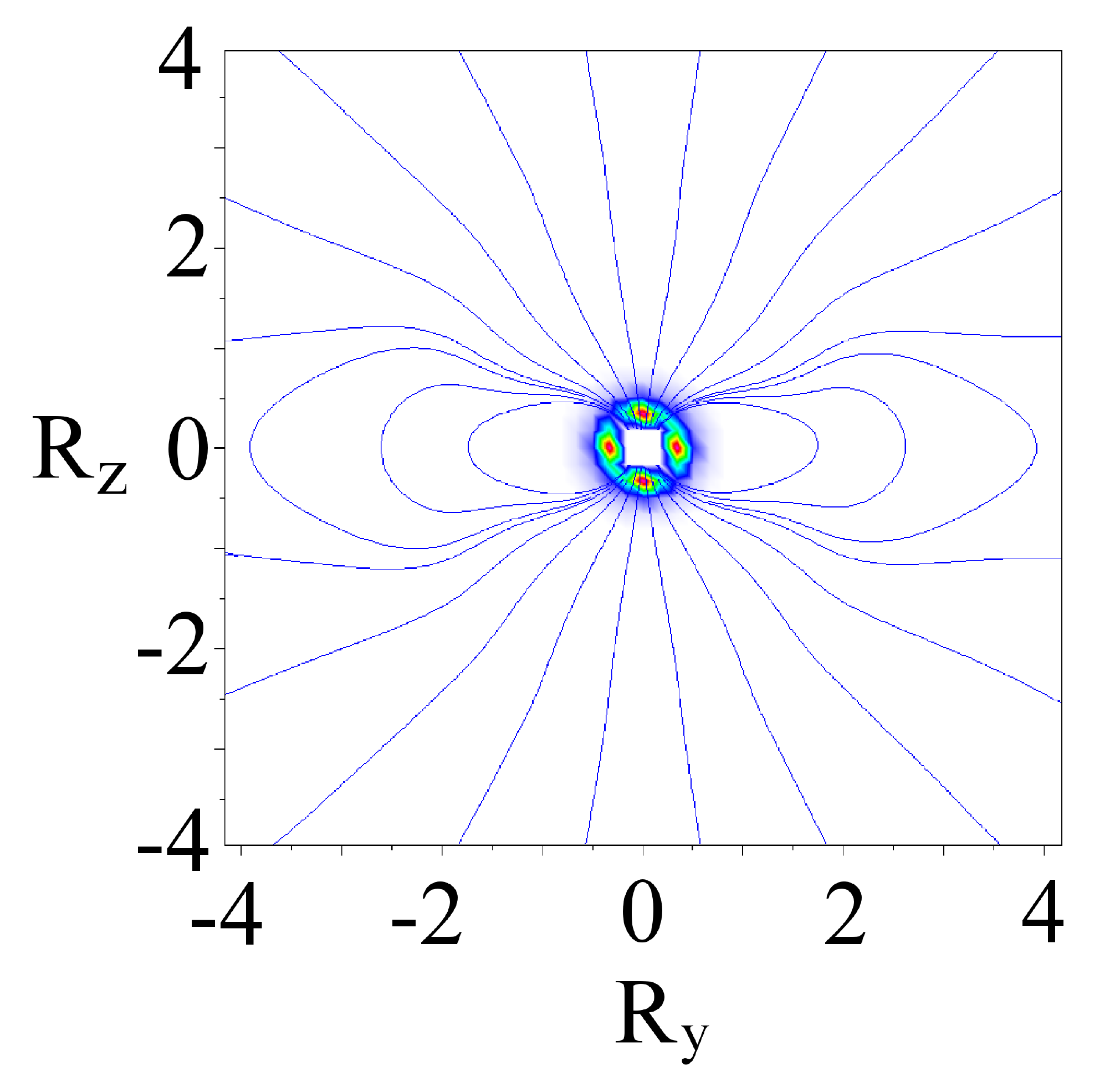,height=4.0cm} \\
\epsfig{file=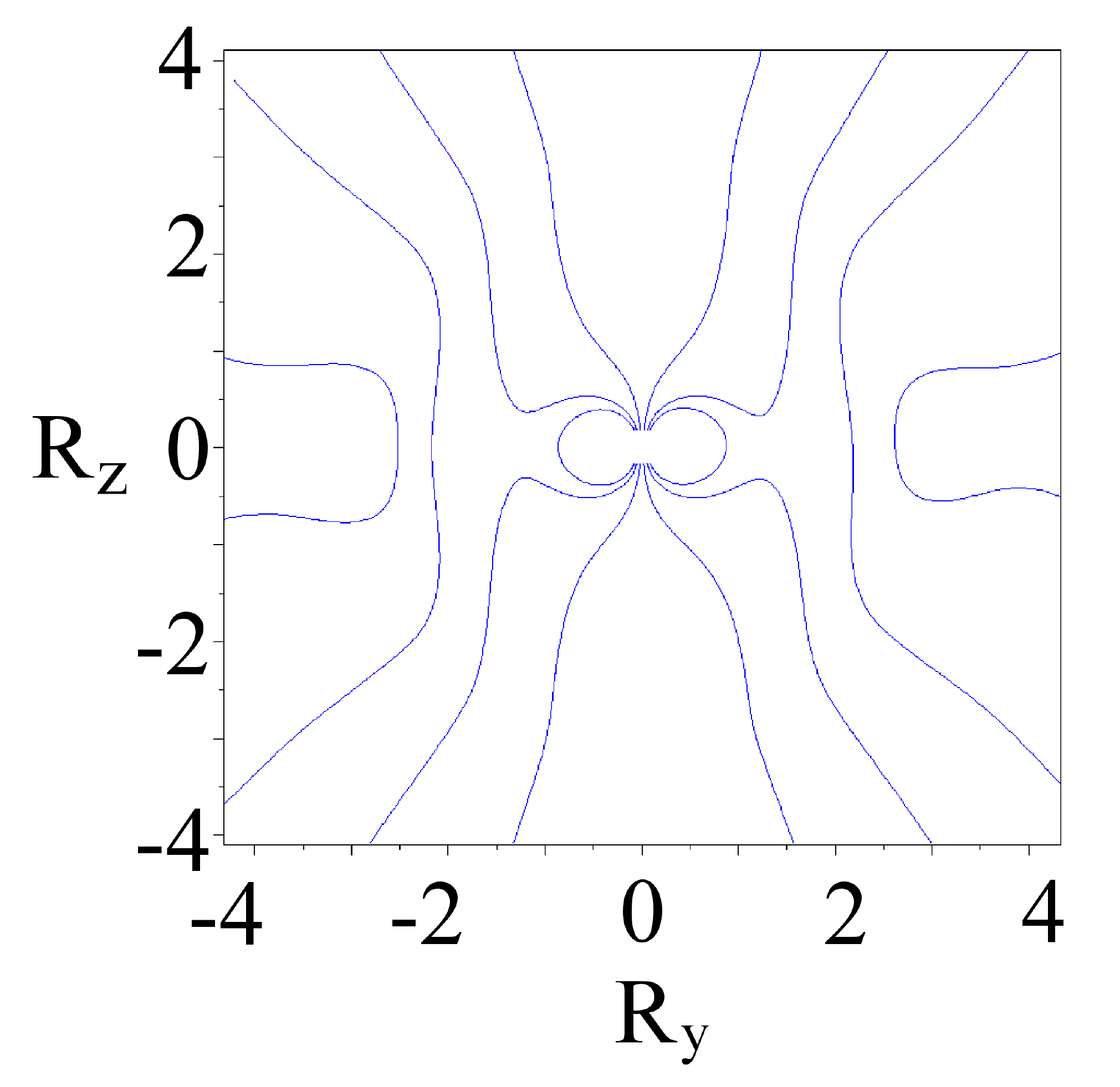,height=4.0cm} 
\epsfig{file=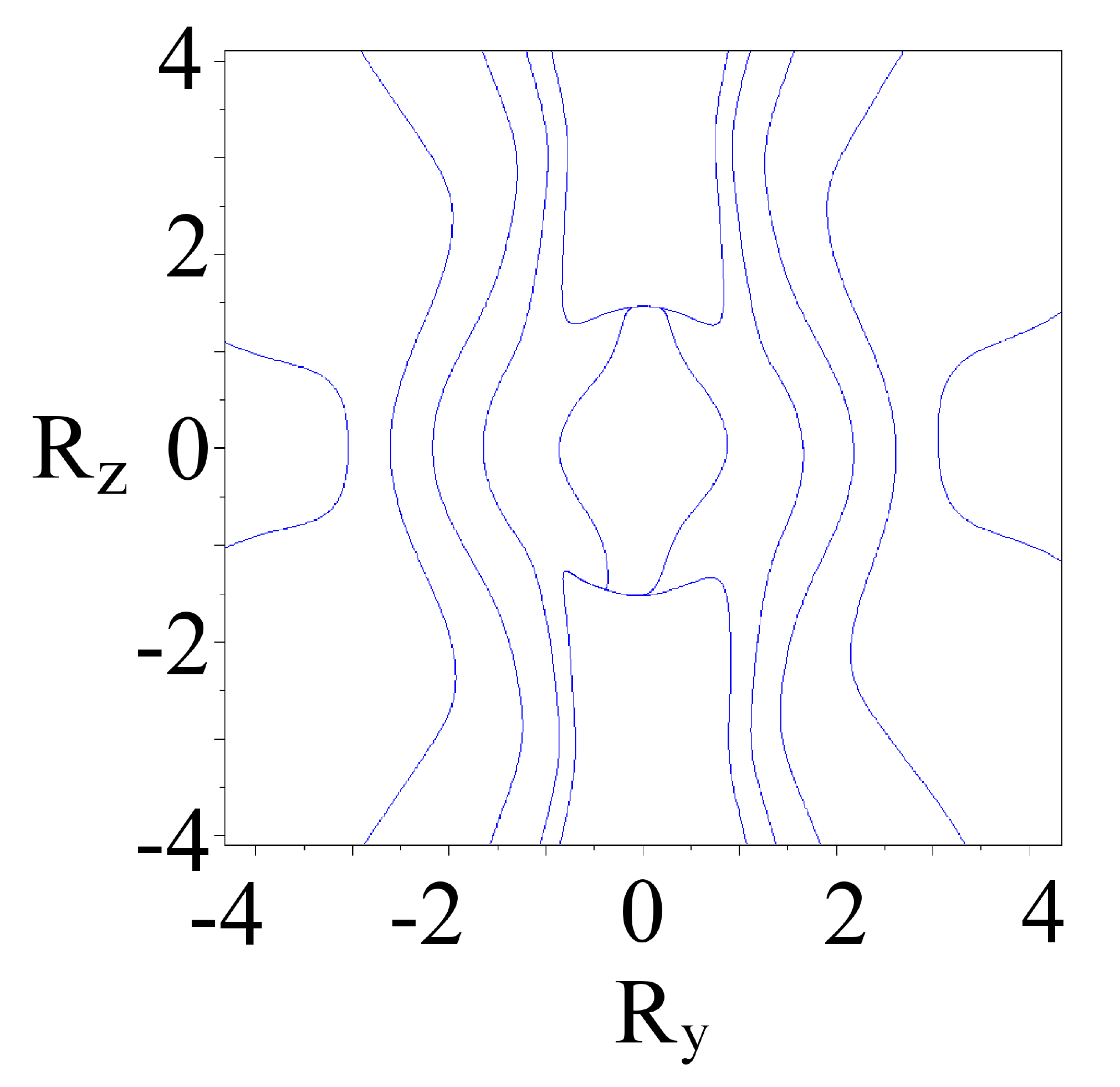,height=4.0cm} \\
\epsfig{file=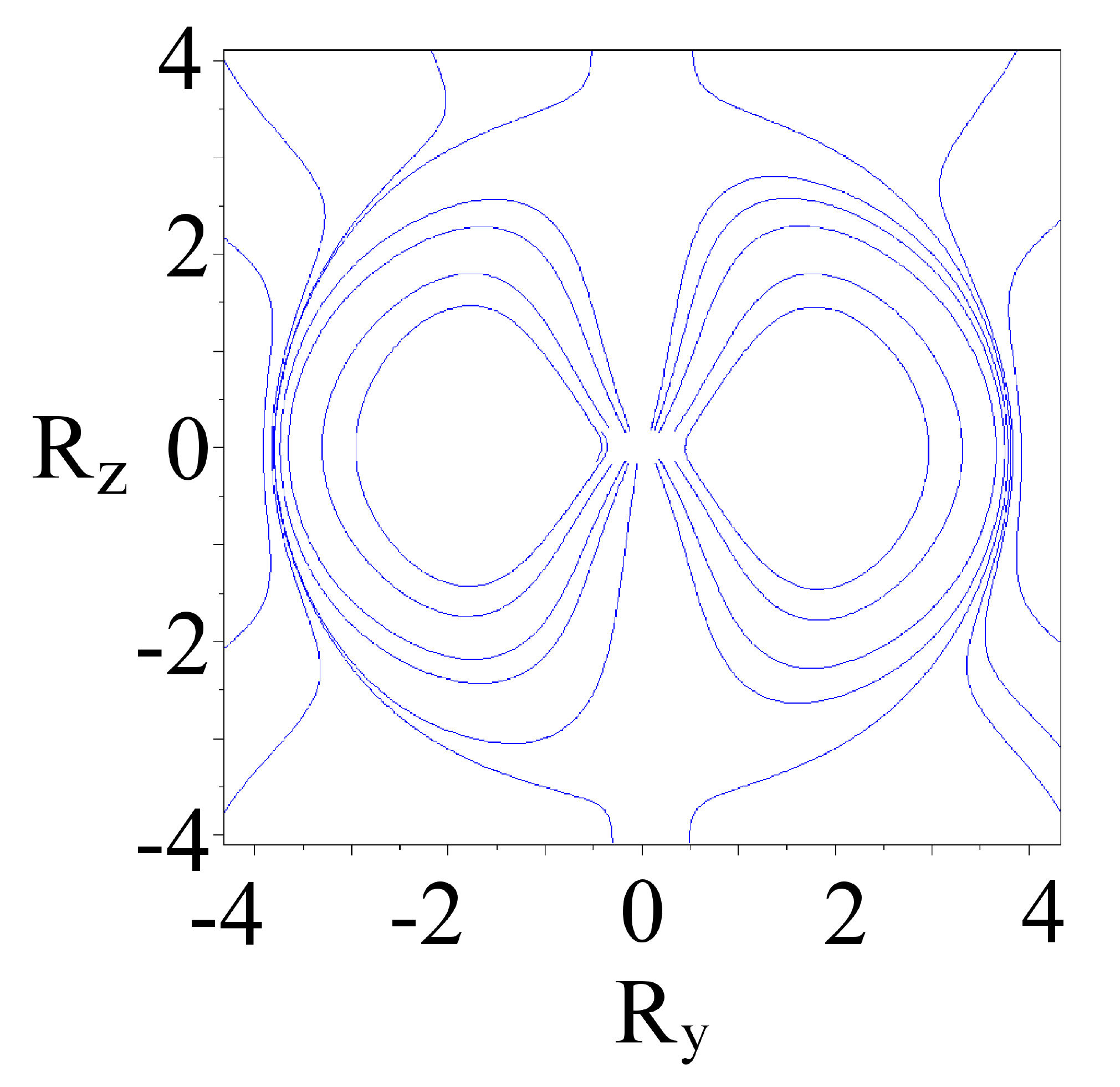,height=4.0cm} 
\epsfig{file=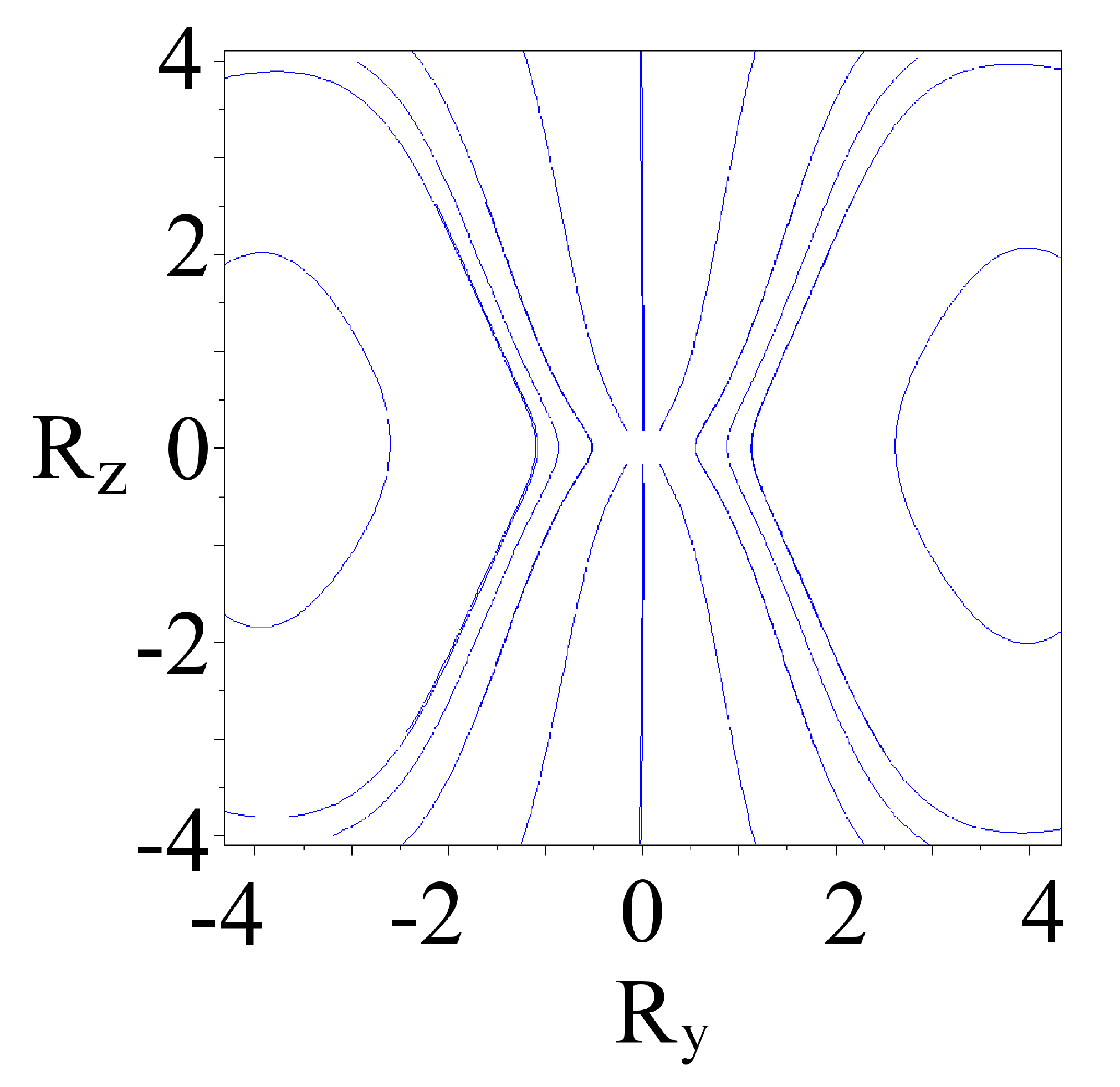,height=4.0cm} \\
\caption{{\it Non-rotating, unstable star (electrovacuum)}. 
The fluid density (colors in the central region) and the
magnetic field lines (blue) at 
$t=(0.03,\,0.06,\,0.12,\,0.125,\,0.13,\,0.15)\,{\rm ms}$.
The magnetic field lines are dragged by the star during the collapse,
stretching the magnetic field lines near the equatorial plane until that
the fluid is swallowed by the black hole. Afterward, the EM dynamics is
mostly described by the Quasi Normal Modes of the system.
} \label{fig:collapse_norot2}
\end{center}
\end{figure}

\begin{figure}
\begin{center}
\epsfig{file=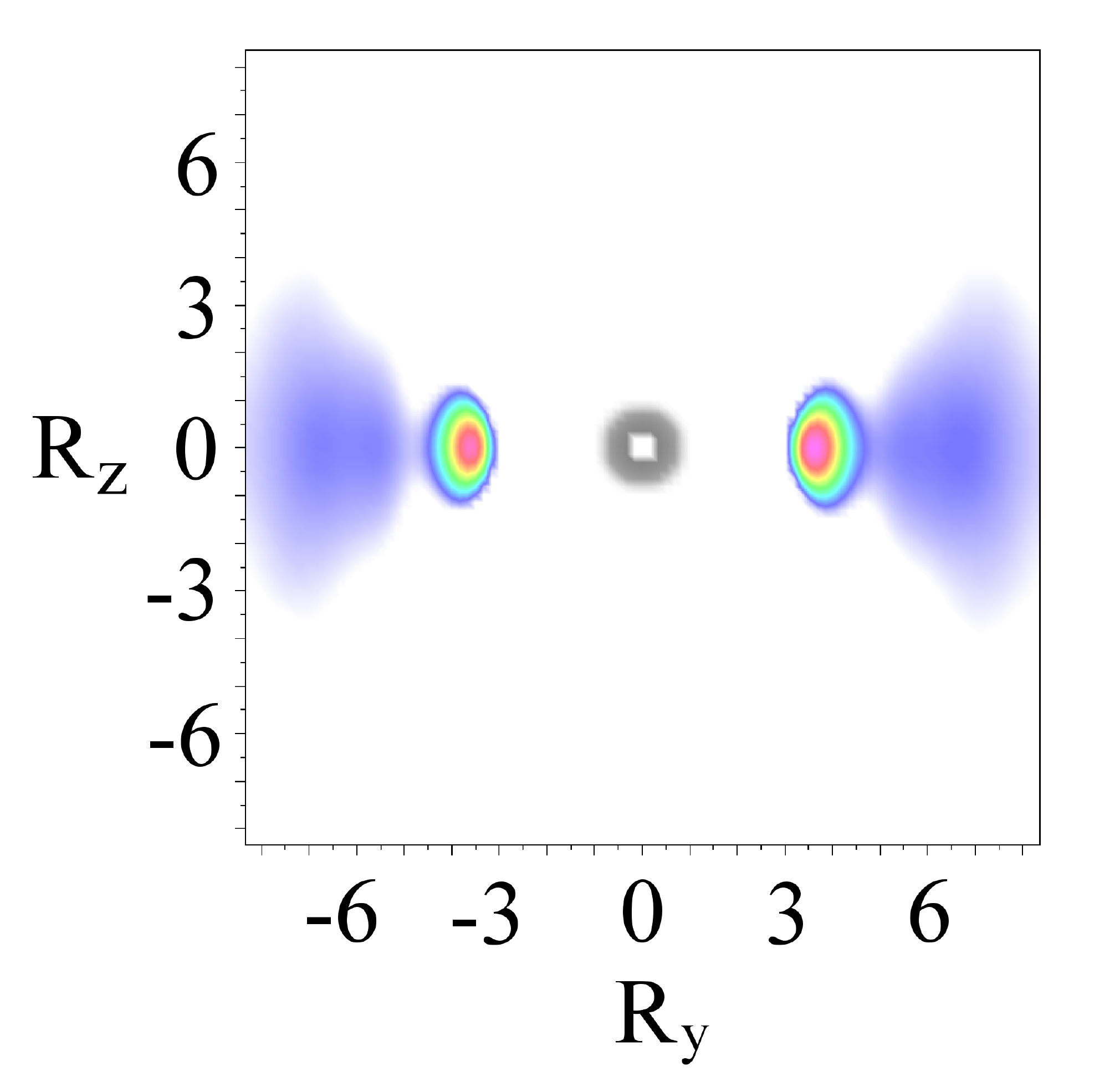,height=4.0cm} 
\epsfig{file=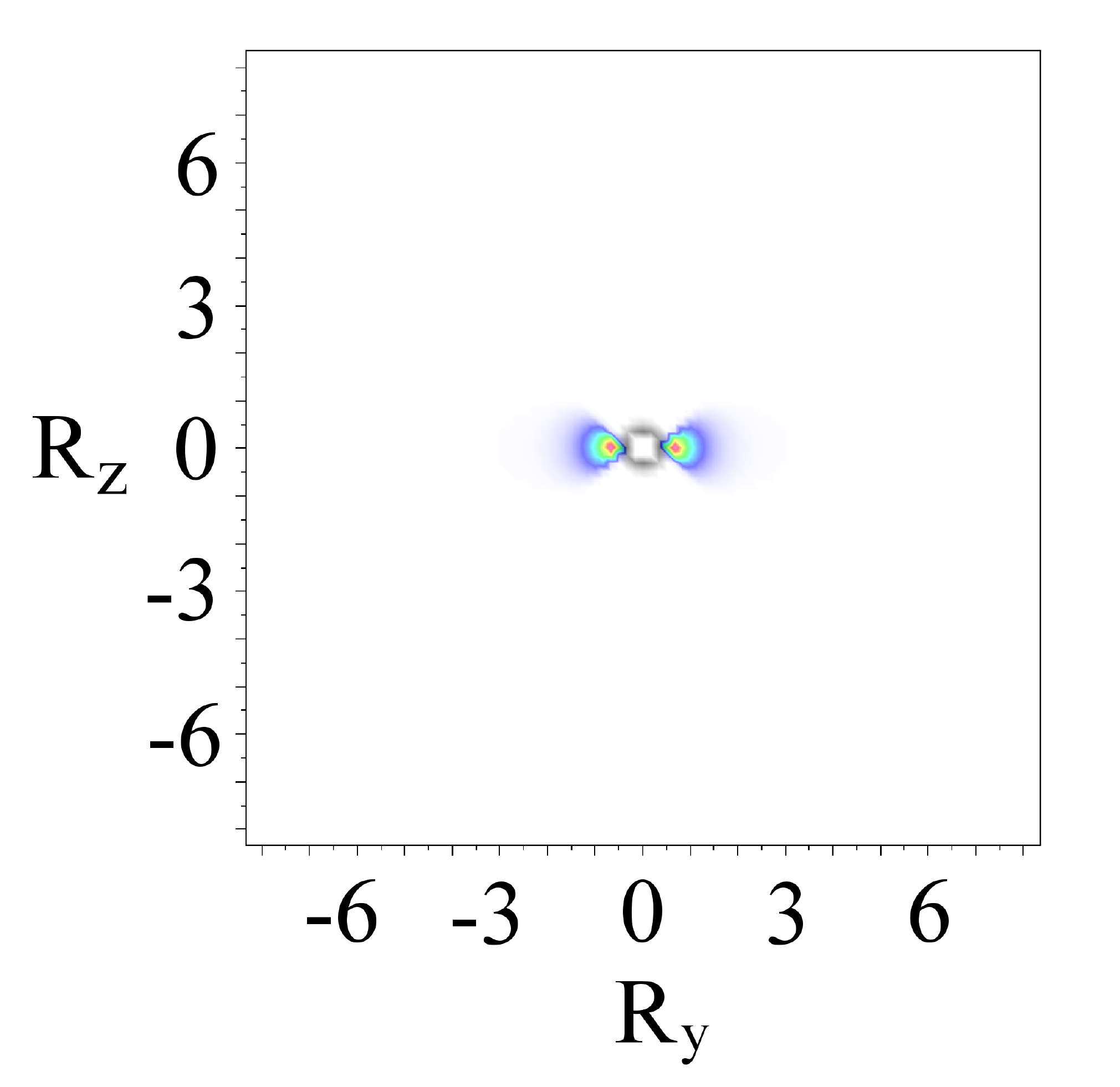,height=4.0cm} \\
\epsfig{file=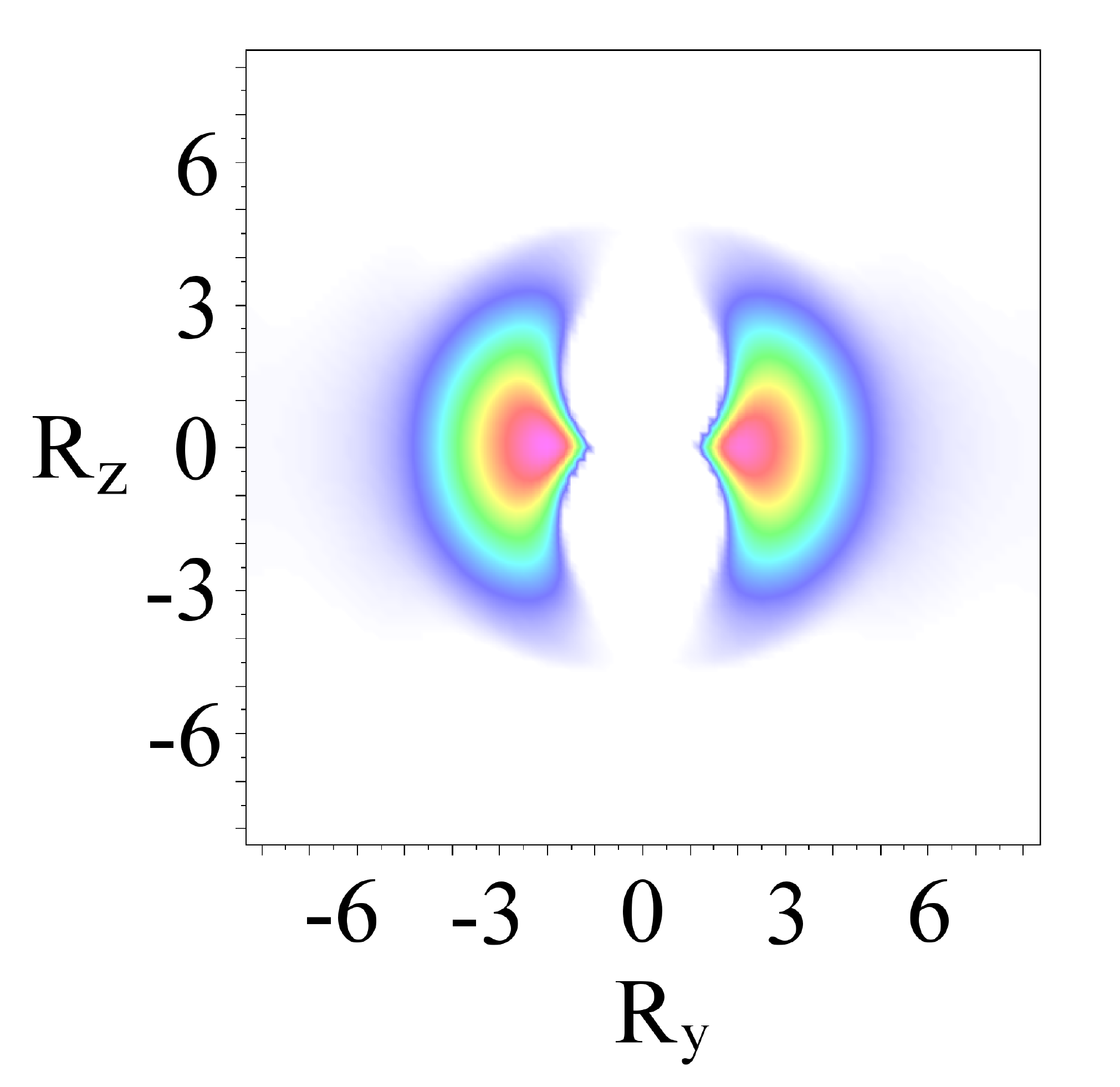,height=4.0cm} 
\epsfig{file=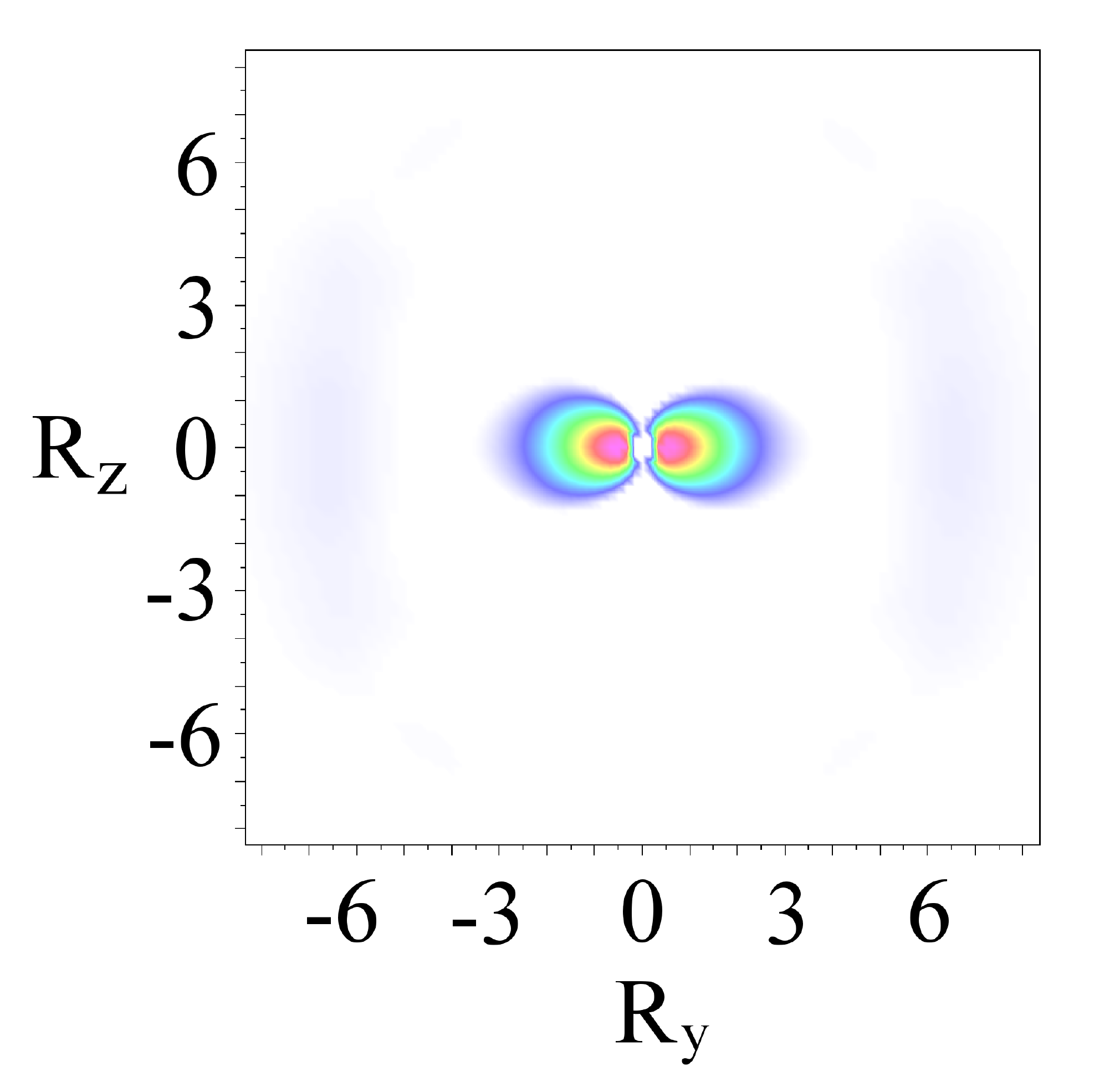,height=4.0cm} \\
\caption{{\it Non-rotating, unstable star (electrovacuum)}. 
Electrically-dominated regions ($E^2 > B^2$) are marked in color,
at times $t=(0.04,\,0.13,\,0.22,\,0.31)\,{\rm ms}$.
Here currents arise in the force-free case, and $E^2 < B^2$ is maintained
through gradual dissipation, resulting in a fundamental different with the
electrovacuum evolution.  As described by the simple Newtonian model,
these regions form near the equatorial plane and close to the collapsing
star. As time progresses, they propagate outward in bursts.  The grey 
zone in the center represents the star.
} \label{fig:collapse_norot3}
\end{center}
\end{figure}

The EM field evolution shows qualitative differences
in the force-free and electrovacuum runs,
especially after the black hole forms (Fig.~\ref{fig:collapse_norot2}). 
Starting from a common dipole structure at the beginning of the collapse, 
the magnetic field in the force-free case 
becomes radially stretched near the magnetic equator, and maintains
a consistent sign.  In the electrovacuum evolution, the stellar magnetic 
field disconnects more readily from the exterior.   Changes in the
connectivity of the magnetic field follow the appearance of
zones where $E^2 > B^2$, as is demonstrated
in Figs.~\ref{Bsq_Esq} and \ref{fig:collapse_norot3}.

\begin{figure}
\begin{center}
\epsfig{file=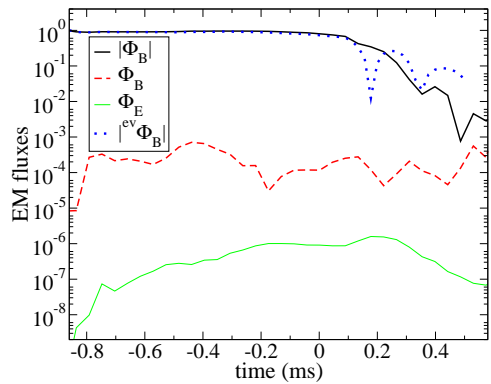, width=2.7in} 
\caption{{\it Non-rotating, unstable star}. The absolute value of the
magnetic flux in the electrovacuum case as a function of time,
and the electric and magnetic fluxes in the force-free case. 
These quantities are integrated over central spheres with radius $r=1.5 R_s$ and
normalized with respect to the initial integral of $|\Phi_B (t=0)|$. 
The total signed fluxes remain very small throughout the simulation, and the
unsigned magnetic flux  decreases as the black hole swallows all the matter which
anchors the magnetic field.}
\label{fig:collapse_EMfluxes}
\end{center}
\end{figure}

For a more quantitative discussion, we consider in 
Fig.~\ref{fig:collapse_EMfluxes} the electric and magnetic fluxes, computed 
over a surface located at $r=1.5 R_s$, as a function of time. 
As expected, both the total (signed) fluxes remain small throughout
the simulation, indicating that essentially no spurious magnetic/electric charges are 
created during the collapse.  The unsigned magnetic flux decays
exponentially in both the electrovacuum and 
force-free cases.  Interestingly, the decay rate in the former case 
matches the $l=1,m=0$ quasi-normal mode 
for electromagnetic perturbations of a Schwarzschild black hole with
mass $M=1.63 M_{\odot}$~\cite{Berti:2009kk}.

The decay rate of the magnetic flux in the force-free run is
roughly twice the electrovacuum result.
One could translate the
measured e-folding time $t_E$ into an ``effective reconnection speed,'' 
$V_{\rm rec} \approx GM c^{-2} {t_E}^{-1}$ by observing
that reconnection takes place within $r \simeq (1 \rightarrow 2) r_H$ (with $r_H$ the horizon radius).
Our results indicate $V_{\rm rec} \simeq 0.14 c$.

This result may appear paradoxical at first sight, and deserves some
comment.  Electrovacuum magnetic fields effectively reconnect 
(by converting to electric-dominated fields) at the speed of light, 
as we discuss in more detail in Sec.~\ref{s:quasi}.
Why then is the decay faster in the force-free case?  
The answer appears to reside in the self-inductance of the black hole.
As magnetic field lines reconnect through the equator, they generate
a strong toroidal EMF.  In a vacuum, this EMF sources a magnetic flux 
of the opposing sign.  When conducting matter is present, there is
no oscillation in the sign of the flux threading the hole.  Rather,
reconnection is a monotonic process and, after the formation of an
x-point, the magnetic field lines interior to the x-point fall through
the horizon.

\subsubsection{Electromagnetic Output}

In order to estimate the efficiency with which EM energy is radiated
to infinity, we have scaled the time-integrated luminosity to 
the peak electromagnetic energy contained in the magnetosphere.  
This peak energy is reached approximately at the formation
of the apparent horizon. In the absence of rotation, it is
$C_{\rm peak} \sim 2$ times the initial dipole energy, as determined
by the conservation of magnetic flux. The measured radiated energy is, 
in turn, a fraction $\epsilon_{\rm rad}$ of $C_{\rm peak} E_{\rm dipole,0}$.
Numerically, $E_{\rm dipole,0} \approx (2\pi/3)(B_{\rm pole}^2/8\pi) 
R_s^3 = (1/12) B_{\rm pole}^2 R_s^3$, and one finds $E_{\rm dipole,0} = 
1.4 \times 10^{47} B_{\rm pole,15}^2 \, {\rm erg}$ for a (initial) polar field 
$B_{\rm pole}$ and radius $R_s \approx 12~{\rm km}$.  Hence
\begin{eqnarray}\label{energy_peak_magnetosphere}
   E_{\rm rad} \approx
   1.4\times 10^{47} \, C_{\rm peak}\, \epsilon_{\rm rad} \,B_{\rm pole,15}^2 \, {\rm erg}.
\end{eqnarray}

\begin{figure}
\begin{center}
\epsfig{file=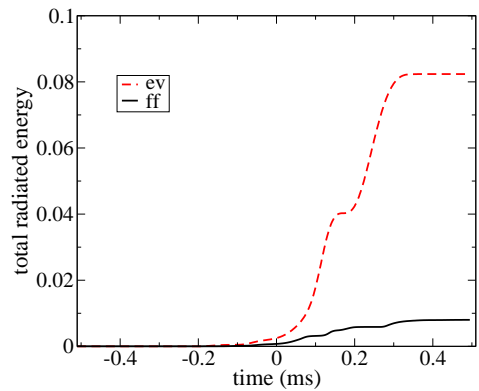, width=2.7in} 
\caption{{\it Non-rotating, unstable  star}. Time integral of the
electromagnetic luminosity, normalized with respect to the peak EM energy of
the magnetosphere (i.e., around the formation of the black hole), in both the
force-free and electrovacuum simulations. In the electrovacuum case, the 
net EM output is $\sim 8\%$, ten times larger than that radiated
in the force-free case.} 
\label{fig:collapse_energyrad_nonrotating}
\end{center}
\end{figure}

The radiative efficiency is very small in the force-free case, 
$\epsilon_{\rm rad} = 0.008$; the rest of the peak energy is swallowed 
by the black hole.  This is illustrated in 
Fig.~\ref{fig:collapse_energyrad_nonrotating}, which shows how 
$\epsilon_{\rm rad}$ grows with time in both the force-free and 
electrovacuum runs.  
The radiated energy is $E_{\rm rad} \approx 10^{45} \,B_{\rm pole,15}^2 \, 
{\rm ergs}$, from Eq.~(\ref{energy_peak_magnetosphere}),
expressed in terms of magnetar-strength magnetic fields.  Most of this is 
radiated in a short interval $\approx 1$ ms surrounding the collapse, 
with an average luminosity $L \approx 10^{48} B^2_{\rm pole,15}$ erg s$^{-1}$.

The radiative efficiency measured at $r=1.5 R_s$ is an order of magnitude higher in the
 electrovacuum case. This indicates that a larger proportion of the electromagnetic energy
 falls into the black hole in the force-free run, instead of escaping to infinity.  
An important check of this result is to measure the dissipation at current sheets in 
both the force-free and electrovacuum simulations.  In both cases, we find that the 
integral of $E.J$ is a small fraction of the total EM energy in the magnetosphere. 
(The constraint $E^2 < B^2$ is maintained in the force-free case by applying a 
small enhanced resistivity; hence the reduction in electric field energy appears as $E.J$ dissipation.)

An important feature of this radiation in both cases is its predominantly 
dipolar structure in energy flux (i.e., $L \propto \sin^2(\theta)$).
In the force-free case, energy is radiated in a rather continuous manner and it propagates 
outwards with a velocity $v=0.89\,c$, equivalent to a mild Lorentz factor of $W=2.2$.  
For the electrovacuum case the energy is radiated mainly in two long bursts
instead of the several,  and with shorter periods, resulting in the force-free case. 

\subsubsection{Late Force-Free Evolution and Magnetic Reconnection}

At later times, when the fluid has completely fallen into the black hole,
the field lines that were dragged toward the horizon reconnect
near the equatorial plane in a few sequential bursts,
expelling most of the remaining magnetic flux.
This behavior appears to be associated with the formation of x-type singularities in the
magnetic field.  Since the field is stretched radially and then reconnects near
the horizon, the resulting electromagnetic pulse bunches up in the radial direction, which explains the
structure seen in the last panel of Fig.~\ref{fig:collapse_norot1}. 

X-point reconnection appears to happen easily in force-free plasmas;
in ohmic plasmas it is associated with inhomogeneities in the electrical resistivity (e.g see~\cite{2006PhPl...13b2312B}).  In the absence of a detailed 
microphysical model for the resistivity, fine details such as these should be 
treated with caution, and the calculation should be viewed as illustrative.



\subsubsection{Vacuum versus Force-Free EM Field Evolution
in Axisymmetric, Non-Rotating Collapse}
Some important features of the vacuum evolution of the electromagnetic field 
around a collapsing star can be understood by neglecting
the effects of spacetime curvature, and by considering a simplified trajectory
for the surface of the star.  If the collapse starts at a finite time, then
an initially potential magnetic field evolves into a hybrid structure that 
consists of an inner potential magnetic field that matches the surface boundary
condition as determined by the conservation of magnetic flux; and
a transient electromagnetic wave that propagates into the original field structure.

Of especial interest is the appearance of zones within this wave structure that
are dominated by the electric field.  If the collapse continues for a long
time (the final stellar radius $R_s$ is small compared with the initial radius),
then this zone where $E^2 > B^2$ extends over a wide range of radius.  Since
a realistic magnetosphere may contain enough free charges to limit the growth
of $E\cdot B$, this provides a nice example of how the nearly force-free
evolution of an electromagnetic field can lead to strong dissipation.

In the absence of rotation, a spherically symmetric collapse implies
a radial fluid velocity inside the star.  As long as the stellar surface
contracts with a speed $-\dot R_s \ll c$, the magnetic field near it
is approximately potential.  For a pure multipole of order $\ell$,
\begin{equation}
\vec B(r) = B_s(t) R_s(t) \vec\nabla\left[ {P_\ell(\cos\theta)\over
   (r/R_s)^{\ell+1} } \right]
\end{equation}
(with $B_s$ the magnetic field at the star's surface).
A toroidal electric field $E_{\phi} = -(\dot R_s/c)\, B_{\theta}$ is present
at the surface of the star, assuming its interior to be perfectly conducting.
The junction condition at the surface ensures the continuity of $E_\phi$.
The surface magnetic field therefore increases in accordance with 
simple flux conservation,
\begin{equation}\label{eq:fluxcon}
{\partial B_s\over\partial t} + \dot R_s{\partial B_s\over\partial r}\biggr|_{R_s}
= -2{\dot R_s\over R_s}B_s.
\end{equation}
At a fixed radius $r$, the magnetic field grows weaker:  
in the case of a simple dipole, the stellar
magnetic moment scales as $\mu(R_s) = B_s(R_s) R_s^3 = \mu_0 (R_s/R_{s0})$.

If the star were to reach infinite density at a finite time 
$t_{\rm col}$, then a strong toroidal electric field would develop at 
$r > c(t_{\rm col}-t)$.  The inner potential zone would shrink along with 
the star as $t$ approaches $t_{\rm col}$.  

The external electromagnetic field can then be obtained by rescaling the
radius, $r\rightarrow \xi \equiv r/R_s(t)$, and transforming derivatives
according to $\partial_t X(r,t) \rightarrow  [\partial_t + (\dot R_s/R_s)
\partial_\xi]X(\xi,t)$.  It is simplest to solve for the vector potential
$A_\phi$, from which the poloidal magnetic field and toroidal electric
field are derived.  The boundary condition at the surface of the star is
\begin{equation}
R_s(t) A_\phi[R_s(t),\theta] = {B_s(t) R_s^2(t)\over \ell} {dP_\ell\over d\theta}
= {R_{s0} A_{\phi 0}(\theta)\over\ell} {dP_\ell\over d\theta}.
\end{equation}
From Eq.~(\ref{eq:fluxcon}), $R_{s0} A_{\phi 0}(\theta)$ is constant.   
Substituting $A_\phi(r,\theta) = R_{s0} A_{\phi 0}(\theta) g(\xi,t)$ into the wave
equation
\begin{equation}
\partial_t^2(r A_\phi) = \partial_r^2(r A_\phi) - {\ell(\ell+1)\over r^2}rA_\phi,
\end{equation}
and adopting a collapse law $R_s(t) \propto (t_{\rm col}-t)^\alpha$ (here
$\alpha = 2/3$ for pressureless collapse from a large radius), we find 
\begin{eqnarray}\label{eq:diffeq}
&&\partial_\tau^2g + {1\over\alpha}\partial_\tau g + 2\xi\partial_\tau\partial_\xi g =
\left[{c^2\over \dot R_s^2} - \xi^2\right]\partial_\xi^2 g \cr
&& \quad\quad\quad -\left(1+{1\over\alpha}\right)\xi\partial_\xi g - {c^2\over \dot R_s^2}
  {\ell(\ell+1)\over \xi^2}g.
\end{eqnarray}
Here $\tau = \int \alpha dt/(t_{\rm col} -t)$ is a dimensionless time
coordinate;  hence $R_s(\tau) = R_{s0} e^{-\tau}$.  The electromagnetic field is constructed
from the solution to Eq.~(\ref{eq:diffeq}) using
\begin{eqnarray} \label{eq:similarl}
B_r(r,t) &=& {R_{s0}\over r^2}\, {g(r/R_s,t)\over\sin\theta}\partial_\theta\left(\sin\theta A_{\phi0}\right); \cr
B_\theta(r,t) &=& -{R_{s0} A_{\phi 0}\over r^2}\, (\xi \partial_\xi g)_{\xi = r/R_s}; \cr
E_\phi(r,t) &=& -{\dot R_s\over c}\,{R_{s0} A_{\phi 0}\over rR_s}\,
                 \left( \partial_\tau g + \xi \partial_\xi g \right)_{\xi = r/R_s}.
\end{eqnarray}

One recovers the usual potential solution $A_\phi(r,\theta) = 
A_\phi[R_s(t),\theta]\,(r/R_s)^{-(\ell+1)}$ where $\xi \ll c/|\dot R_s|$.
A self-similar solution is also available in the case of collapse at
a uniform speed, $\alpha = 1$, if the collapse starts at a very large
initial radius.  Then one can take $\partial_\tau g = 0$ and the 
electromagnetic field is a function only of $r/R_s$.  In this case, 
the magnetic field can retain a dipolar form out to large distances 
$r \gg c(t_{\rm col}-t)$ from the star.  Restricting to $\ell= 1$ gives
\begin{eqnarray} \label{eq:similar}
B_r(r,t) &=& {R_s\over R_{s0}} {2\mu_0\cos\theta\over r^3}; \cr
B_\theta(r,t) &=& {R_s\over R_{s0}} {\mu_0\sin\theta\over r^3}; \cr
E_\phi(r,t) &=& {r\over c(t_{\rm col}-t)} B_\theta(r,t).
\end{eqnarray}
Although the magnetic field is identical to that sourced by a stationary
dipole of magnitude $\mu(R_s) = \mu_0 (R_s/R_{s0})$, the electric field energy
dominates outside a distance $\sim (c/|\dot R_s|)R_s(t)$ from the star.  
The similarity solution is accurate out to the larger distance
$\sim (c/|\dot R_s|)\, R_{s0}$, where $R_{s0}$ is the stellar radius at the 
beginning of the collapse.  

\begin{figure}
\begin{center}
\epsfig{file=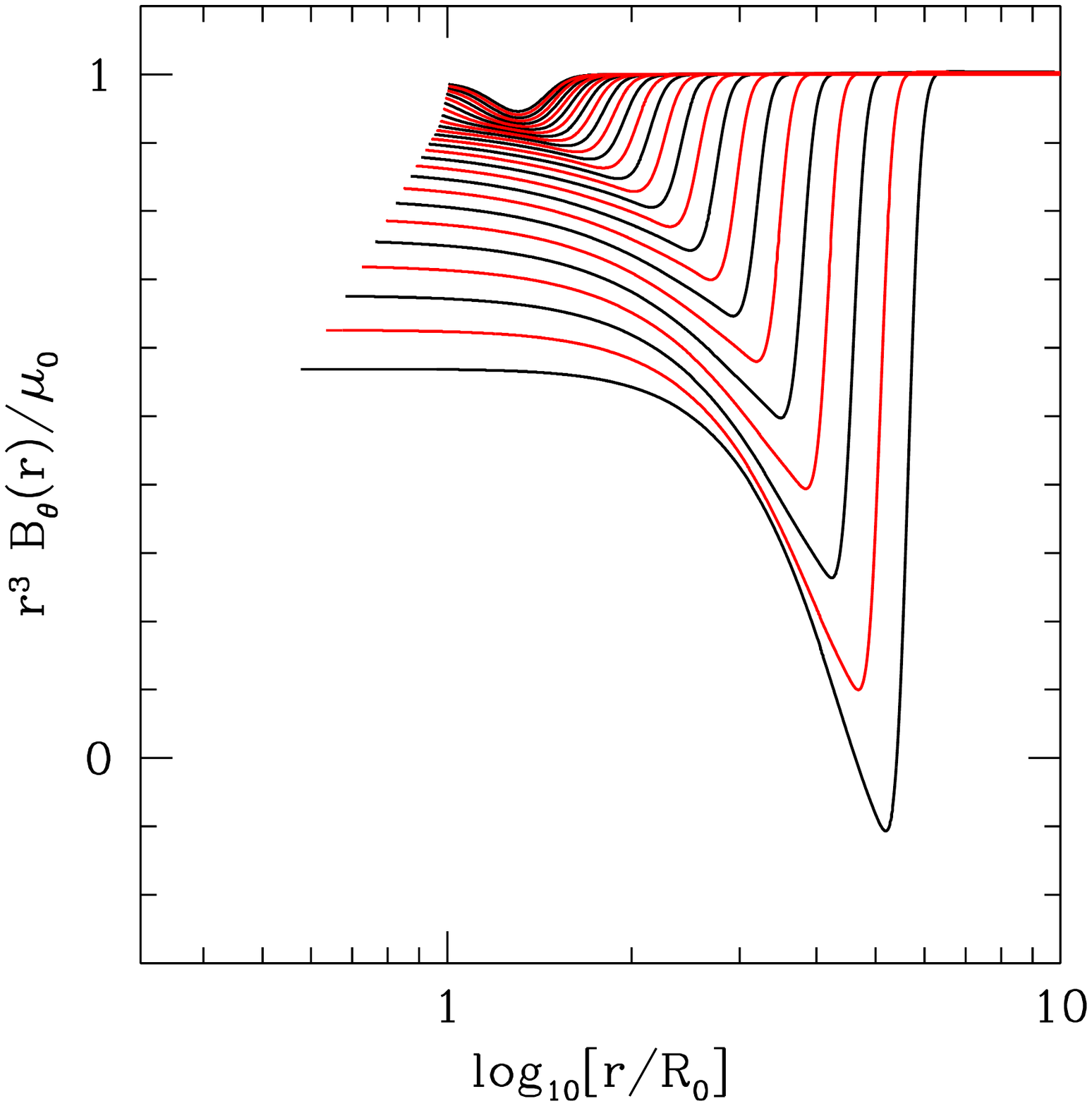,width=2.1in}
\epsfig{file=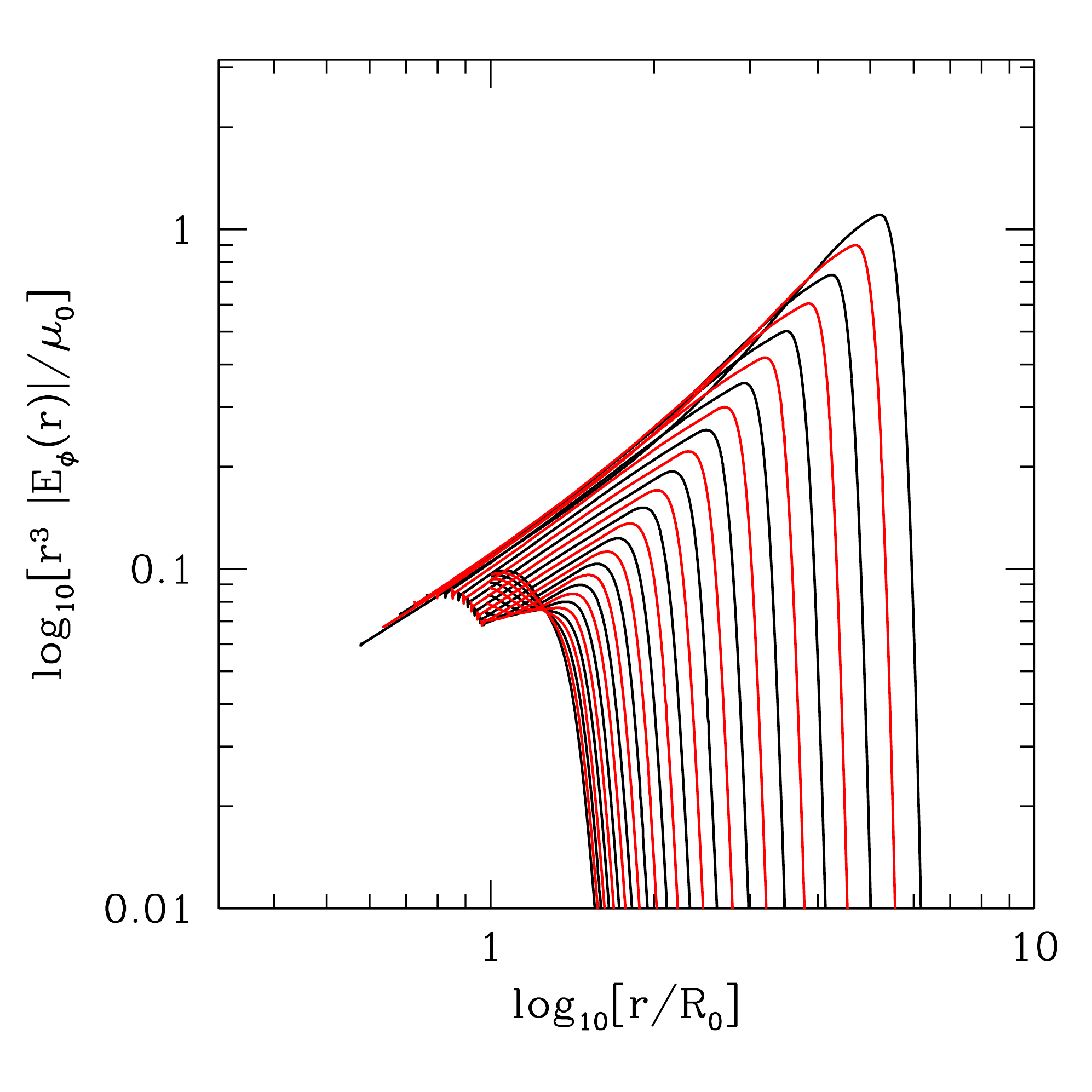,width=2.1in}
\caption{
{\it Evolving vacuum EM field around a star collapsing to
half its initial size (uniform $dR_s/dt$).}  
Top panel: the inner magnetic field tracks the instantaneous 
dipole of the collapsing star;  at a fixed radius, $B \propto
\mu(R_s) \propto R_s/R_{s0}$.  Time progresses top to bottom.
Bottom panel:  The zone of rising $E_\phi$ closely follows
the similarity solution of Eq.~(\ref{eq:similar}).  Time progresses
left to right.  (Two colours are employed for clarity.)
} \label{1D_EM_Profile}
\end{center}
\end{figure}

\begin{figure}
\begin{center}
\epsfig{file=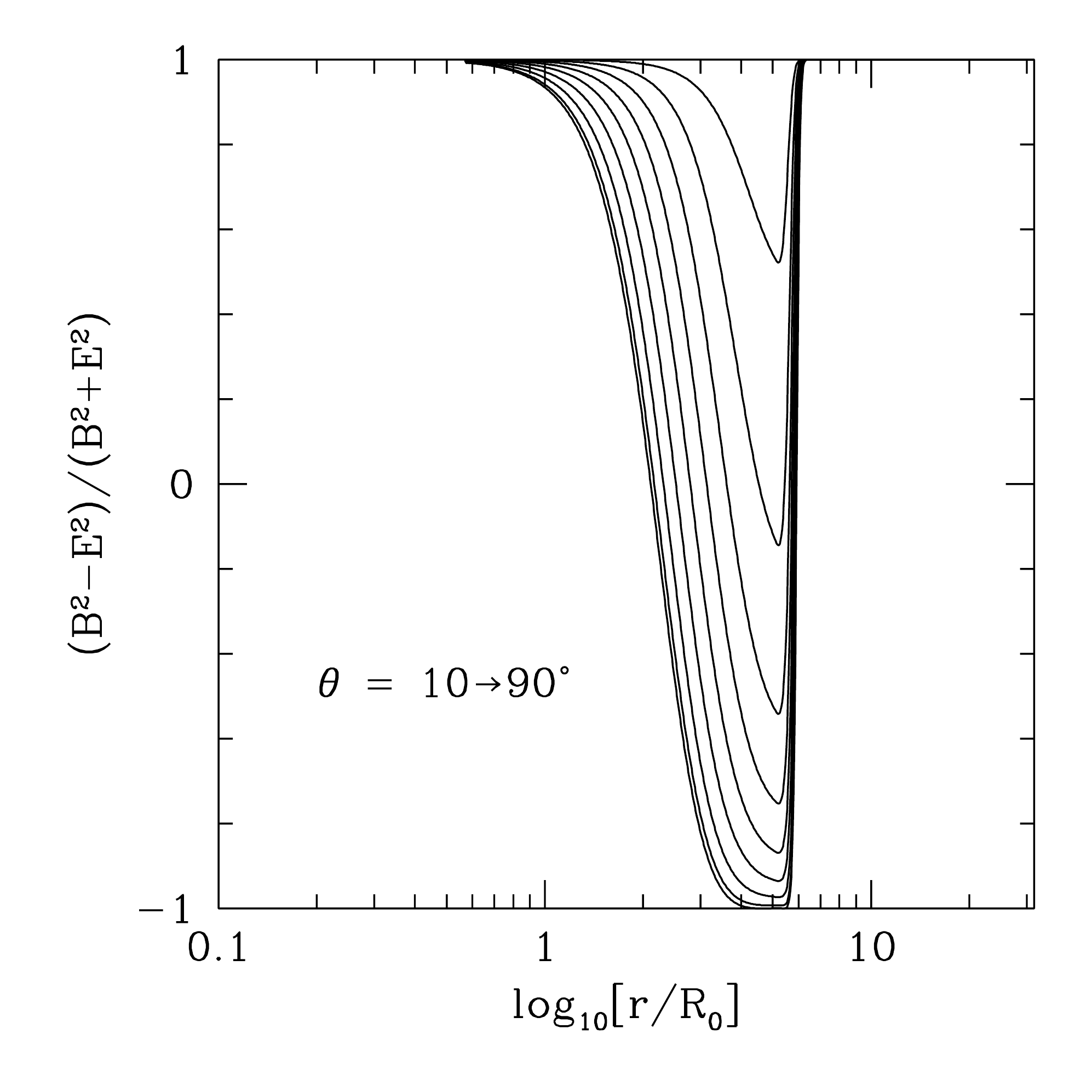,width=2.5in}
\caption{
{\it Relative strength of vacuum electric and magnetic fields around 
collapsing star (uniform $dR_s/dt$).}  Different lines illustrate the obtained behavior for
different values of $\theta$ (from $10^o$, top, to $90^o$, bottom). 
The zone where $E^2 > B^2$ is more extended near the magnetic equator.
} \label{Bsq_Esq}
\end{center}
\end{figure}

To understand how this inner solution for the electromagnetic field matches
onto the initial potential magnetic field that was present prior to the
collapse, or to consider cases other than constant $\dot R_s$, one must calculate
the full time-dependent solution to (\ref{eq:diffeq}).  We have done this by
evolving $g(\xi,\tau)$, $\partial_\tau g(\xi,\tau)$ and $\partial_\xi g(\xi,\tau)$
using a centered discretization of Eq.~(\ref{eq:diffeq}) and employing
a small Kreiss-Oliger dissipation value $O(10^{-3}$) to damp high-frequency noise in each of
these variables.  The resulting electromagnetic field profile is plotted in
Fig.~\ref{1D_EM_Profile}, with distance normalized to the initial radius of the
star.  One observes in  Fig. \ref{Bsq_Esq}
the emergence of an extended zone with $E^2 > B^2$,
as is expected from the similarity solution of Eq.~(\ref{eq:similar}).
The magnetic field is relatively stronger in the polar regions, since 
$A_\phi \propto \sin\theta \rightarrow 0$ at small $\theta$.

\begin{figure}
\begin{center}
\epsfig{file=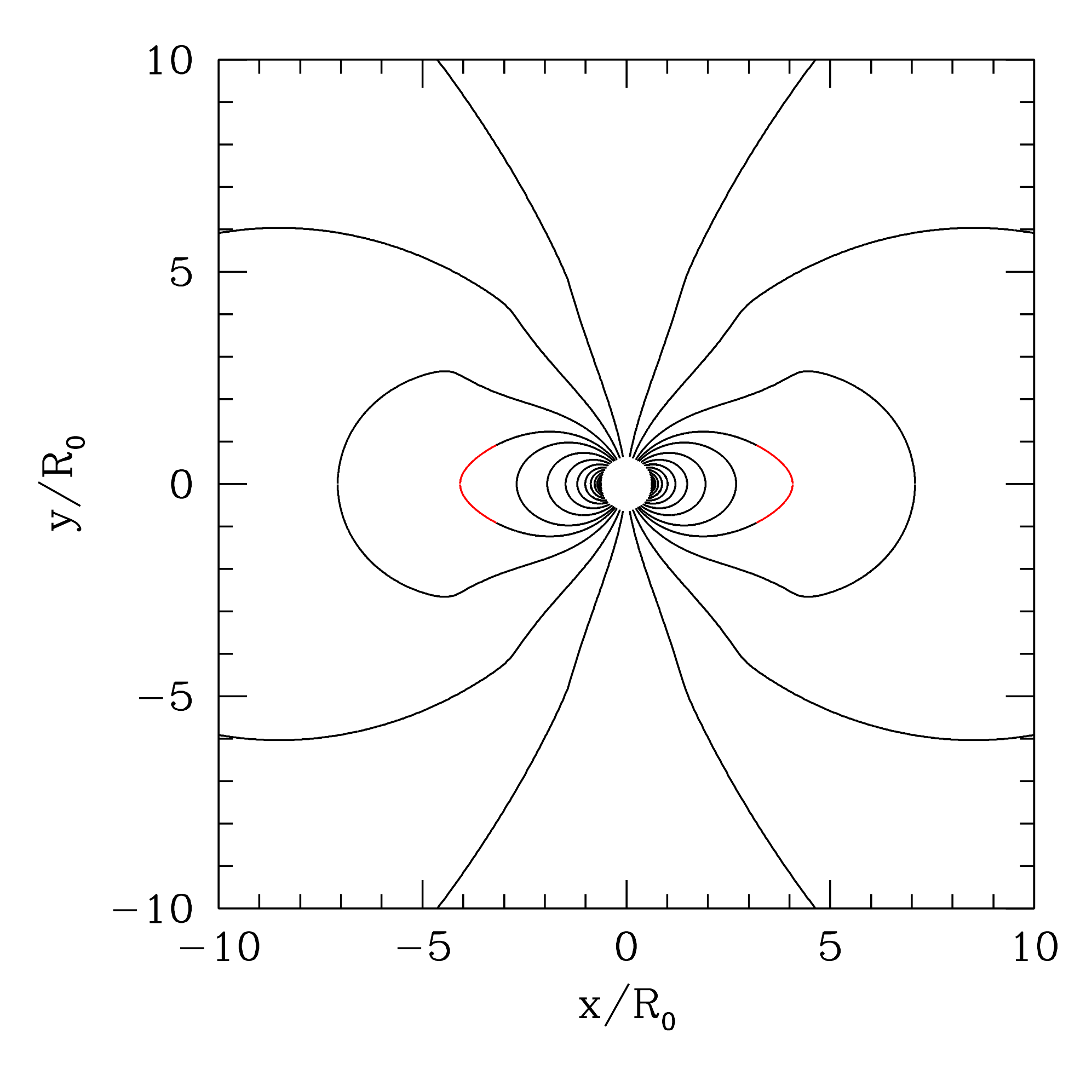,height=2in,width=2.3in}
\epsfig{file=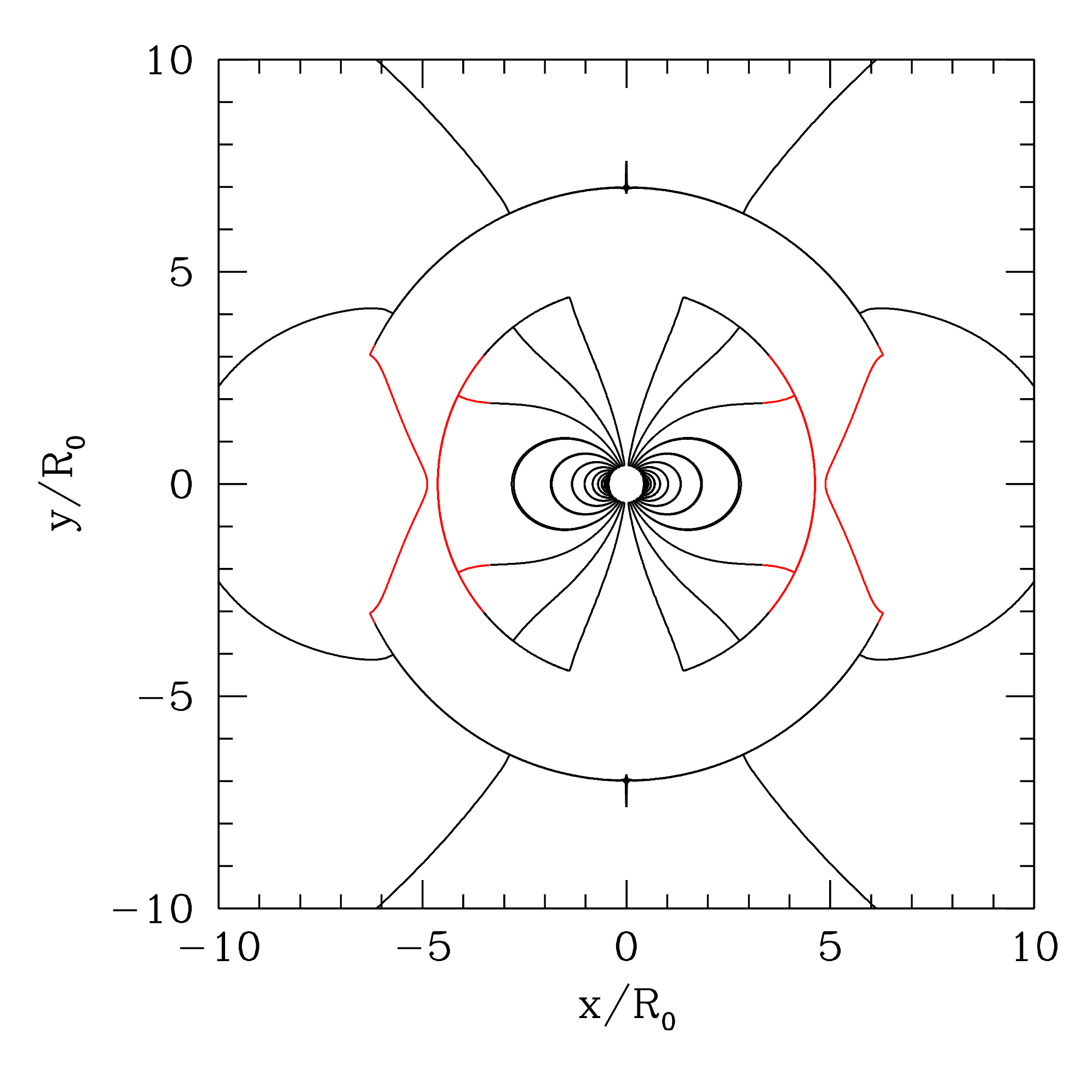,height=2in,width=2.3in}
\caption{
{\it Connectivity of the magnetic field around a collapsing star 
(uniform $dR_s/dt$).}  The dipolar magnetic field that is anchored in the
star disconnects from the surrounding, undisturbed, potential field when
the amplitude of the output wave becomes large enough that $B_\theta$
reverses sign.  Note that $E^2 > B^2$ (the red sections of the field lines)
        in the outer part of the inner
zone with a dipolar magnetic field line profile.
These two frames display the solution at the two latest times
shown in Fig.~\ref{1D_EM_Profile} in which $B_\theta$ becomes negative.
} \label{1D_EM_Field_lines}
\end{center}
\end{figure}

Eventually the amplitude of the outgoing wave disturbance becomes large
enough that $B_\theta$ changes sign.  After this happens, the magnetic field 
remains dipolar inside the radius $\sim (c/|\dot R_s|) R_{s0}$,
but disconnects from the external zone of undisturbed potential field.
This transition is illustrated in Fig.\ref{1D_EM_Field_lines} using two 
snapshots corresponding to the two most extended field profiles in
Fig.~\ref{1D_EM_Profile}.  

The first part of the collapse leads to a re-arrangement of
the magnetic field outside the star, while electromagnetic energy flows inward:
the Poynting flux,
\begin{equation}
S_r = -E_\phi B_\theta c \simeq -{\dot R_s\over c}{A_{\phi 0}^2(\theta)R_{s0}^2\over R_s(t)^4}
{(\partial_\xi g)^2\over\xi}
\end{equation}
is negative inside $r \sim (c/|\dot R_s|)R_{s0}$.  After the horizon forms and reaches
the surface of the star, the compression of the magnetic field stops,
and some of the trapped magnetic field can be radiated to infinity.  
The angular distribution of this radiated energy reflects the symmetry
of the initial field. In the case of a dipole, the energy is carried away
by closed magnetic loops and is mainly channeled through the magnetic equator.

We note that the appearance of zones with $E^2 > B^2$ in the electrovacuum
solution implies a fundamental difference with the 
alternative force-free solution.  In this case, $E\cdot B = 0$ throughout
the collapse, so consistent force-free evolution is obtained only by
removing energy from the electric field.  From an MHD
perspective, such a transition signals the appearance of nearly luminal
plasma motions, where the inertia of even a small residue of entrained matter
can become important.  

This simple example shows that there can be profound differences in the macroscopic
structure of the electromagnetic fields when even a small amount of conducting matter
is present.  In the complete absence of free charges, macroscopic zones where
$E^2 > B^2$ are the natural consequence of the time-evolution of an EM field that,
initially, is purely magnetic.

\subsubsection{$E^2 > B^2$ in the Vector QNM of a Black Hole.}\label{s:quasi}

The quasi-normal mode (QNM) behavior observed shows two remarkable features:  an oscillation
in the sign of the magnetic flux threading each hemisphere, and the
appearance of an equatorial zone with $E^2 > B^2$.  These two features can
be related to each other using a simple planar analogy.

Consider an initial magnetic field configuration
${\bf B} = B_0 \hat x$ ($-B_0\hat x$) for $y > 0$ ($< 0$).  
This is familiar from studies of conducting fluids, where in the case of
uniform resistivity one finds a slow flow of matter to a thin current sheet 
at $y = 0$.  The sheet thickness is limited by the flow of matter along the
sheet to large $|x|$ (Sweet-Parker reconnection).  But now
we are interested in the case where the conducting matter is absent, 
and the electromagnetic field evolves according to the vacuum wave equation.
One finds, instead, a growing zone of pure {\it electric} field 
${\bf E} = - B_0\hat z$  for $|y| < ct$, which maintains a uniform sign 
across the initial magnetic null surface.  A pure magnetic field is
converted to a pure electric field.  

In the case of EM fields localized around a black hole, this conversion
of magnetic to electric fields occurs at the magnetic equator, where
the poloidal field lines merge together.  The toroidal electric field that
is created sources a poloidal magnetic flux through the horizon of the 
opposing sign to the pre-existing flux.  In this way, a continuing 
interconversion of magnetic and electric fields can be maintained.

\subsection{Rotating stellar collapse}

\begin{figure}
\begin{center}
\epsfig{file=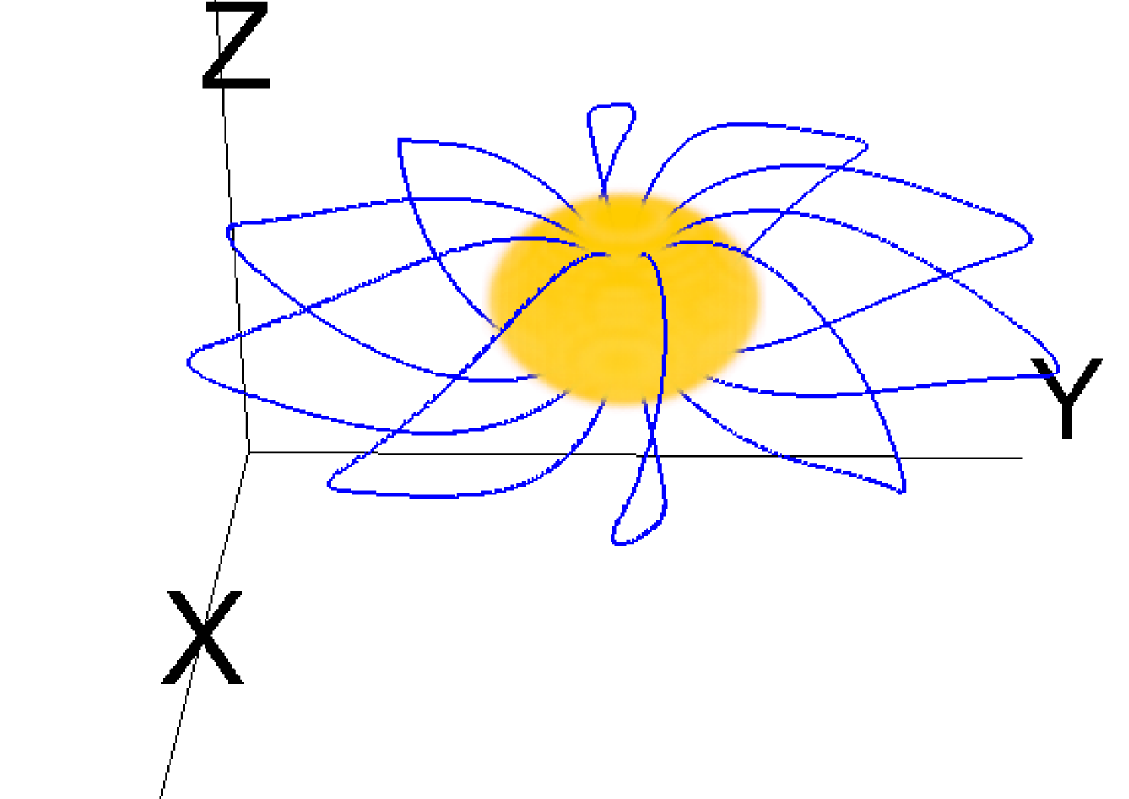,height=2.5cm}
\epsfig{file=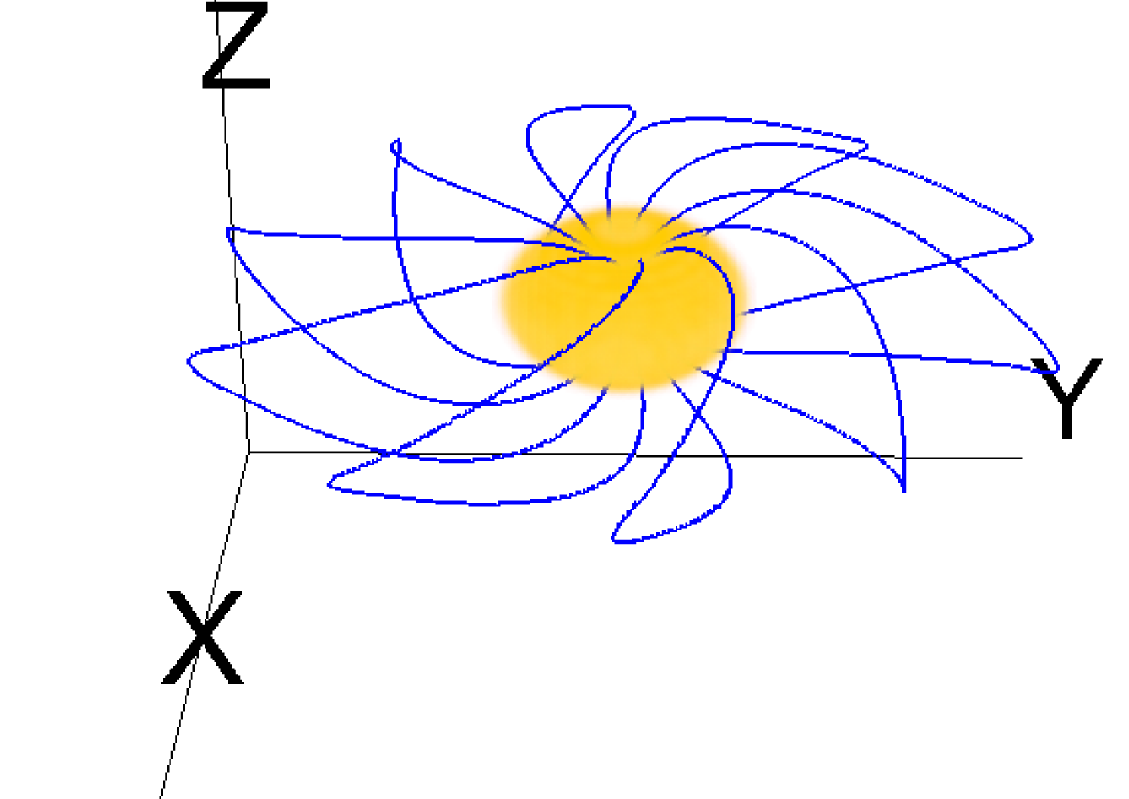,height=2.5cm} \\
\epsfig{file=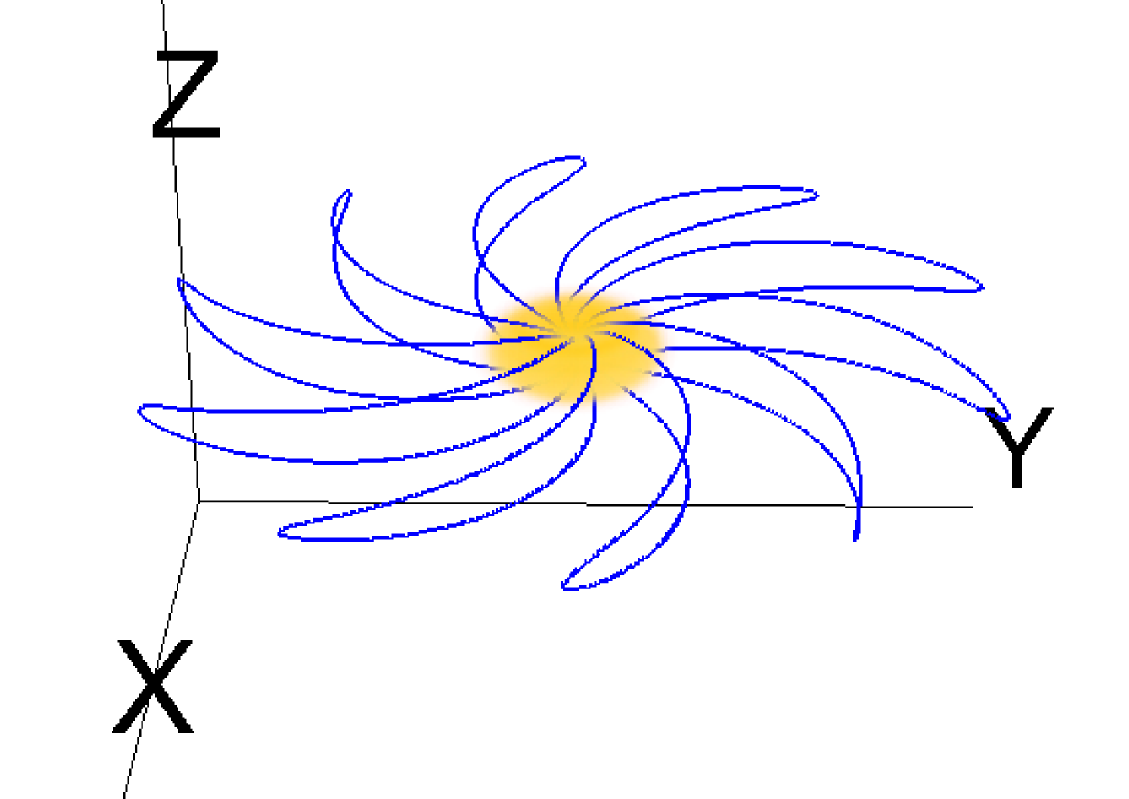,height=2.5cm}
\epsfig{file=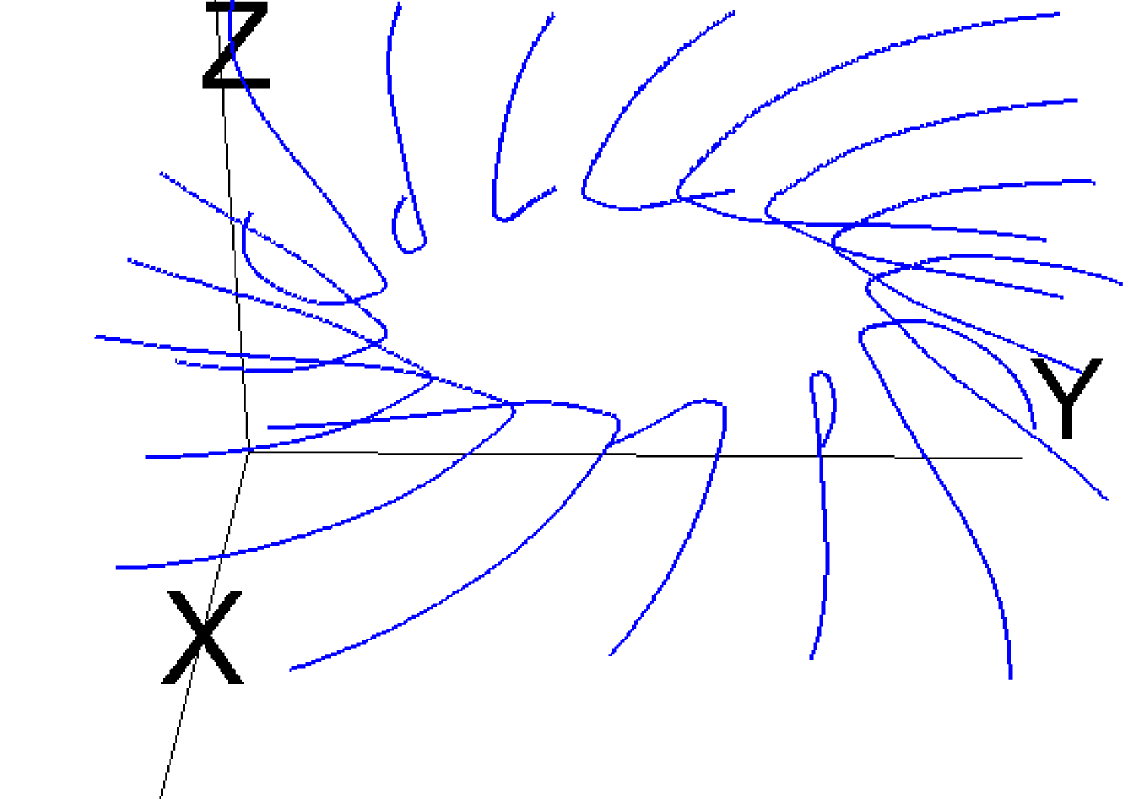,height=2.5cm} \\
\end{center}
\caption{{\it Rotating, unstable star (force-free)}. 
Magnetic field configuration (blue lines) and fluid density
(marked in red) at times 
$t=(-0.47,\,-0.17,\,-0.01,\,0.12)\,{\rm ms}$.
As the collapse proceeds the increasing spin rate of the star
pulls the external magnetic field in the toroidal direction.
}\label{fig:collapse_rot1}
\end{figure}

The collapse of a rotating, magnetized star produces
an interesting generalization of the relativistic wind problem for a 
stationary star (e.g~\cite{Goldreich:1969sb,Spitkovsky:2006np}).  As in the case
of spherical collapse, qualitatively new effects are introduced after the
formation of a horizon. We have chosen an unstable, rotating model star
with a mass $M=1.84 M_{\odot}$ and equatorial radius $R_s = 10.6~{\rm km}$.
The star rotates with a period $T=0.78 {\rm ms}$, so that the light cylinder
is initially located at $R_{\rm LC} = 37~{\rm km} \approx 3.5\, R_{s}$.
The numerical domain and resolution are identical to those employed in the
non-rotating case.

To remove unphysical transients we evolve the force-free equations 
for a couple of periods with both geometry and matter fixed, and afterwards all equations are
evolved. This approach ensures the force-free fields
relax to a configuration consistent with the physical scenario considered.
The expected field configuration emerges during this startup phase, with  
a closed, corotating magnetosphere extending out to the light cylinder (LC).
The Goldreich-Julian (G-J) current structure is present, with an
outflow along the polar field lines balanced by a return flow through
an equatorial current sheet.

 \begin{figure}
 \begin{center}
 \epsfig{file=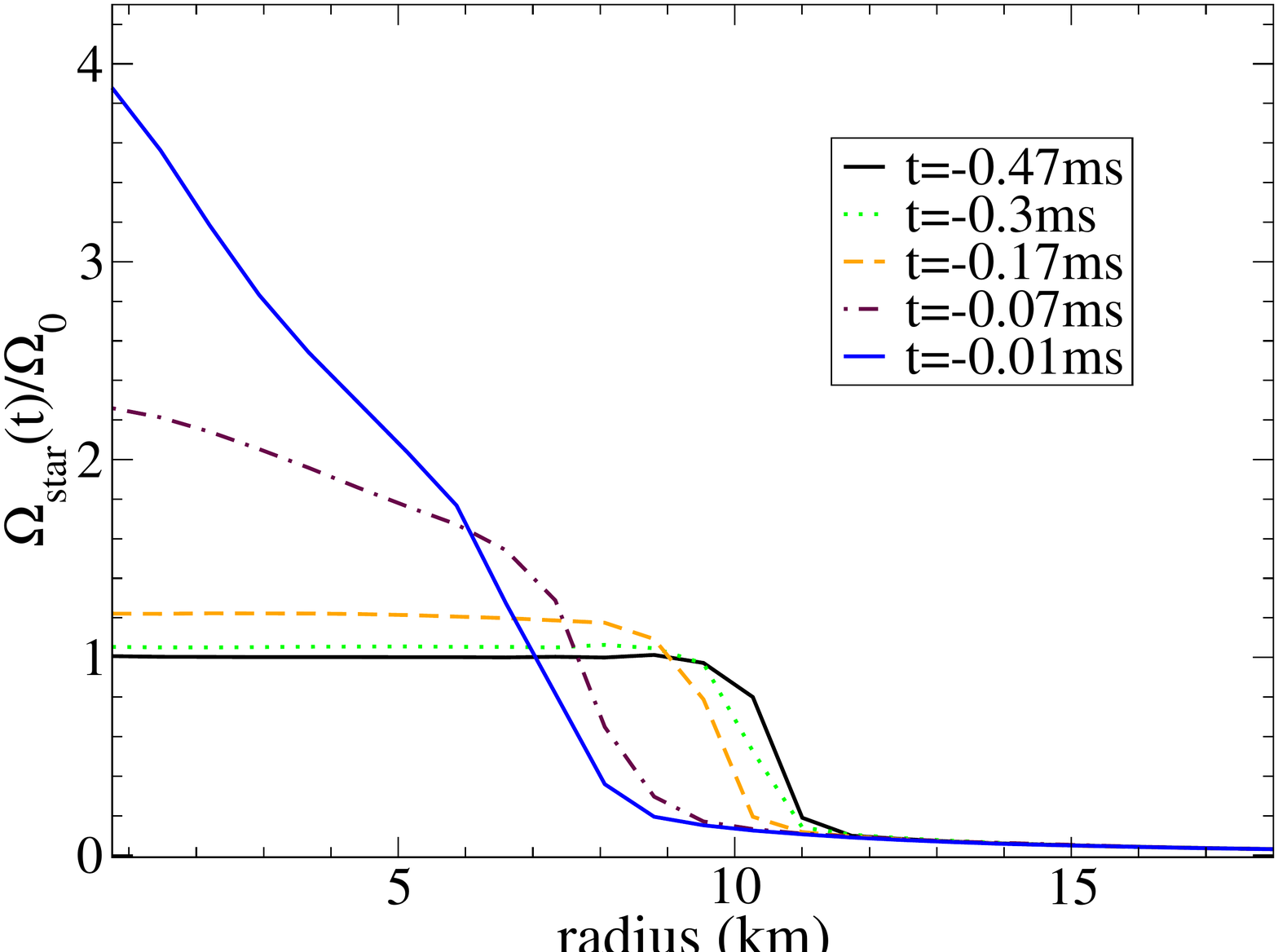,height=2.0in} 
 \caption{ {\it Rotating, unstable star (force-free)}.
Star's angular rotational velocity, measured at the equator, during the collapse. }
 \label{fig:collapse_omegastar}
 \end{center}
 \end{figure}

The different stages of the collapse are represented in 
Fig.~\ref{fig:collapse_rot1}, while that the angular velocity of
the star is displayed in Fig.~\ref{fig:collapse_omegastar}.  
As the star contracts, and its rotational 
frequency increases, the instantaneous LC 
approaches the star\footnote{I.e., the one associated 
with the instantaneous rotation frequency of the star.}. 
Differential rotation develops in the magnetosphere, due to the
lack of causal contact between the star and the LC, and the magnetic 
field is wound in the toroidal direction.  Furthermore, 
the deepening of the gravitational potential forces significant changes
in the magnetic field profile, by pulling the field lines more tightly
toward the star.

The poloidal magnetic field strengthens due to flux freezing in the star, just as in 
the non-rotating case, but now most of the EM energy is in the toroidal component.
Near the poles, the field lines twist around, generating a cone-like structure.
In general, the magnetic field preserves a stretched dipolar
topology for a longer time than in the non-rotating case, up to the point that all
the fluid is swallowed by the black hole. 

\begin{figure}
\begin{center}
\epsfig{file=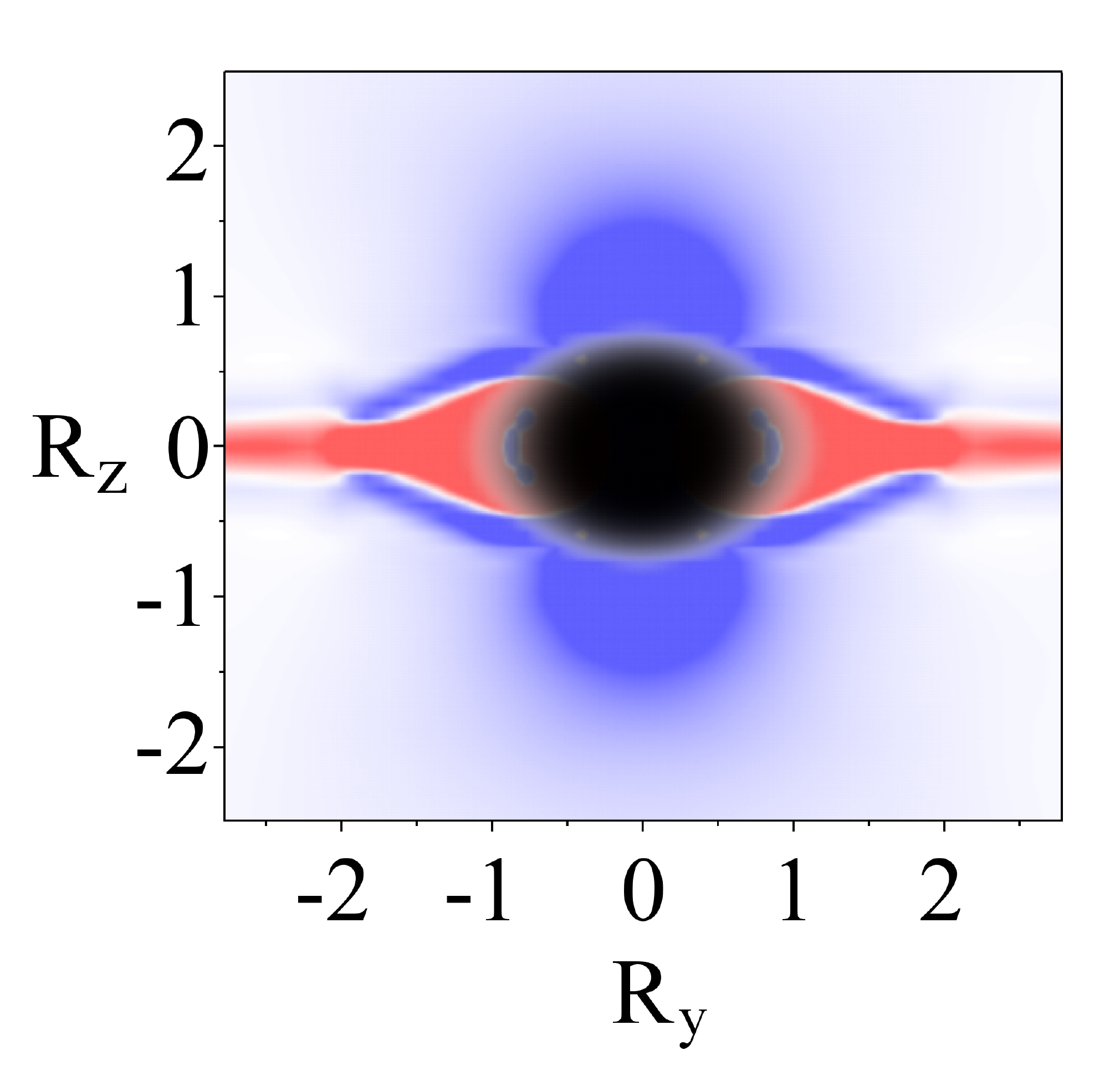,height=4.0cm}  
\epsfig{file=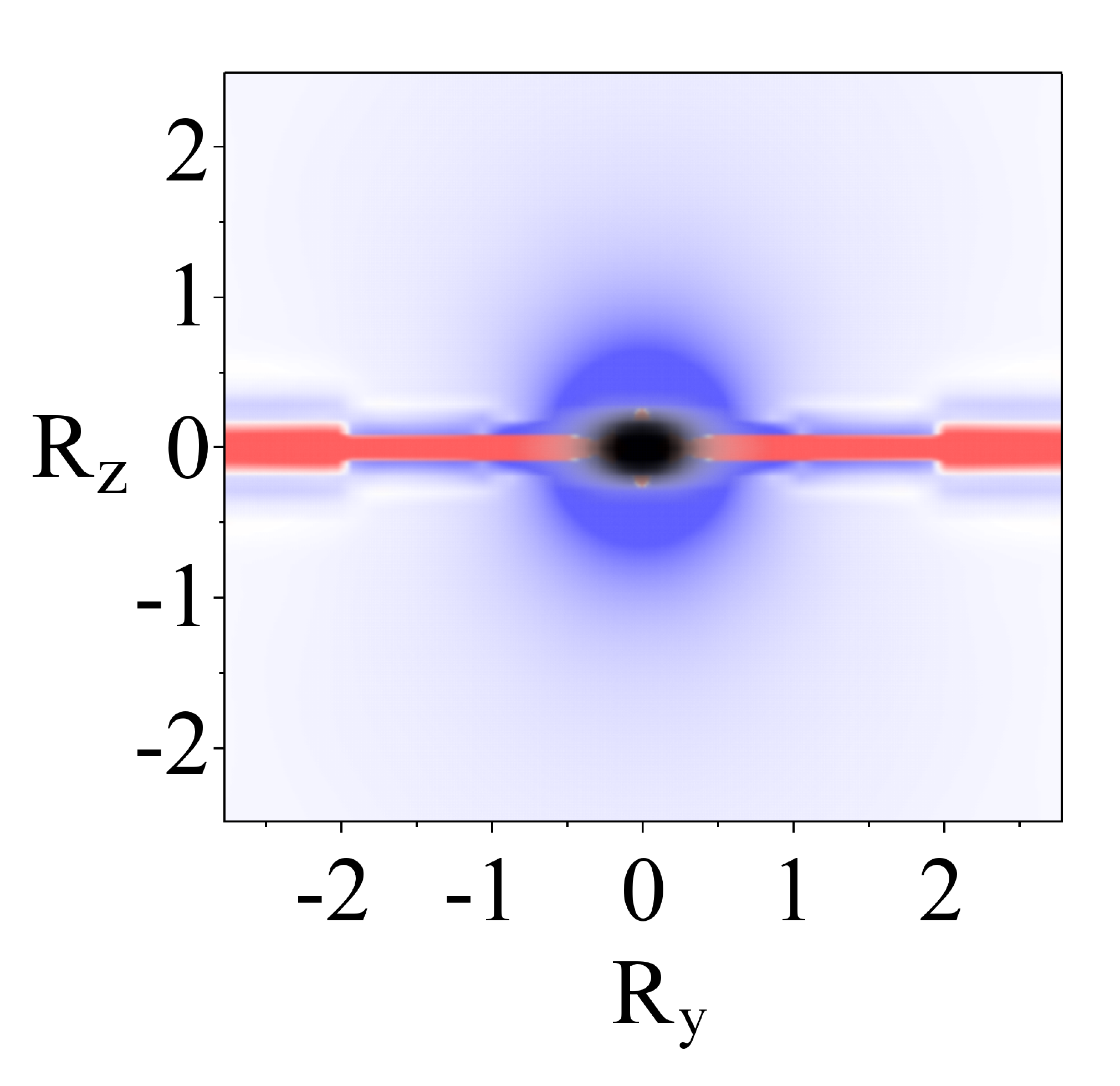,height=4.0cm} \\
\epsfig{file=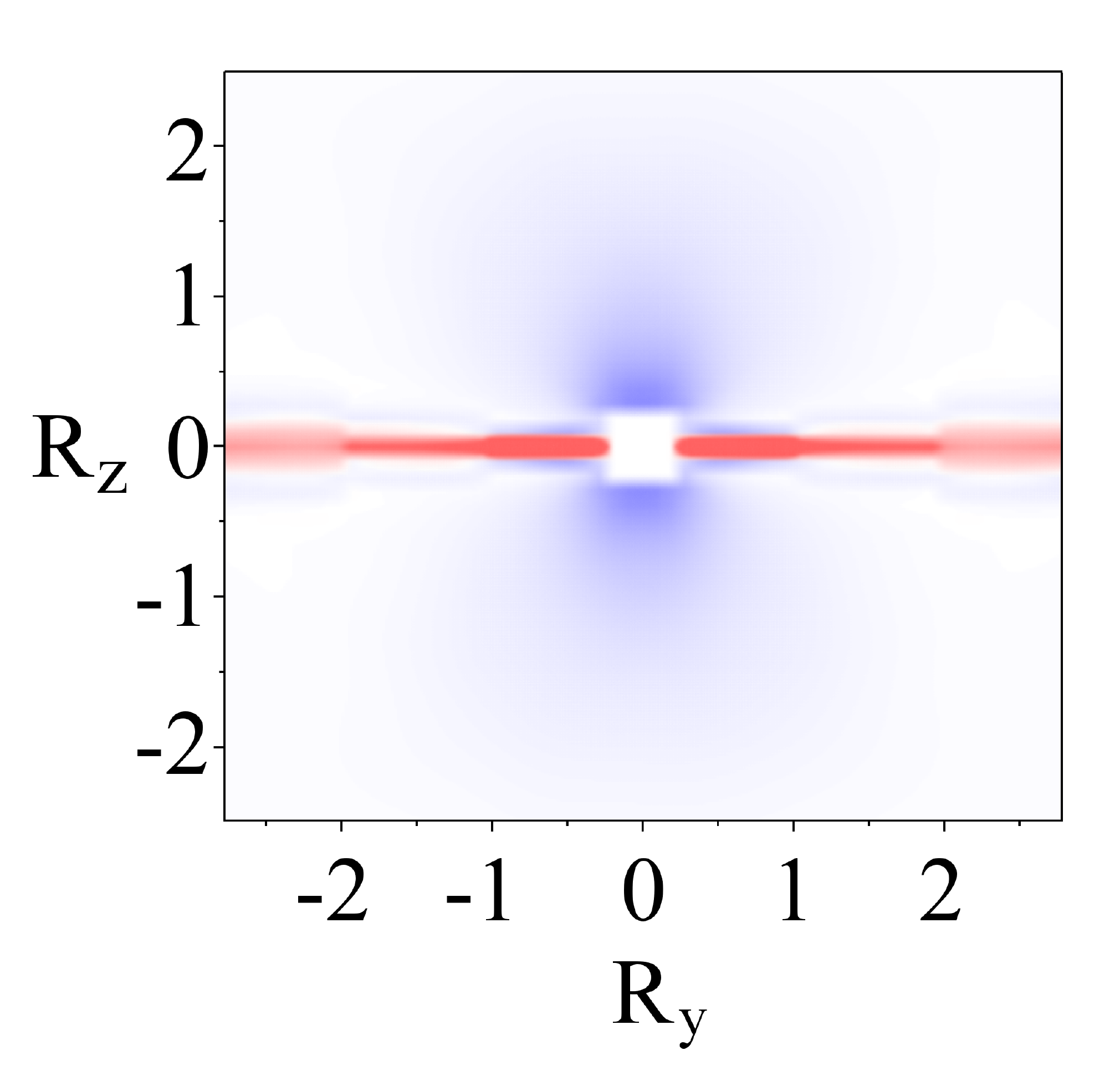,height=4.0cm}  
\epsfig{file=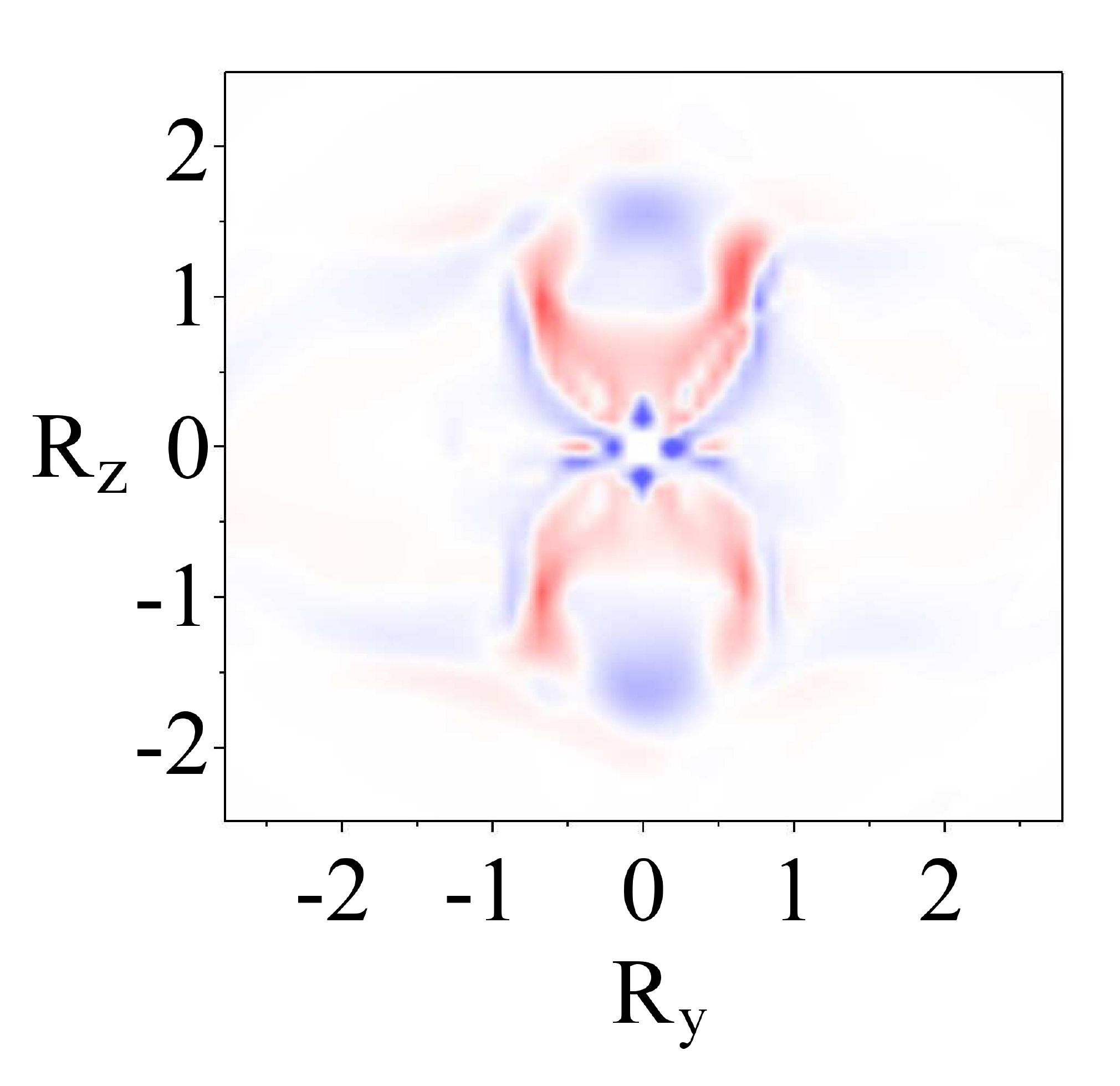,height=4.0cm}  \\
\caption{{\it Rotating, unstable star (force-free)}. 
Charge density (blue: positive and red: negative) is plotted
at $t=(-0.17,\,-0.01,\,0.12,\,0.35)\,{\rm ms}$.  The central fluid
region marked in black disappears as the horizon emerges.
} \label{fig:collapse_q_rotating}
\end{center}
\end{figure}

\begin{figure}
\begin{center}
\epsfig{file=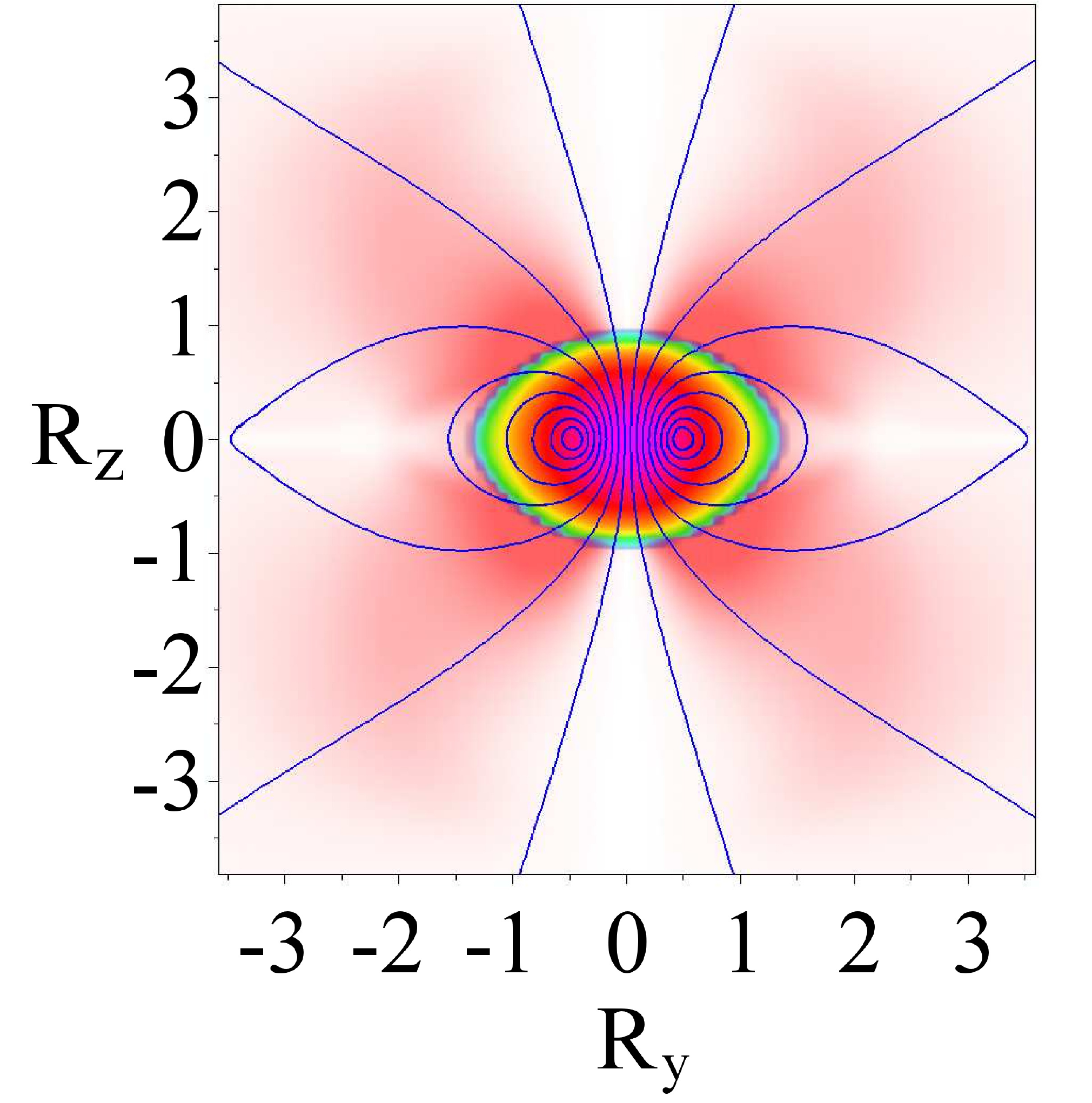,height=4.0cm} 
\epsfig{file=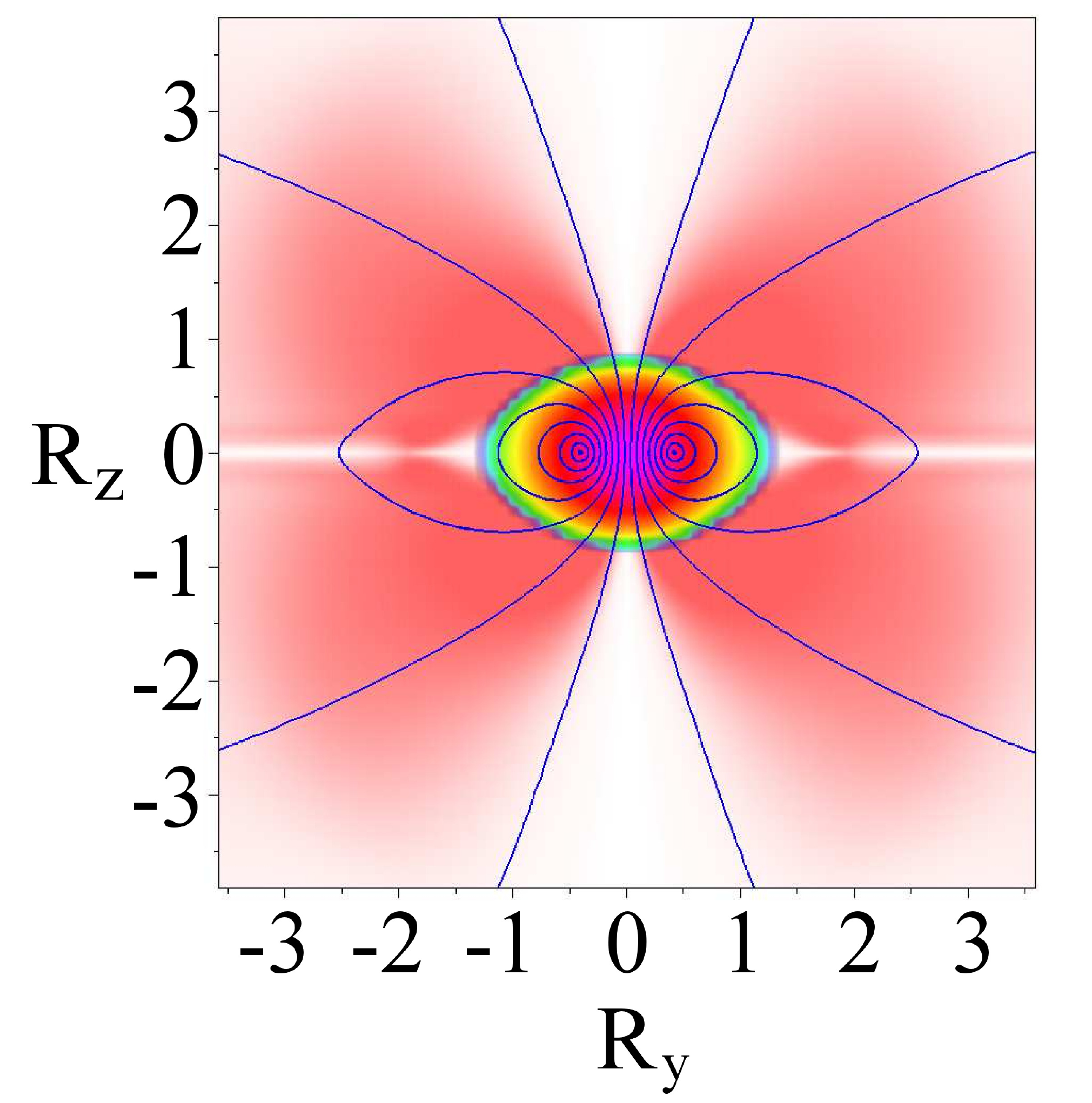,height=4.0cm} \\
\epsfig{file=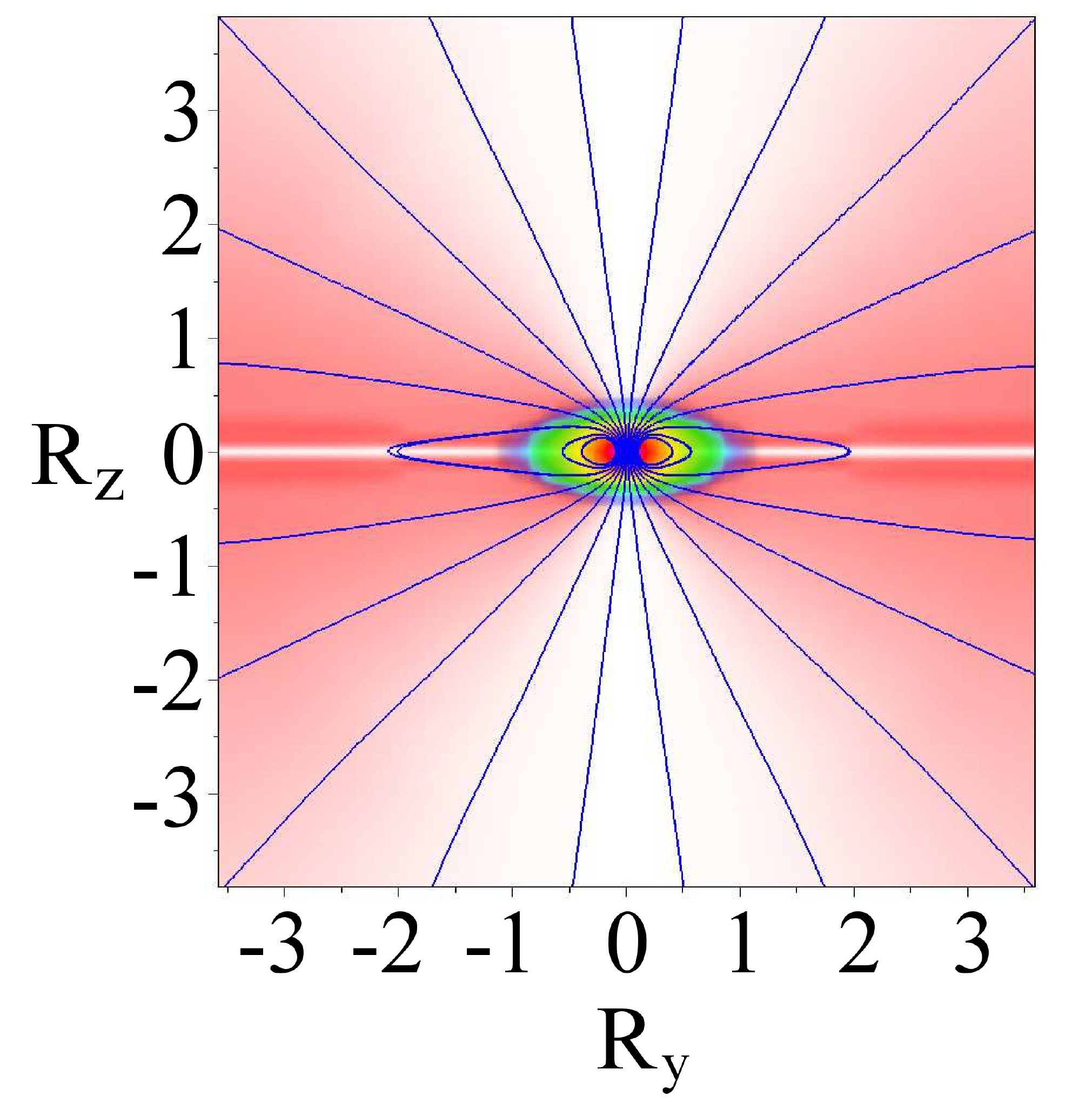,height=4.0cm} 
\epsfig{file=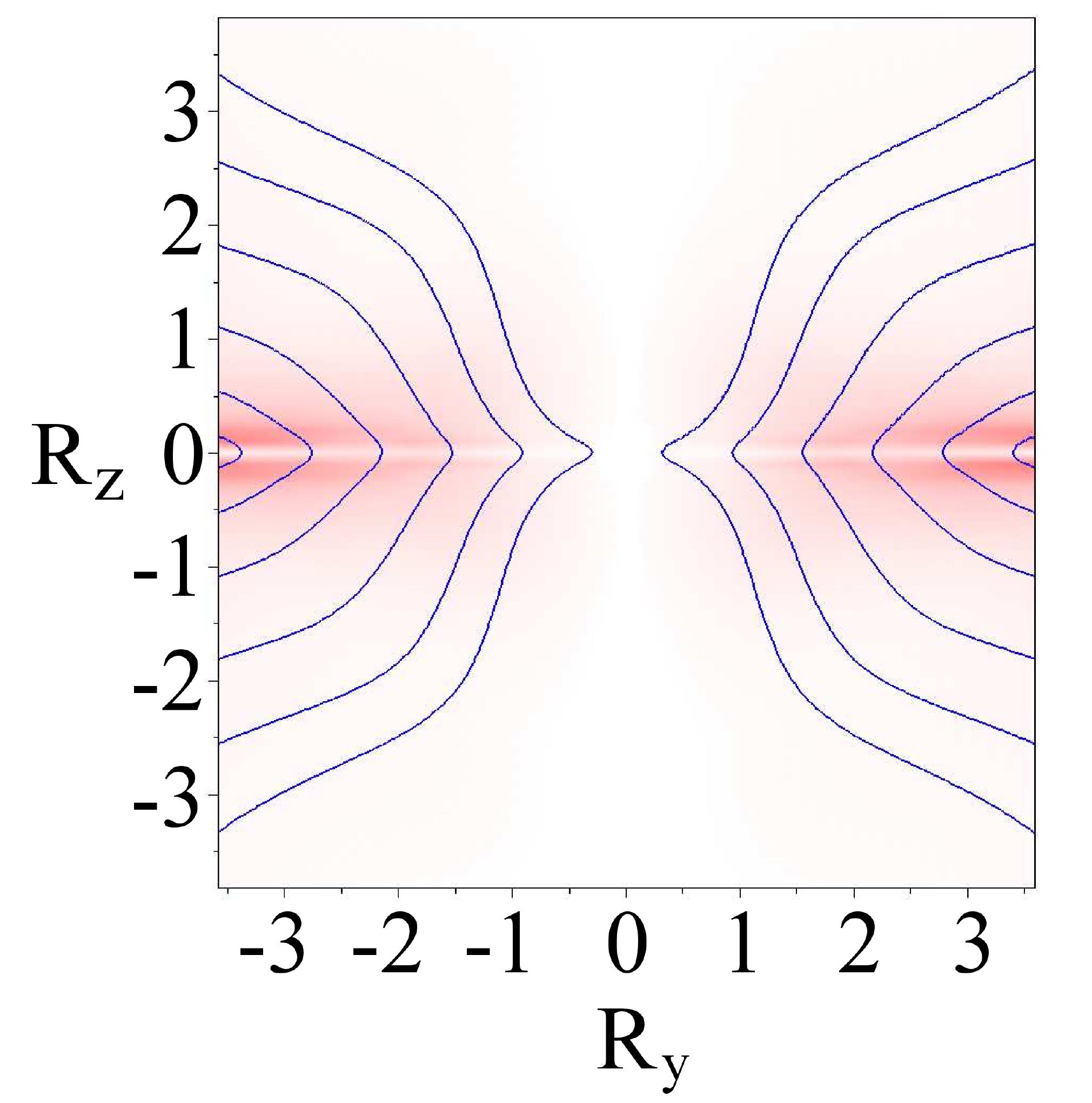,height=4.0cm} \\
\caption{{\it Rotating, unstable star (force-free)}. 
Radial Poynting flux in red at $t=(-0.3,\,-0.17,\,-0.01,\,0.12)\,{\rm ms}$. 
The central colored zones mark the stellar fluid.  The evolution of the
poloidal magnetic field (blue lines) in the equatorial regions is qualitatively similar
to that observed in the non-rotating case.
} \label{fig:collapse_Bphi2_rotating}
\end{center}
\end{figure}

As the black hole forms inside the star, and the fluid falls inward, many of the
features in the EM field evolution are preserved from the non-rotating runs.
Snapshots of the
field profile, charge density and Poynting flux, taken at various times close
to horizon formation, are displayed in Figs.~\ref{fig:collapse_q_rotating} 
and~\ref{fig:collapse_Bphi2_rotating}.
Just as in the non-rotating case, the topology of the magnetic field lines changes dramatically:
the y-point structure in the magnetic field disappears from the equatorial regions,
and a current sheet extends inward to the horizon. This current sheet is subject
to spasmodic episodes of reconnection, the details of which may depend on the prescription
for the electric resistivity and its variation with radius.   

In less than a
millisecond, the magnetic flux threading the horizon has almost completely vanished.
This is qualitatively similar to the rapid evolution of a black hole interacting with a dipolar
force-free configuration, as seen by~\cite{Lyutikov:2011tk}. 
In contrast with what is argued in~\cite{Lyutikov:2011vc}, magnetic reconnection prevents
the emergence of a relatively long-lived split-monopole configuration.  In Sec.~\ref{s:resist}
we discuss whether force-free or ideal MHD calculations more accurately describe
the global structure of the current sheet, and the implications of the slower reconnection 
that is seen in the MHD calculation of \cite{Lyutikov:2011tk}.

\begin{figure}
\begin{center}
\epsfig{file=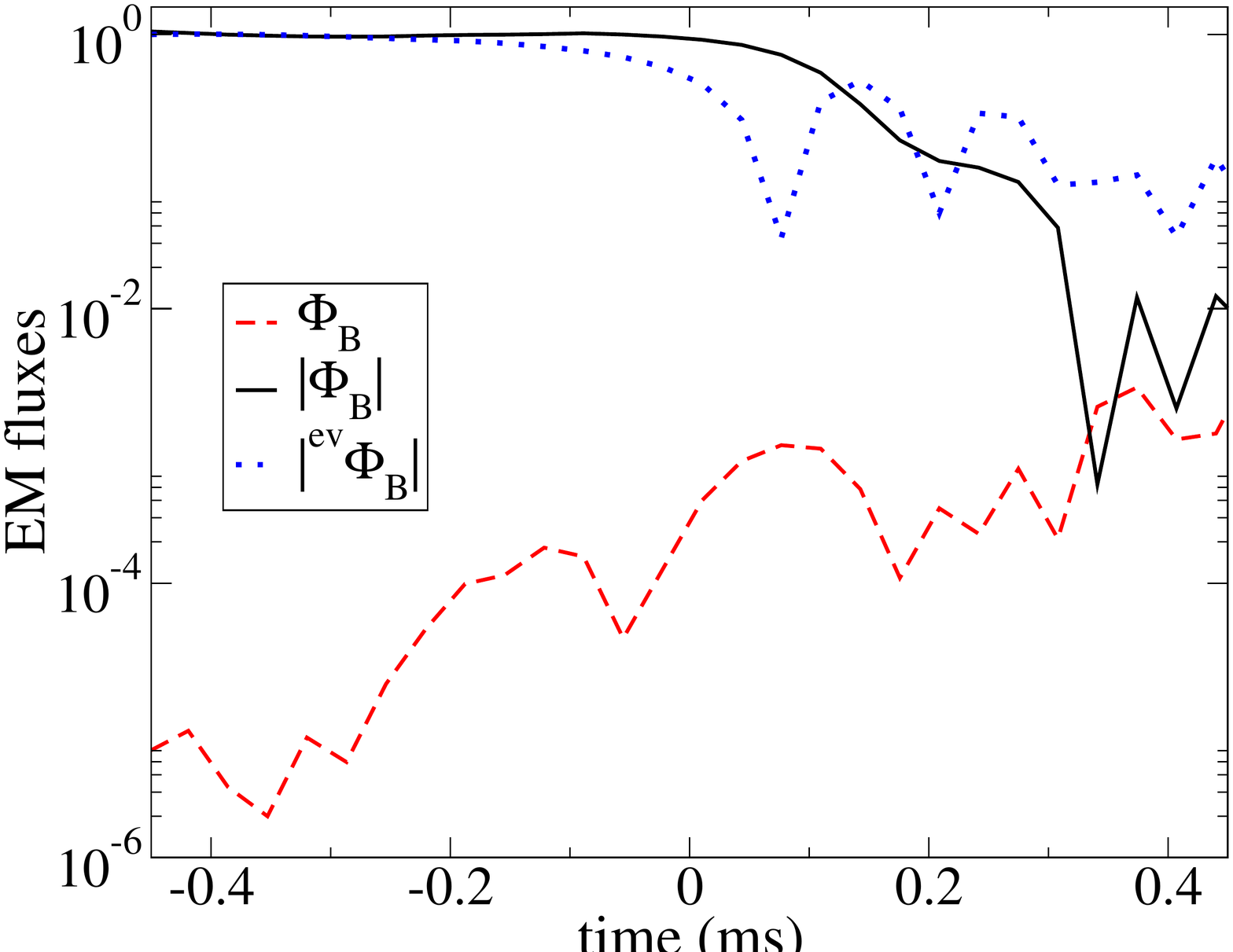, width=2.7in} 
\caption{{\it Rotating unstable star (force-free)}. The magnetic
flux as a function of time, computed at $r=1.5 R_s$, normalized with respect to
the initial value $|\Phi_B (t=0)|$. The unsigned magnetic flux
decreases after the black hole formation, similar to the non-rotating case.}
\label{fig:collapse_EMfluxes_rotating}
\end{center}
\end{figure}

The evolution of the magnetic flux during the collapse is displayed in
Fig.~\ref{fig:collapse_EMfluxes_rotating}, again computed on a surface 
located at $r=1.5 R_s$.  The total (signed) flux is small through
most of the simulation, and only at late times does it becomes comparable 
to the absolute value of the magnetic flux.  In the electrovacuum case, 
the final decay is governed by the main quasi-normal modes of the 
rotating black hole.  One observes faster decay in the force-free run, 
just as as in the non-rotating case, possibly due to the greater
self-inductance of the vacuum black hole.

\begin{figure}
\begin{center}
\epsfig{file=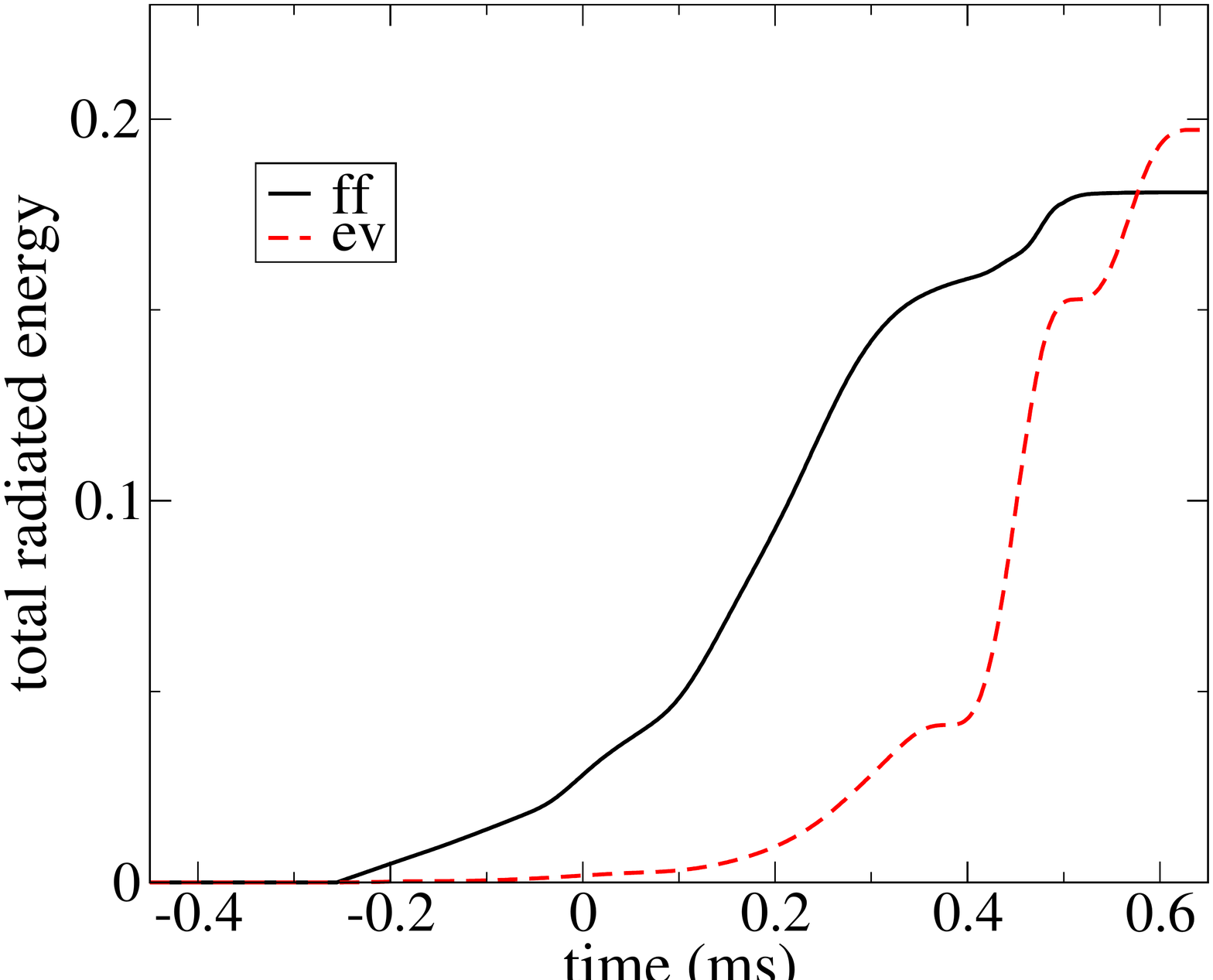, width=2.7in} 
\caption{{\it Rotating, unstable star}. Time integral of the
electromagnetic luminosity, normalized to the peak EM energy of 
the magnetosphere, in both the force-free and electrovacuum cases.
The EM output depends weakly on rotation in the electrovacuum calculations,
whereas in the force-free case the output is much larger
}  
\label{fig:collapse_energyrad_rotating}
\end{center}
\end{figure}

 \begin{figure}
 \begin{center}
 \epsfig{file=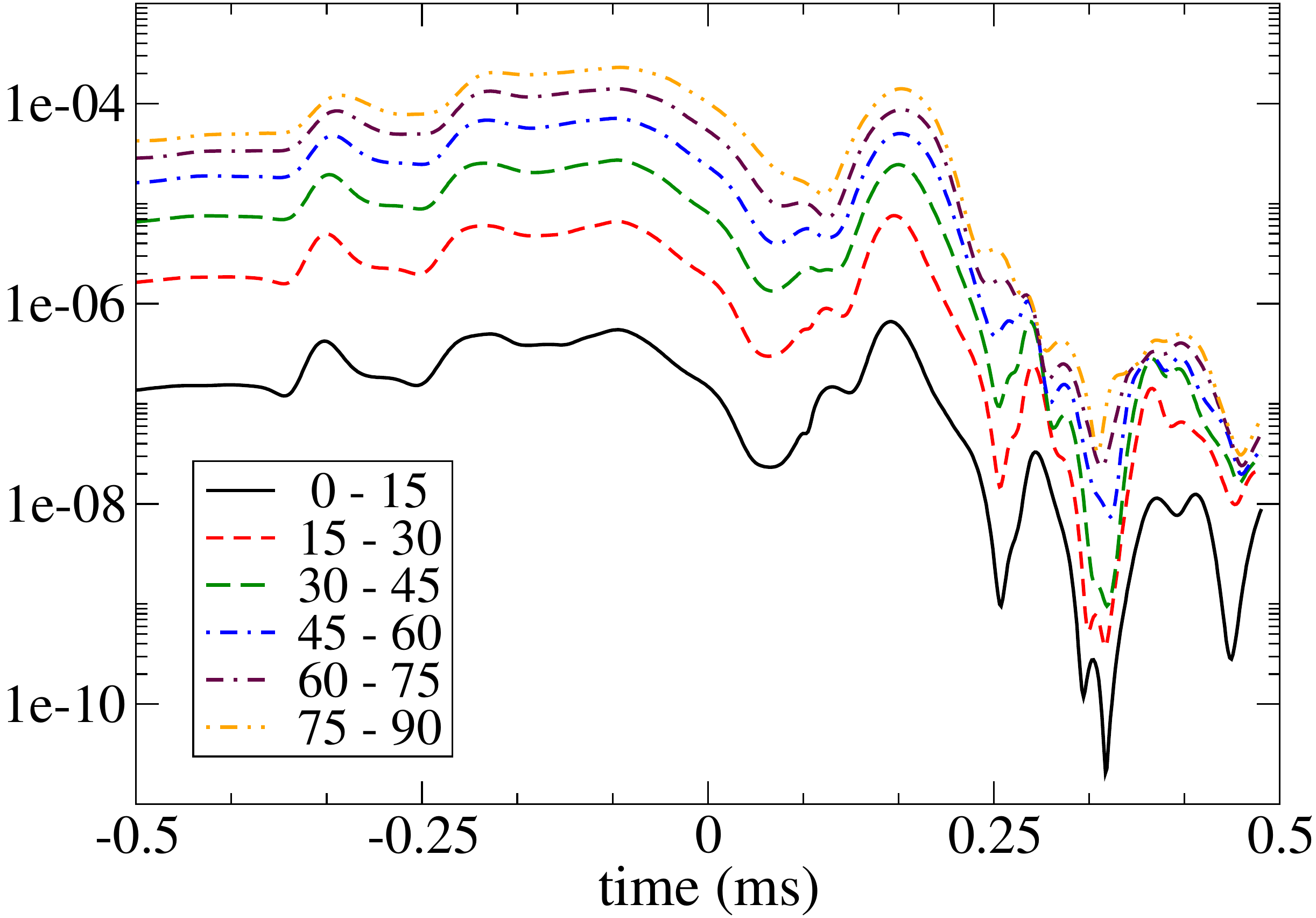,width=3.5in}
 \caption{ {\it Rotating, unstable star (force-free)}. The electromagnetic flux
within annuli defined symmetrically in the northern and southern
hemispheres by concentric cones having apertures in $[i 15^o, (i+1) 15^o] $
($i=0..5$).  The highest flux is within $\theta \in [-50^o,50^o]$.  
Here, as in the non-rotating case, the radiated energy decays 
faster than what would be expected from a quasi-normal mode behavior,
due to the effects of reconnection. }
 \label{fig:collapse_fluxangles_rotating}
 \end{center}
 \end{figure}

 \begin{figure}
 \begin{center}
 \epsfig{file=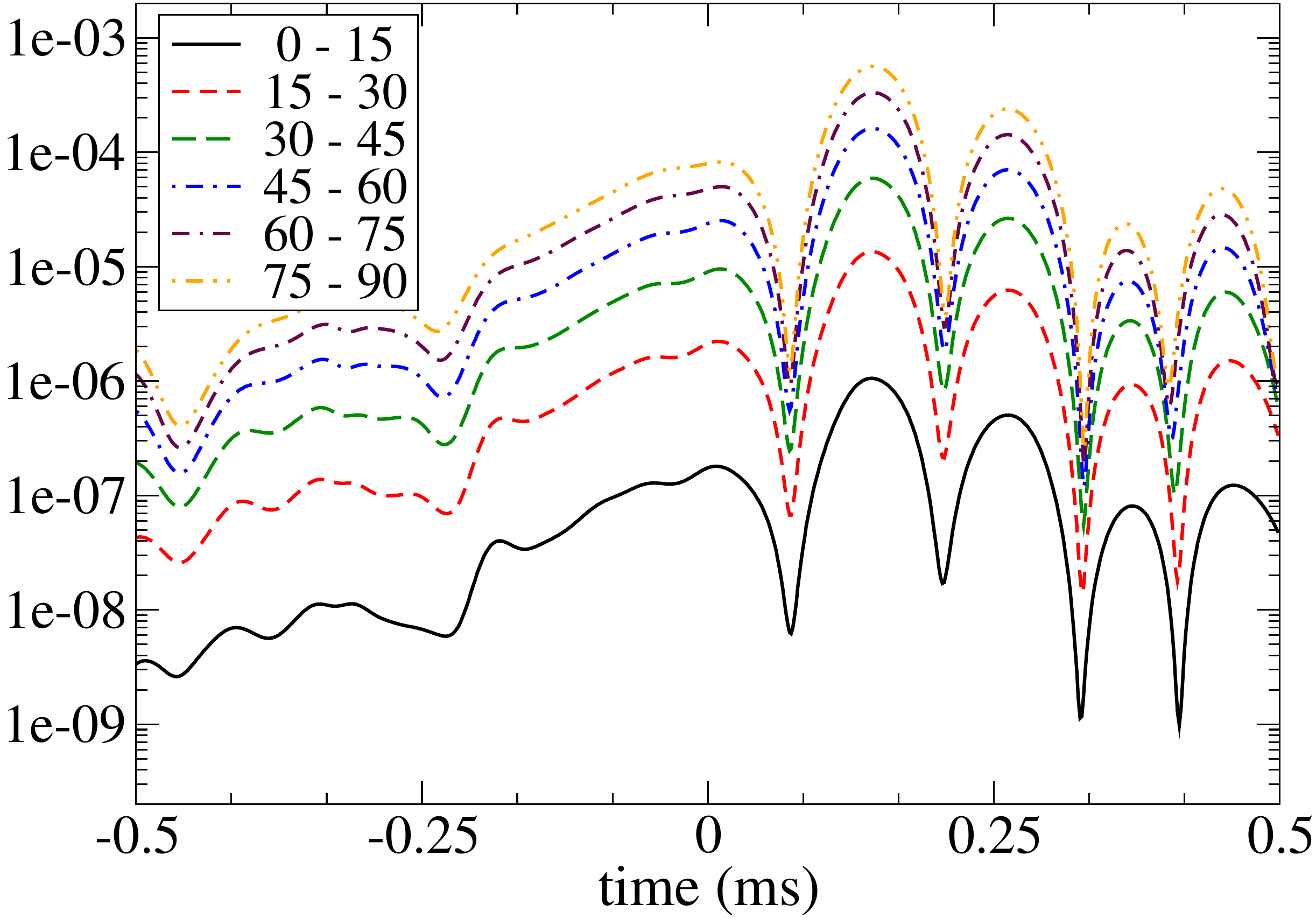,width=3.5in} 
 \caption{ {\it Rotating, unstable star (electrovacuum)}.  The electromagnetic
flux within annuli defined symmetrically in the northern and southern
hemispheres by concentric cones having apertures in $[i 15^o, (i+1) 15^o] $
($i=0..5$). The radiated energy decays exponentially with a rate consistent with that
expected from a quasi-normal mode behavior.}
 \label{fig:collapse_fluxangles_rotating_ev}
 \end{center}
 \end{figure}

The radiated electromagnetic energy is displayed in 
Fig.~\ref{fig:collapse_energyrad_rotating}, rescaled again 
with respect to the energy peak in the magnetosphere.

The simulations indicate that around $20\%$ of the energy
stored in the magnetosphere is radiated during the collapse
in the force-free case, similarly to the electrovacuum case. This
is in clear contrast to the non-rotating case, where a much
larger energy is radiated in the electrovacuum solution than in the force-free one.
We note that although the energy of the magnetosphere is similar in 
the non-rotating and rotating cases (we find $C_{\rm peak} = 1.5$), 
the inclusion of rotation leads to a $20-$fold enhancement in the
EM energy radiated by a collapsing force-free magnetosphere:
$\epsilon_{\rm rad} = 0.18$.  Hence Eq.~(\ref{energy_peak_magnetosphere})
gives $E_{\rm rad} \approx 1.3\times 10^{46} \,B_{\rm pole,15}^2 \, {\rm erg}$,
resulting in a strong average luminosity of
$L \approx 1.3 \times 10^{49} B^2_{\rm pole,15}$ erg s$^{-1}$ during the collapse.

The distribution of the radiated energy is essentially quadrupolar, since
the Newman-Penrose scalar $\Phi_2$ has angular dependence
mainly determined by an $(l=2,m=0)$ mode. 
This radiation propagates outwards with a velocity $v=0.88 c$, 
which is very similar to what was obtained in the non-rotating case. 
As displayed in Fig.~\ref{fig:collapse_fluxangles_rotating}, most of the 
energy is radiated more efficiently near the angle $\theta=\pm 50^o$ with a peak
intensity shortly after the formation of the black hole
(For comparison purposes,
Fig.~\ref{fig:collapse_fluxangles_rotating_ev} illustrates the behavior in the
electrovacuum case). 
Subsequently this emitted radiation fades away rapidly as the magnetic flux is
radiated away.


\section{Implications for Astrophysical Transients:  I.  Gamma-Ray Bursts}
\label{sec:astrophysicsI}

We now consider the implications of our simulation results for high-energy transient phenomena.  
The very luminous gamma-ray bursts (GRBs) are generally believed to result from the formation of a stellar-mass black hole
by sudden gravitational collapse, either in the core of a massive 
star \cite{woosley:1993,paczynski:1998}, or after the merger of
 a neutron star
with another compact object~\cite{eichler:1989,narayan:1992}.   
A rapidly rotating magnetar is an interesting alternative 
\cite{duncan:1992,usov:1992,thompson:2004}.  Our focus here is on ultraluminous EM outflows: 
the same general mechanism is strongly favored in other contexts such as pulsar 
synchrotron nebulae \cite{rees:1974} and AGN jets \cite{blandford:1977}.  

Our calculations focus on the {\it transition} between a magnetar and a black hole,
 as a key part of 
the engine that drives a GRB.  
The implication is that a build-up of magnetic flux in the surface layers of
a rapidly rotating neutron star (formed, e.g., in a binary merger) can driven an electromagnetic outflow 
during the collapse of the star, independently of any Blandford-Znajek process operating afterward.  
As we explain here, this has some advantages over an EM wind operating before the collapse, in that this
brief EM transient is likely to be significantly cleaner.  We also compare the net EM output with that expected
from a post-collapse jet, in the situation where the surface layers of the neutron star, and not the surrounding
disk, are the dominant zone for magnetic flux generated by a dynamo process.

Our calculations have revealed strongly dissipative processes at work in the EM
field surrounding the collapsing star.  These include the reconnection of field lines near the magnetic equator,
large-amplitude oscillations in the field, and (in vacuum calculations) the formation of extended zones where $E^2 \simeq B^2$, 
which in the presence of conducting matter imply relativistic motions of the entrained particles transverse to the magnetic field.
The ejection of loops of magnetic field is 
observed in our non-rotating simulations.  The dynamical evolution of these loops has interesting implications for 
the gamma-ray flares from gravitationally stable magnetars (see Sec.~\ref{sec:magnetar}).

Although our calculations take into account the presence of plasma implicitly by enforcing 
$E\cdot B = 0$, adding the associated force-free current and enforcing $|E|<|B|$, we expect that strong heating will occur in practice.  Where the magnetic field reaches
$\sim 10^{15}-10^{16}$ G, the implied temperature is above $\sim 1$ MeV.  Such a high density of electrons and positrons is
created that macroscopic zones of non-vanishing $E\cdot B$ -- as are required by
most pulsar discharge models (e.g. \cite{ruderman:1975}) -- cannot be maintained.  Effective heating can occur
by other channels:  for example, long-wavelength gradients in the magnetic field are converted efficiently to internal energy 
if the magnetic field becomes turbulent, so that a wide spectrum of wave motions is formed that extends down to
very small scales~\cite{thompson:1998}.   

The relativistic outflow that is emitted by a rapidly collapsing star should therefore be quite hot.  Large-scale magnetic fields
and a relativistic photon-electron-positron plasma will both contribute substantially to the energy flux.  Magnetar outbursts
provide a fairly direct example of this phenomenon.

\subsection{EM Output Before, During, and After Collapse}

The star that collapses to form a black hole passes through three distinct phases:  a pre-collapse phase during which it
emits a magnetized wind if it rotates; the dynamic collapse phase; and, if an orbiting disk is present -- as it is following a neutron
star merger -- an accretion phase that is accompanied by a relativistic jet.  Since our current simulations focus on the intermediate
step, it is worth examining its relative contribution to the total EM output of the star.

A large enough EM output $\Delta E_{\rm collapse}$ is measured in our rotating force-free simulation to 
power some short GRBs, especially if some account is made for beaming.  Here the requirement is that
the surface magnetic field is strong enough to hold off any accretion flow.
At an accretion rate $\dot M$, this implies a polar magnetic field stronger 
than\footnote{A more stringent condition is that the magnetosphere is able to hold off 
the accretion flow out to the corotation radius, as was considered recently by \cite{piro:2011}
in the case of spherical core collapse around a pre-existing, rotating magnetosphere.} 
\begin{eqnarray}\label{eq:bmag}
B_{\rm pole} &\sim& 2\left[{\dot M V_{\rm c}(R_{\rm NS})\over R_{\rm NS}^2}\right]^{1/2} \cr
&\sim& 7\times 10^{15}\,\left({\dot M\over M_\odot~{\rm s}^{-1}}\right)^{1/2}\quad{\rm G}.
\end{eqnarray}
Here $V_c$ is the circular speed, approximated as Keplerian.  In this section we take a stellar mass $M_{\rm NS} = 2.6\,M_\odot$ 
and a radius $R_{\rm NS} = 15~{\rm km}$, as appropriate to a hot and rapidly rotating neutron star.
The dipole field energy before the collapse is 
\begin{equation}\label{eq:edipole0}
E_{\rm dipole,0} = {1\over 12}B_{\rm pole}^2 R_{\rm NS}^3 = 1.5\times 10^{49}\,\left({\dot M\over M_\odot\,{\rm s^{-1}}}\right)\quad{\rm erg}.
\end{equation}

We estimate $\Delta E_{\rm collapse} \sim 0.3\,E_{\rm dipole,0}$, given that the energy released is about 0.2 times 
the peak magnetic energy, which in turn is $C_{\rm peak} \sim 1.5$ times $E_{\rm dipole,0}$.  
From Eq.~(\ref{eq:edipole0}) we obtain 
\begin{equation}\label{eq:dele}
\Delta E_{\rm collapse} \sim 5\times 10^{48}\,\left({\dot M\over M_\odot\,{\rm s^{-1}}}\right)\quad {\rm erg}. 
\end{equation}
The precise numerical value depends non-linearly on the initial specific angular momentum $J/M_{\rm NS}$ through
the factor $\epsilon_{\rm rad} C_{\rm peak}$:  faster initial spins imply stronger winding of the magnetic field during
the collapse.

The magnetic field of an isolated neutron star acts as a couple between its reservoir of rotational energy and a dissipative outflow.  Even in the case 
of a (gravitationally stable) magnetar, the magnetic energy begins to dominate the rotational energy only at an advanced age, as the star spins down.
Therefore the output of the pre-collapse phase could be very large compared with the release of EM energy during
the collapse.  Comparing the spindown energy radiated over a time $\Delta t$ with the external magnetic energy 
(\ref{eq:dele}) gives
\begin{equation}\label{eq:lsd}
{L_{\rm sd} \Delta t\over E_{\rm dipole,0}} = 0.6 \left({P_{\rm NS}\over {\rm ms}}\right)^{-3}
\left({R_{\rm NS}\over15~{\rm km}}\right)^3 \left({\Delta t\over P_{\rm NS}}\right).
\end{equation}
Here we have substituted the spindown power of an aligned, force-free rotator \cite{spitkovsky:2006},
\begin{equation}\label{eq:lsd_dipole}
L_{\rm sd} = {1\over 4}B_{\rm pole}^2 R_{\rm NS}^2 c \left({\Omega_{\rm NS} R_{\rm NS}\over c}\right)^4,
\end{equation}
where $\Omega_{\rm NS} = 2\pi/P_{\rm NS}$. An isolated, rotating star would radiate energy equal to 
$E_{\rm dipole,0}$ in a few milliseconds..

In the context of binary neutron star mergers, one requires a magnetosphere to emerge from the very strong shear layer
near the surface of the merger remnant (see Section~\ref{s:dynamo} for further discussion of how this could happen).  
If the neutron star is formed hot, a lengthy pre-collapse spindown phase would cause significant difficulties with the application to short GRBs,
because the wind generated during the pre-collapse phase is heavily loaded with nucleons and $\alpha$ particles that are driven
outward by charged-current absorption of electron-type neutrinos near the neutrinosphere \cite{duncan:1986,metzger:2007,dessart:2009}.
The connection between short GRBs and neutron star mergers would also be disfavored if this pre-collapse
outflow lasted longer than $\sim 300$ ms, given a characteristic short GRB lifetime of 0.03-0.3 seconds \cite{kouveliotou:1993}.
On the other hand, the survival of the merger remnant for $\sim 100-300$ ms would have the advantage of allowing stronger amplification 
of the magnetic field before the remnant collapses.  The possibility of longer-lived merger remnants for some configurations has been raised by recent simulations 
that employ a realistic, finite-temperature EOS \cite{sekiguchi:2011}.

Although two merging neutron stars are expected initially to have magnetospheres, the torus formed by the tidal disruption of
the lighter star has a pressure vastly exceeding that of a typical pulsar dipole.   The torus would, therefore,
suppress a magnetosphere around the newly formed merger remnant.  The neutrino-driven wind that emerges from the polar regions
of the remnant will comb out the magnetic field, but this field need not initially be coherent across the star or dynamically important.
By the same token, if a torus were entirely absent, then the magnetic field threading the star would dissipate rapidly after the
black hole forms, and the post-collapse phase would contribute negligibly to the output of the star.

It is useful to express the EM power in terms of the ``open'' magnetic flux that connects the surface of the star
to the outflowing wind.  In the case of an isolated star, this is the flux extending beyond the light cylinder,
\begin{equation}\label{eq:phiop}
\Phi_{\rm open}(\Omega_{\rm NS}) \simeq \pi B_r(R_{\rm LC}) R_{\rm LC}^2 = \Phi_{\rm NS} \left( R_{\rm NS}\over R_{\rm LC} \right),
\end{equation}
where $\Phi_{\rm NS} = B_{\rm pole} \cdot \pi R_{\rm NS}^2$ is the dipolar magnetic flux threading the star.
One can then re-write Eq.~(\ref{eq:lsd_dipole}) as
\begin{equation}\label{eq:lsd2}
L_{\rm sd} = {1\over 4\pi^2 c }\left({\Phi_{\rm open}\Omega_{\rm NS} }\right)^2.
\end{equation}

After the star forms a distinct magnetosphere, the accretion torus can continue to influence the wind power by modifying
the open magnetic flux.  Let us suppose that the magnetic pressure dominates the torus ram pressure out to an equatorial
distance $R_A > R_{\rm NS}$.   Approximating the magnetosphere by a dipole, a fraction
\begin{equation}
{\Phi_{\rm open}\over\Phi_{\rm NS}} \sim {R_{\rm NS}\over R_{\rm A}}
\end{equation}
of the stellar flux is trapped by the torus.  In this situation, it is still possible for $\Phi_{\rm open}$
to exceed Eq.~(\ref{eq:phiop}), because the torus can extend inside the stellar light cylinder,
\begin{equation}
{\Phi_{\rm open}\over\Phi_{\rm open}({\rm no~torus})} \sim {R_{\rm LC}\over R_A}
= 3.3\,\left({R_{\rm A}\over 15~{\rm km}}\right)^{-1}\left({P_{\rm NS}\over {\rm ms}}\right).
\end{equation}
During our simulations of the collapse of an isolated star, we observe that the
magnetic field lines are strongly twisted, so that most of the closed magnetic flux opens out.  Then
\begin{equation}
\Phi_{\rm open} \rightarrow \Phi_{\rm open}(\Omega_H),
\end{equation}
where $\Omega_H$ is the angular velocity of the horizon.

A torus also plays an important role after the black hole forms by trapping a certain fraction of the magnetic flux
that threads the star.  If the magnetic field is strong enough to hold off the torus from the star before the collapse,
then it will have a similar effect after the collapse.  We therefore calculate the Blandford-Znajek power emerging from the horizon 
by assuming a uniform flux density out to some radius $R_A'$.  From Eq.~(8.65) of \cite{frolov:1998}, one gets,
\begin{eqnarray}\label{eq:lbz}
L_{\rm BZ} &\sim& {2\over 15}\left[{\Omega_F(\Omega_H-\Omega_F)\over\Omega_H^2}\right]
\left({\Omega_H R_H\over c}\right)^2\,R_H^2\langle B\rangle^2 c\cr
&\alt& {1\over 30\pi^2 c} \left({\Phi_H\Omega_H}\right)^2 ,
\end{eqnarray}
where $\langle B\rangle$ is the flux density threading the region interior to the torus, and $\Phi_H = \pi\langle B\rangle R_H^2$.  
The difference in the normalizations of Eqs.~(\ref{eq:lsd2}) and~(\ref{eq:lbz}) largely reflects 
the fact that the torque on the black hole is maximized when the magnetic field has an angular velocity $\Omega_F = \Omega_H/2$.  

We can now relate Eq.~(\ref{eq:lbz}) to the EM power generated before the collapse.
The spin angular momentum is approximately conserved during the collapse, $J \simeq I_{\rm NS}\Omega_{\rm NS}$,
and we also set $M_{\rm BH} = M_{\rm NS}$.  Then
\begin{equation}
{J\over GM_{\rm BH}^2/c} = \varepsilon_I\left({P_{\rm NS}\over {\rm ms}}\right)^{-1}\,\left({R_{\rm NS}\over 15~{\rm km}}\right)^2
\left({M_{\rm BH}\over 3\,M_\odot}\right)^{-1},
\end{equation}
where $\varepsilon_I = I_{\rm NS}/M_{\rm NS}R_{\rm NS}^2 \sim 0.3$.  The angular frequency of the black hole is
$\Omega_H R_H/c = J/M_{\rm BH}R_{\rm S}c = Jc/2GM_{\rm BH}^2$,  and is related to $\Omega_{\rm NS}$ pre-collapse by
\begin{equation}\label{eq:omratio}
{\Omega_H\over\Omega_{\rm NS}} \simeq 0.3\left({R_{\rm NS}\over R_{\rm H}}\right)^2\left({\varepsilon_I\over 0.3}\right).
\end{equation}
where $R_{\rm Sch} = 2GM_{\rm BH}/c^2$ is the Schwarzschild radius.
One expects $\langle B\rangle \sim B_{\rm pole}$ after the collapse if most of the magnetic flux threading the star is
open before the collapse, and the pre-collapse magnetosphere is limited in size.
The proportion of the trapped flux threading the hole is
\begin{equation}\label{eq:phiratio}
{\Phi_H\over\Phi_{\rm open}} = \left({R_H\over R_A'}\right)^2 \sim \left({R_H\over R_{\rm NS}}\right)^2.
\end{equation}

We can now show that the EM power is suppressed immediately following the collapse.  Substituting equations (\ref{eq:omratio}) 
and (\ref{eq:phiratio}) into (\ref{eq:lbz}) gives
\begin{equation}\label{eq:lbz_2}
{L_{\rm BZ}\over L_{\rm sd}} \simeq 0.01,
\end{equation}
with a coefficient $(R_A/R_{\rm NS})^2\,(R_{\rm NS}/R_A')^4 (\varepsilon_I/0.3)^2$.
Additional power will flow along the magnetic field lines that thread the ergosphere, but this portion of the black hole magnetosphere
will mix with the accretion flow and may be less strongly magnetized.  

The interesting conclusion here is that the rapid twisting up of the magnetic field during the collapse can generate
a larger EM output than a fairly extended jet emission after the collapse.  It is worth summarizing the three main sources of this
result:  i) before the horizon forms, the EM power is proportional to $\Omega^2$ rather than $\sim \Omega^2/4$;  ii) the magnetic
flux remains pinned in the star for a few rotation periods during the collapse (before the onset of the black hole), and then springs out to fill a larger volume; and iii)
the relation between rotation frequency and angular momentum is enhanced by a factor $\sim \varepsilon_I^{-1} \sim 3$ prior to the collapse;
in other words, relativistic gravity has the effect of softening the growth of the rotation frequency as the star collapses.

A reduction in the trapped flux (due to outward diffusion of the magnetic field into the torus) would, in this situation,
initially {\it increase} the Blandford-Znajek power flowing from the horizon.  The pressure of the trapped field approximately balances
the ram pressure of the accretion flow at some point outside the horizon.  Therefore a reduction in the trapped flux allows the
flow to reach closer to the black hole, and attain higher pressures, before being interrupted.  Although we are considering the
flux originating in a dynamo process before the collapse (Sec.~\ref{s:dynamo}), continued flux generation in the torus by the 
magnetorotational instability \cite{balbus:1991} could play a role in modulating the jet power.

\subsubsection{Numerical Comparison}
Let us consider the EM energy that would be radiated following a binary NS merger, if the remnant 
survives long enough to form a magnetosphere (Sec.~\ref{s:dynamo}).  Taking $\Delta E_{\rm collapse} \sim 0.3E_{\rm dipole,0}$, 
and normalizing to an accretion rate $0.1 M_\odot$ s$^{-1}$ through Eq.~(\ref{eq:edipole0}), 
gives $\Delta E_{\rm collapse} \sim 5\times 10^{47}$ erg.
We expect that the gain factor $\Delta E_{\rm collapse}/E_{\rm dipole,0}$ depends non-linearly on the pre-collapse rotation rate,
since the winding of the magnetospheric field results from a competition between differential rotation and torsional
wave motion.  In addition, $\Delta E_{\rm collapse}$ depends indirectly on the torus
mass and pressure through the strength of the magnetic field that is required to hold off the torus material from the neutron
star surface.   The larger accretion rate in a collapsar environment implies a larger transient energy.

It is possible to make a direct comparison with the Blandford-Znajek jet that follows the collapse, if the magnetic flux
that threads the black hole is left behind by the collapsing magnetar.  Combining equations (\ref{eq:lsd}) and (\ref{eq:lbz_2}),
and taking pre-collapse rotation period and radius $0.8$ msec and $15$ km, one finds that an energy $\sim 0.3 E_{\rm dipole,0}$ 
would be radiated by a BZ jet over $\sim 20$ ms.  Equivalently, a Blandford-Znajek jet from a $\sim 3\,M_\odot$ 
black hole with $Jc/GM_{\rm BH}^2 \sim 0.7$ would generate a power (\ref{eq:lbz}) $L_{\rm BZ} \sim 1\times 10^{50}\, B_{\rm 15}^2
$ erg/s.  
A recent binary merger simulation with GRMHD and dynamical gravity \cite{Rezzolla:2011da} presented a polar 
magnetic field $\sim 7\times 10^{14}$ G developing from a  much weaker seed field in the torus.  The equation of state used gave 
a 
fast collapse to a BH (within $\sim 10$ ms), and therefore precluded the surface shear dynamo that we have conjectured. (Note that this polar field
could be affected by numerical resistivity, e.g.~\cite{obergaulinger:2010}).

\subsubsection{Baryon Poisoning}

Comparing the output $\Delta E_{\rm collapse}$ with Eq.~(\ref{eq:dele}), one
sees that the pre-collapse star would release a comparable energy within $\sim 100$ ms after forming a magnetosphere.  
However, it is well known (e.g. \cite{metzger:2007}) that such an outflow would bear a much higher density of nucleons 
than a Blandford-Znajek jet from a black hole, due to the absorption of electron-type neutrinos and anti-neutrinos.  
We expect that this nucleon loading would be strongly suppressed in the dynamical collapse phase, due to i) the 
relatively short duration of the emission; ii) the strongly wound field geometry ($B_\phi/B_P \agt 5$); and iii) redshifting
effects.  Combining these effects suggests a significant suppression $\alt 0.01\times 0.1 \sim 10^{-3}$ in the nucleon loading, 
so that the mass ejected would essentially be that present in the magnetosphere before the collapse.

\subsection{Emergence of a Magnetosphere via Dynamo \\ Action in a Surface Shear Layer}\label{s:dynamo}

The immediate aftermath of a binary neutron star merger is distinguished from disk accretion onto a black hole, in that
the velocity shear is strongest where the orbiting material makes a transition from centrifugal to hydrostatic support.  
Another feature which distinguishes the merger remnant from ordinary accreting neutron stars (X-ray pulsars) is
that it does not initially have an ordered magnetosphere.  

Long after the first stage of the merger is complete, the velocity shear provides a tremendous source of free energy
for amplifying a magnetic field.  The power dissipated in the material settling onto the neutron star surface is
\begin{eqnarray}\label{eq:shearpower}
L_{\rm shear} &\sim& {1\over 2}\dot M\left[V_c^2(R_{\rm NS}) - \Omega_{\rm NS}^2 R_{\rm NS}^2\right]\cr
&\sim&  3\times 10^{52}\,\left({\dot M\over 0.1M_\odot~{\rm s}^{-1}}\right)
f_{\rm shear}\quad{\rm erg~s^{-1}},\nn
\end{eqnarray}
where
$f_{\rm shear} \equiv 1-[\Omega_{\rm NS}R_{\rm NS}/V_c(R_{\rm NS})]^2$.
Note that the surface shear becomes more radially concentrated with time: as differential rotation is erased in the interior of the
merger remnant, the surface shear is maintained by continuing accretion. The inner part of the
shear layer develops positive $d\Omega/dr$ and the magnetorotational instability is extinguished.

A strong {\it feedback mechanism} is present which allows rapid magnetic field
growth, but causes this growth to saturate once the star is able to form a magnetosphere that holds off the accretion
flow.  When the magnetosphere is present, the accreting material follows the magnetic field
and reaches the star at the same angular velocity. Shearing of the magnetic field in the outer layers of the
star is therefore turned off.  This effect is clearly demonstrated in the 3D accretion simulations of \cite{romanova:2011}.

Only a tiny fraction of the accretion energy $L_{\rm shear} \Delta t$ must be converted to a poloidal magnetic field
to hold off the accretion flow:  substituting (\ref{eq:bmag}) into the dipole energy (\ref{eq:edipole0}) gives
\begin{eqnarray}
{E_{\rm dipole}\over L_{\rm shear}\Delta t} &=& {2\over 3 f_{\rm shear}\Delta t}
\left({R_{\rm NS}^3\over GM_{\rm NS}}\right)^{1/2}\nn
&=& 6\times 10^{-4}\left({\Delta t\over 100~{\rm ms}}\right)^{-1}f_{\rm shear}^{-1}.
\end{eqnarray}
over a duration $\Delta t$.

Low-mass X-ray binaries provide a nice example of systems
where this feedback appears to operate.  The millisecond radio pulsars that are descended from them have
magnetic fields that are just strong enough ($B \sim 10^8-10^9$ G) to hold off an Eddington-level accretion flow onto
a neutron star -- but not much stronger. The absence of persistent pulsations in the majority of LMXBs then requires
that the magnetic field be aligned with the angular momentum of the accretion flow.

Now let us consider how the magnetic field evolves in the surface shear layer.  The field present initially in the merging stars is rapidly
amplified by a Kelvin-Helmholtz instability \cite{price:2006,obergaulinger:2010}, or the magnetorotational instability \cite{balbus:1991}.
Rapid growth of the magnetic field on large scales is sensitive to the speed of magnetic reconnection in the fluid, and requires
three-dimensional motions.  The two-dimensional wrapping of a magnetic field by Kelvin-Helmholtz vortices does not generate
net flux; and the initial growth length of the MRI is very small compared with the scale height of the torus.  MRI growth is fastest on
a scale $k^{-1} \sim B/(4\pi\rho)^{1/2}\Omega \sim 10^{-4}(B/10^{12}~{\rm G})\,r$ in a torus of mass $\sim 0.01\,M_\odot$ \cite{balbus:1991}.

The hydrostatic structure of the inner shear layer makes it easier for the magnetic field to be pinned and retained than it would be in the
surrounding torus.  The magnetic field threading the shear layer is wound up, and since $d\Omega/dr > 0$, the mean toroidal flux density
grows at least in a linear manner.   If a hot, rotating, and massive neutron star ($M_{\rm NS} \sim 2.6-3\,M_\odot$) can
survive collapse for more than $\sim 100$ ms, as is suggested by recent simulations of \cite{sekiguchi:2011}, then this toroidal field becomes
quite strong.  Even if the seed poloidal field is as weak as $B_{P,0} \sim 10^{13}$ G --
within the range of pulsar fields -- then the toroidal field reaches
\begin{eqnarray}
  B_\phi &\sim& [\Omega_c(R_{\rm NS})-\Omega_{\rm NS}]\Delta t\, B_{P,0}\cr
 &=& 1\times 10^{16}~\left({\Delta t\over 100~{\rm ms}}\right)\left({\Omega_c(R_{\rm NS})-\Omega_{\rm NS}\over
10^4~{\rm s^{-1}}}\right)\nn
&&\quad\quad\quad\quad\times \left({B_{P\,0}\over 10^{13}~{\rm G}}\right)\quad{\rm G}.
\end{eqnarray}
Here $\Omega_c(R_{\rm NS})$ is the angular frequency at the surface of the torus.

Exponential magnetic field growth becomes possible when the wound-up field can rise buoyantly through the shear layer.
The large dissipated power (\ref{eq:shearpower})
will generate a strongly positive entropy gradient in the inner part of the shear layer.  Where the material is
convectively stable, only magnetic fields stronger than $\sim (GM_{\rm NS} M_{\rm shear}/R_{\rm NS}^4)^{1/2}                                       
\sim 1\times 10^{17}\, (M_{\rm shear}/0.1~M_\odot)^{1/2}$ G can directly overcome the pressure of the overlying material.
But even in this case, the intense flux of electron-type neutrinos allows a magnetic field
stronger than $\sim 10^{15}$ G to rise buoyantly on the Alfv\'en timescale, by erasing gradients in
entropy and electron fraction that impede buoyancy \cite{thompson:2001a}.  The buoyancy time is
$t_A \sim \ell_P (4\pi\rho)^{1/2}/B_\phi \sim 1~(B_\phi/10^{16}~{\rm G})^{-1}(\rho/10^{14}~{\rm g~cm^{-3}})^{1/2}~{\rm ms}$
across a pressure scale height $\ell_P \sim R_{\rm NS}/4 \sim 3-4~{\rm km}$.  Thus an exponential feedback loop appears quite likely
in this situation.

One observes that the large-scale dipole magnetic field that emerges may be sensitive to the seed magnetic field,
in the sense that a {\it minimal} seed field is required for the linearly wound field to reach the buoyancy threshold.
Nonetheless, it is also clear that magnetar-strength magnetic fields do not require magnetar-strength seed fields
in the presence of persistent surface shear.

The strength of the magnetosphere that eventually emerges results
from a competition between the finite rate of amplification and the diminishing
shear stress at the surface of the star.  At an age of $\sim 10 (100)~{\rm ms}$, a poloidal magnetic field stronger than
$\sim 10^{16}$ G ($10^{15}$ G) will begin to apply a strong, negative feedback on the surface shear.
Once the torus mass drops to the point that accretion is mediated mainly by magnetic stresses, the
accretion rate is $\dot M_0 \sim M_{T,0}/t_{\rm diff,0} \sim 10\,(\alpha/0.1)(M_{T,0}/0.1~M_\odot)\,M_\odot$ s$^{-1}$.  Here
$M_{T,0}$ and $t_{\rm diff,0} \sim 10(\alpha/0.1)^{-1}~{\rm ms}$ are the initial torus mass and diffusion time in this
viscous phase, and $\alpha$ is the viscosity coefficient.  The accretion rate drops as the torus material spreads outward,
as $\dot M \sim \dot M_0(t/t_{\rm diff,0})^{-4/3}$, as long as the torus remains geometrically thick and conserves its
angular momentum (e.g. \cite{metzger:2008}).

\subsection{Galactic Magnetars and Delayed Collapse}\label{s:delay}

We now consider the observational imprint of a magnetar if it forms by the accretion of a thin layer of strongly sheared
material, but is (initially) gravitationally stable.  We focus on the rate of energy loss and the spin history of the star.

There is a substantial reduction in the energy release by spindown, compared with a star that initially rotated as a solid body,
due to the internal rearrangement of angular momentum.  Approximate solid-body rotation is attained on the poloidal Alfv\'en
timescale $R_{\rm NS}(4\pi\rho)^{1/2}/B_{\rm pole}$, which is generally much shorter than the magnetic-dipole spindown time
\begin{equation}
t_{\rm sd} = {I_{\rm NS} P_{\rm NS}^2 c^3\over  2\pi^2 B_{\rm pole}^2 R_{\rm NS}^6}.
\end{equation}
Most of the shear energy is then dissipated {\it internally}.

A delayed collapse is possible if the magnetar only slightly exceeds the 
maximum mass for a non-rotating, zero-temperature neutron star.   The loss of rotational support by a magnetic
wind would trigger a collapse on the rotational braking time.

Quite generally, the accretion of a thin layer of strongly sheared material provides
an attractive mechanism for creating magnetars, and so the results of this section
should have a broader application to the Galactic magnetar population,
even if these stars are well below the maximum mass.  Several possible channels are available:  the rotation of the accreting
material could be generated by an instability of the supernova shock \cite{thompson:2000,blondin:2007,iwakami:2008,fernandez:2010};
or the neutron star could be exposed to material with very strong vorticity in a merger event, e.g. when it merges
with a companion CO white dwarf, or with the core of an evolving Be star.  

The spindown power that is released by the magnetar is then reduced significantly
compared with a star which rotates uniformly with the same angular velocity as the surface material.  The final spin period resulting from
the accretion of a layer of mass $M_{\rm shear}$ and rotation period $P_{\rm shear}$ onto a neutron star of total mass $M_{\rm NS}$ is
\begin{eqnarray}
P_{\rm NS} &=& \left[{(2/3)M_{\rm shear} R_{\rm NS}^2\over I_{\rm NS}}\right]^{-1} P_{\rm shear}\cr
           &=& 10.5\,j_{\rm shear}^{-1}\left({M_{\rm NS}\over 2\,M_\odot}\right)\quad{\rm ms},
\end{eqnarray}
where $j_{\rm shear} \equiv (M_{\rm shear}/0.1\,M_\odot)(P_{\rm shear}/{\rm ms})^{-1}$.
As the star spins down, it deposits a rotational energy
\begin{equation}
{1\over 2}I_{\rm NS}\Omega_{\rm NS}^2
= 2.5\times 10^{50}\,j_{\rm shear}^2\left({M_{\rm NS}\over 2\,M_\odot}\right)^{-1}\left({R_{\rm NS}\over 10~{\rm km}}\right)^2\quad{\rm erg}
\end{equation}
in the surrounding shock wave.
For a magnetar of polar field $B_{\rm pole}$, the corresponding spindown luminosity is fairly modest,
\begin{eqnarray}
L_{\rm sd} &=& 1.4\times 10^{45}\,\left({B_{\rm pole}\over 10^{15}~{\rm G}}\right)^2\,\left({P_{\rm NS}\over 10~{\rm ms}}\right)^{-4}\nn
&&\quad\quad\quad\quad\quad\times\left({R_{\rm NS}\over 10~{\rm km}}\right)^6\quad{\rm erg~s^{-1}},
\end{eqnarray}
and the star enters the extended spindown phase after a time 
\begin{eqnarray}
t_{\rm sd} &=& 2\,\left({B_{\rm pole}\over 10^{15}~{\rm G}}\right)^{-2}\,
     \left({P_{\rm NS}\over 10\,{\rm ms}}\right)^2\left({M_{\rm NS}\over 2\,M_\odot}\right)\nn
&&\quad\quad\quad\quad\quad\quad\quad\times\left({R_{\rm NS}\over 10~{\rm km}}\right)^{-4}\quad{\rm day}.
\end{eqnarray}
In this situation, only a very strong internal magnetic field ($B_{\rm toroidal} \agt 10^{17}$ G) would induce sufficient triaxiality in the star
that the gravity wave torque competed with the external electromagnetic torque \cite{cutler:2002}.

\section{Plasmoid Dynamics and Radiation: Implications for Magnetar Flares}
\label{sec:magnetar}

Escaping loops of magnetic field are observed in our non-rotating simulations.   These loops are formed by magnetic reconnection, which means
their size and rate of formation are sensitive to the treatment of resistivity, and to the disregard of stresses associated with bulk plasma flow
by the adoption of the force-free equations.  An outgoing plasmoid (containing plasma and a closed magnetic field)
has a well-defined center of mass:  we find that its bulk motion is 
measurably smaller than the speed of light, $V_{\rm bulk} \sim 0.9 c$ and $\Gamma_{\rm bulk} \sim 2$, 
at a distance $r \sim 10$ times the pre-collapse neutron star radius $R_{\rm NS}$.   This is significantly less relativistic
than the bulk motion expected for a gas of freely expanding particles released by the collapsed star ($\Gamma_{\rm bulk} \agt r/R_{\rm NS} \sim 10$).  

After addressing each of these issues, we make contact with the giant gamma-ray flares of the Galactic magnetars.  These
appear to involve the ejection of an energetic plasmoid ($\sim 10^{44}-10^{46}$ erg), 
but due to the shearing of the magnetic field lines rather than the collapse of the star \cite{thompson:1995}.

\subsection{Dependence of Reconnection Rate on Resistivity Model}\label{s:resist}

When magnetic field lines of an opposing sense are stretched out and forced into contact,
the rate at which they reconnect is sensitive to the microscopic model of resistivity.  In an ohmic
plasma with a uniform resistivity, a long current sheet forms and reconnection is very slow;
fast reconnection with an x-point geometry depends on a local maximum in the resistivity
\cite{baty:2006}.  In some contexts, such as the Solar corona, the microscopic explanation 
for this behavior may be provided by the Hall terms in the conductivity \cite{1998JGR...103.9165S}.
These are relatively less important if the plasma is loaded with $e^+/e^-$ pairs, as would be 
expected in the magnetosphere of a collapsing magnetar.  In a fluid, small-scale hydromagnetic turbulence 
appears to greatly accelerate the reconnection rate \cite{kowal:2009}; but whether such a process 
can operate in a low-$\beta$ plasma of astrophysical dimensions is not yet determined.  

In the present work we find evidence for relatively fast reconnection of magnetic field lines, 
$V_{\rm rec} \sim 0.1~c$, as calculated in the force-free approximation (Sec.~\ref{subsec:nonrotating}).
This results from a change in topology of the field lines (the formation of an x-point).  It is not
 due to strongly enhanced dissipation in an extended equatorial current sheet: $E.J$ dissipation is
 measured to be small in both the rotating and non-rotating cases.  Such a concern arises in 
force-free calculations of pulsar magnetospheres, where the magnetic field is less dynamic and 
is anchored in the star.  There the absence of fluid pressure support in the current sheet leads 
to a rapid collapse of the magnetic field toward the sheet, unless explicitly 
compensated \cite{Spitkovsky:2006np,McKinney:2006sc}.  In the present case, after the black hole 
forms the inflow of magnetic flux toward the equator can continue in a more dynamic manner into 
the star, or out to the computational boundary.

A recent treatment of reconnection in the magnetosphere of a stationary black hole 
\cite{Lyutikov:2011tk} illustrates a slower field decay when employing an ideal MHD treatment, as compared to a force-free approach.
(Resistivity in the MHD case arises through the numerical approximation.) Such a slower decay is also observed
at late stages after the formation of the black hole in our simulations when comparing force-free and ideal MHD
fields\footnote{We employ however a different $\Gamma$ and our implementation does not cope
with as large magnetizations as that in~\cite{Lyutikov:2011tk}.}; 
though by this time the field strengths are orders of magnitude below their peak values.
A comparison of force-free and MHD reconnection calculations which explores more general
resistivity models, and their influence on the reconnection geometry, remains to be developed.

\subsection{Magnetic Reconnection Delayed \\ by Plasma Outflow}\label{s:rec_delay}

Reconnection is usually studied in a context where the plasma flow speed along the 
magnetic field is a small fraction of the Alfv\'en speed:  for example, in a steady
MHD wind, the conservation of angular momentum implies a slow drift of particles
along the spiral magnetic field outside the Alfv\'en critical point.  However, in
some contexts, such as magnetar flares, the flow speed can approach the speed of
light.  Similarly, in our collapse simulations we see large-amplitude motions 
on the magnetic field loops threading the neutron star, which suggest strong plasma heating.
The backreaction of the outflowing plasma on the reconnection of field lines is
not taken into account.

Even when the magnetic field has a tendency to reconnect through an x-point,
reconnection will be {\it delayed} until the kinetic pressure $\sim U\beta^2$
of the outflowing pair-photon plasma (with thermal energy density $\sim U$) drops
below the Poynting flux that would flow toward the current sheet in the absence
of plasma flow.  One requires
\begin{equation}
U\beta^2 \alt 0.1 {V_A\over c} {B^2\over 8\pi},
\end{equation}
where the coefficient on the right-hand-side is appropriate for fast x-point reconnection.

The magnetic field lines are stretched outward by the plasma flow beyond an Alfv\'en radius \cite{thompson:1998}
\begin{eqnarray}\label{eq:ralf}
{R_A\over R_{\rm NS}} &=& \left({B_{\rm pole}^2R_{\rm NS}^2c\over 4L_\gamma}\right)^{1/4} \nn
    &=& 16\,\left({B_{\rm pole}\over 10^{15}~{\rm G}}\right)^{1/2}
\left({L_\gamma\over 10^{47}~{\rm erg~s^{-1}}}\right)^{-1/4}
\end{eqnarray}
given $R_{\rm NS} \sim 10~{\rm km}$.  Note that, at the peak of the outflow,
the pressure of the pair-photon fluid is comparable to the pressure of the stretched magnetic field lines at $r = R_A$, and 
outside this radius grows as $\sim (r/R_A)^2$  with respect to the split-monopole field pressure. For example, in a magnetar giant flare,
the $\sim 0.1$ s width of the main gamma-ray pulse is comparable to the time for magnetic and elastic stresses to
rearrange the stellar interior;  but it is $\sim 300$ times longer than the flow time out to the Alfv\'en radius (\ref{eq:ralf}),
and therefore much longer than the timescale for x-point reconnection at a speed $\sim 0.1 c$.

\subsection{Radio Afterglow from Strongly Magnetized Outflows}\label{s:radio}  

A magnetically-dominated plasma that is ejected from a collapsing magnetar (or a nearby Soft Gamma Repeater)
can be a much stronger source of synchrotron emission than a shocked plasma of comparable energy density.
Existing calculations of radio afterglows of short GRBs (\cite{nakar:2001} and references therein), as well as calculations of the
radio afterglow of SGR giant flares \cite{granot:2006}, focus on synchrotron emission by a population of non-thermal electrons
that are accelerated at a shock that leads the outflow.  For the bulk of GRBs, one infers
efficiencies of conversion $\varepsilon_e, \varepsilon_B \sim 0.1$ of bulk kinetic energy to non-thermal electrons and to magnetic
fields downstream of the forward shock.  These moderate values of $\varepsilon_e$ and $\varepsilon_B$ are consistent with the
broadband tails of radio--X-ray emission that follow the brief, bright gamma-ray phase.  

The synchrotron emission from a magnetically-dominated plasmoid will be proportionately much brighter, by up to a factor $\sim 100$,
for two reasons.  First, and most obviously, $\varepsilon_B$ now approaches unity.   Second, if the magnetic energy density dominates 
the thermal energy density, then a cascade process (involving the creation of high-wavenumber Alfv\'en modes) preferentially 
heats the electrons \cite{quataert:1999}.  If the plasma is very relativistic, damping is mainly due to charge-starvation of 
the waves, at wavenumbers where the amplitude of the fluctuating current density begins to exceed $en_e c$ \cite{thompson:1998,thompson:2006}. 
(A different damping mechanism operates at higher ion densities:  the waves are Landau-damped on the parallel motion of the electrons.)
In practice, the relative amplitudes of the bulk synchrotron emission and the shock emission will depend
on the degree of disorder in the plasmoid magnetic field.  But for magnetar flares, indirect evidence that bulk synchrotron emission dominates
comes from a rapid initial drop in radio flux that is consistent with the sudden compression of the plasmoid,
followed by rapid adiabatic cooling \cite{granot:2006}.

\subsection{Applications to Magnetar Outbursts}

Magnetar outbursts involve more limited releases of energy that leave the original star intact.  They are triggered when
the footpoints of a $\sim 10^{15}$ G magnetic field are strongly sheared by an elastic instability of a 
neutron star crust, thereby generating a hot plasma and an intense burst of gamma-rays \cite{thompson:1995,thompson:2001b}.  The first, extremely bright, stage 
of a giant magnetar flare lasts only $\sim 100~{\rm ms}$ and bears a considerable resemblance to a `classical' gamma-ray burst,
demonstrating high temperatures ($kT \agt 200$ keV) and a significant non-thermal component to the spectrum~\cite{hurley:2005}.  
(This phase saturates almost all X-ray detectors, but was well-resolved by the Geotail experiment \cite{tanaka:2007}
in the 27 August 1998 and 27 December 2004 flares.)
The duration and luminosity are consistent with the internal rearrangement of the magnetic field in a neutron star, with a strength 
$\sim 4-5\times 10^{15}$ G based on considerations of magnetic field transport and global flare energetics \cite{thompson:1996},
several times stronger than the standard dipole expression for the spindown-derived magnetic field.   

The giant flares appear to involve the ejection of a plasmoid.  
The combination of fast variability and extreme luminosity (up to $\sim 10^{48}$ ergs s$^{-1}$) implies, through the usual arguments 
of gamma-ray opacity \cite{cavallo:1978}, that the emitting plasma has expanded to a much larger volume than that of the neutron star.  
This expanding plasma is an excellent electrical conductor and must carry some of the stellar magnetic field with it.  

This expected property of magnetar giant flares helps to explain two apparently contradictory phenomena.
A straightforward argument based on the theory of thermal fireballs (e.g. \cite{shemi:1990}) shows that the expanding plasma 
is moving with a high Lorentz factor at the radius ($\agt 10^8$ cm) where it becomes transparent to the gamma-rays
\cite{thompson:2001b}.   The observation of variability
on a timescale $\delta t_{\rm var} \sim 4-20~\rm{ ms}$
 in the gamma-ray flux implies strong constraints on the baryon rest energy flux.   The
advected electrons and ions must become transparent to the gamma-rays close enough to the magnetar that the differential light-travel 
time $r/2\Gamma^2c$ across the outflowing plasma is shorter than $\delta t_{\rm var}$.  Given a total outflow luminosity $L$,
this implies that the ion rest energy contributes no more than a fraction
\begin{eqnarray}
{\dot Mc^2\over L} &<& \left({16\pi m_p c^4 \delta t_{\rm var}\over L\sigma_T}\right)^{1/5}\nn
                   &=& 0.1 \left({\delta t_{\rm var}\over 10~{\rm ms}}\right)^{1/5}\left({L\over 10^{47}~{\rm erg~s^{-1}}}\right)^{-1/5}
\end{eqnarray}
of the energy flux.

On the other hand, radio monitoring detected transient emission in the weeks following both
flares~\cite{gelfand:2005,taylor:2005}.  The emission following the 2004 flare was especially bright, as befitting the much greater energy of the burst,
and could be tracked on the sky for more than a year.  The expansion of the radio source implies a transverse velocity
$v \sim 0.7(D/15~{\rm kpc}) c$ \cite{taylor:2005}.
After this, clear evidence is seen for a break in the radio light curve consistent with a transition from uniform expansion
to a Sedov-like phase.  We infer that the measured proper motion is probably the free expansion velocity, as corrected for relativistic aberration.

In principle it is possible for the measured transverse speed to be less than the speed of light, if the intrinsic motion $V$ is nearly luminal
but the motion is directed away from the observer (at some angle $\theta > \pi/2$ with respect to the line of sight to the magnetar):  
$\beta_{\perp,\rm obs} = \beta\sin\theta/(1-\beta\cos\theta)$.  The observed relative brightness of both the radio and gamma-ray emission
argues against this: one measures $E_\gamma \sim 4\times 10^{44}(D/15~{\rm kpc})^2$ in 2004 vs. $E_\gamma \sim 5\times 10^{46}
(D/15~{\rm kpc})^2$ in 1998, and a radio flux $F_\nu \sim 50$ mJy vs. 0.3 $\mu$Jy at $8.5$ GHz 1 week after the flare
\cite{tanaka:2007,frail:1999,gelfand:2005}. 
(A factor $\sim 2$ underestimate of the $\sim 15$ kpc distance to SGR 1806-20 appears unlikely given that the source position is in 
the Galactic plane.)

The energy reservoir that powers the early, subluminal stage of the radio afterglow must be composed of something other
than electrons and positrons emitted by
the star, which would mainly have annihilated during the very brief fireball phase.  A reconnected magnetic field is the most
plausible delayed carrier of energy, especially given the strong limitations on the baryon flux during the $\alt 100~{\rm ms}$
 gamma-ray pulse.
A closed loop of magnetic field carries a finite inertia, and so its center-of-mass frame
will not accelerate with distance from the source as does that of a collimated particle beam.  

The outflowing pair-photon fluid must overcome the tension of the magnetic field lines that are anchored in the star, and so comparable
energy can be put into the stretched field, which is pulled into a split-monopole configuration.   The energy of the stretched field 
is concentrated close to the star, $B^2 r^3 \sim r^{-1}$, although not as strongly concentrated as in a static dipole ($B^2 r^3 \sim r^{-3}$).
In a static plasma, the time for reconnection at the Alfv\'en radius $R_A$ 
is $t_{\rm rec} \sim 3\,(R_A/100~{\rm km})(V_A/0.1~c)^{-1}~{\rm ms}$.  Comparing this expression with the measured
e-folding time of $\sim 30~{\rm ms}$ for the tails of the giant flare pulses \cite{tanaka:2007} suggests that reconnection is gated
by the decrease in pair-photon pressure.  

We conclude that, in a magnetar giant flare, most of the released magnetic energy tends to {\it follow} the pair-photon pulse, with 
the delay between the two components being due to continued shearing of the external magnetic as the interior of the magnetar adjusts 
on the internal Alfv\'en time of $\sim 0.1~{\rm s}$.


\section{Concluding remarks}
\label{sec:concluding_remarks}
Understanding the global behavior of strongly gravitating, dynamical systems, containing dense matter
coupled to ultra-strong magnetic fields, is of key importance for a thorough understanding of possible signals from them.
In this work, we have presented a new approach to this end by combining the ideal 
MHD and force-free approximations in a suitable manner within general relativity.

We expect that this hybrid scheme represents real progress towards 
greater realism. The stellar interior utilizes ideal MHD to faithfully model the neutron
star without the disadvantages incumbent in the low density exterior. Likewise,
the exterior solution uses the force-free approach and so captures the dynamics of
the tenuous plasma. The entire domain
is described by a fully nonlinear and fully dynamic general relativity solution
necessary for strong-field gravity. Indeed, a key aspect of this approach is
its generality. It does not require a prescribed stationary stellar boundary, and can be applied,
for example, to dynamical systems such as collapsing stars and non-vacuum compact binaries.

We have exploited this approach to study stellar collapse in both rotating and non-rotating cases. 
Our studies reveal a rich phenomenology in the magnetosphere as the collapse proceeds. In particular, 
magnetic reconnection plays an important role by inducing strong electromagnetic emission as well 
as the infall of electromagnetic energy into the black hole, which in a short time loses all its `hair'.
When the star starts off rapidly rotating, the energy of the magnetospheric plasma grows significantly during the collapse.
As the star rotates faster, its magnetic field lines do not have time to re-adjust to the increased rotation, and are
strongly wound up out to the initial light cylinder.
It is worth emphasizing that this conversion of dynamical energy into electromagnetic energy does not depend on resistive effects.

Two issues of principle have arisen in performing these calculations.  First, we have shown conclusively that 
the force-free approximation to the evolution of a dilute, relativistic plasma inevitably leads to singularities.  These
singularities are avoided in our calculations by continuously pruning the electric field.  In the simulations that we
have run, this procedure appears to cause limited energy dissipation, but its necessity should be kept in mind.

The second issue of principle regards the dependence of the reconnection geometry on the resistivity model.  Fluid pressure
is responsible for slowing down the rate of reconnection unless an x-point geometry is able to form.
By neglecting fluid pressure, the force-free approximation clearly facilitates the formation of x-points.  
In spite of this, we observe very limited numerical dissipation in current sheets.  Although the plasma that
is represented by our model magnetospheres is, in reality, strongly collisional, it should be kept in mind that
fast x-point reconnection still occurs in collisional plasmas in the presence of hydromagnetic turbulence -- 
as has been demonstrated so far in weakly magnetized plasmas \cite{kowal:2009}.  Some further exploration of the reconnection
geometry is possible in resistive MHD calculations of stellar collapse by varying the spatial dependence of the resistivity.

We have also discussed how the electromagnetic outbursts from collapsing magnetars may have interesting observational effects.
In particular, our results are relevant to binary neutron star merger scenarios, in which a hypermassive neutron star forms 
and collapses to a black hole.  Collapse to a black hole can happen either promptly or after many dynamical times, depending on the masses 
involved and the equation of state describing the stars (see, for instance,~\cite{Hotokezaka:2011dh}).  
Although Kelvin-Helmholtz and magnetorotational instabilities will create ultrastrong magnetic fields ($\agt 10^{14-16}$ G), we have
suggested that the global field (in particular, the amount of magnetic flux threading the merger remnant and eventual black hole) 
could depend strongly on the lifetime of the remnant.  The formation of a distinct magnetosphere would require
tapping only $\sim 10^{-3}-10^{-4}$ of the energy dissipated in shear layer at the remnant surface.  As such a star collapsed,
the magnetosphere would qualitatively follow the behavior outlined here.   The electromagnetic output of the collapse
could compete with the later emission from a Blandford-Znajek jet emanating from the black hole horizon,
and source a powerful electromagnetic counterpart to the gravity wave signal (e.g.~\cite{Metzger:2011bv}).
In addition, the magnetic field dynamics that is revealed in our simulations has a number of interesting implications for
gamma-ray bursts and magnetar flares, as discussed in Secs.~\ref{sec:astrophysicsI} and~\ref{sec:magnetar}.

Beyond the work analyzed here, our approach is readily applicable to other relevant systems and will be applied, in particular, 
to study binary neutron star systems~\cite{PalenzuelaBNSpassive}, and black hole-neutron star systems 
(see e.g.~\cite{McWilliams:2011zi}). It is important to stress, however, that our approach is not
free of ambiguities, in particular with respect to how and where the matching between the 
force-free and ideal MHD regions is implemented. To the extent possible, we have tested
the robustness of our results versus known solutions, which give us confidence in this approach.

%
%
\noindent{\bf{\em Acknowledgments:}}
It is a pleasure to thank  P. Goldreich for insights and discussions during the course
of this work.  We also thank
E. Berti, P. Cerda-Duran, J. McKinney, 
R. O'Shaughnessy, E. Ramirez-Ruiz, E. Quataert, A. Spitkovsky and D. Tsang 
as well as our long time collaborators M. Anderson, E. Hirschmann, P. Motl, D. Neilsen and O. Reula,
for discussions and comments.
This work was supported by the NSF (PHY-0969827 to Long Island University) 
and NSERC through Discovery Grants (to CT and LL). Research at Perimeter Institute is
supported through Industry Canada and by the Province of Ontario
through the Ministry of Research \& Innovation. Computations were
performed at Teragrid and Scinet.

%
%
\appendix

\section{The Transition from Ideal to Force-Free MHD}
\label{app:issues}
Details of the transition from the ideal MHD regime to the force-free
regime merit particular attention, especially in our rotating collapse solutions.
Since we match two different formulations of Maxwell's equations coupled to conducting
matter, it is important to understand how the choice of matching layer affects the
exterior force-free solution.
Of course, an unambiguous test can only be provided by a complete resistive MHD solution that
can handle the strongly magnetized regions outside the star and, in particular, can follow the
large changes in density and rotation that are encountered during the collapse. While work on
this direction is in progress, the simpler approach presented here allows us to obtain a first solution to the
relativistic magnetosphere in this strongly dynamic situation.

The positioning of the matching zone is constrained by competing considerations.
On the one hand, if it sits too close to the surface, then the interior MHD solution that sources
the exterior force-free solution will be unrealistic, due to the density floor that is applied in the
MHD atmosphere. One thus might want to place the transition layer well within the stellar surface.
However, such a deep layer might underestimate the magnetic field strength 
at the base of the magnetosphere, and imply force-free behavior of the magnetic field where
that approximation is not justified

In particular, we have found that placing the transition zone at too low a density implies
an unrealistically large toroidal magnetic field at the base of the force-free zone.
Recall that the region exterior to the star, with its tenuous plasma,
will have a relatively large magnetization, and should be forced to co-rotate with
the interior of the star. Unless prohibitively high resolution is employed, the
atmosphere of the MHD solution has a large enough inertia that this condition
can be violated.  Instead, the magnetic field experiences a non-negligible (and non-physical)
differential rotation.  This effect is easily exacerbated in a collapsing solution.

Within our current approach, we address these issues in two ways, by (i) adopting
an initial configuration of the force-free fields and ideal MHD fields respecting
such co-rotation up to the light cylinder and with a suitable differential rotation
outside that radius; and (ii) choosing a transition layer at an appropriate
distance inside the stellar surface. 

Regarding the initial data, since the rotation induces a toroidal magnetic field
we adopt initial data that has a toroidal magnetic defined as
$B_{\phi} = - \Omega_F R_{\rm cyl} B_r$ and also
$E = - v \times B$ throughout domain. We also choose the MHD atmosphere to 
co-rotate with the star for $R_{\rm cyl} < 2 R_{\rm star}$, and not to rotate otherwise.

\begin{figure}
 \begin{center}
 \epsfig{file=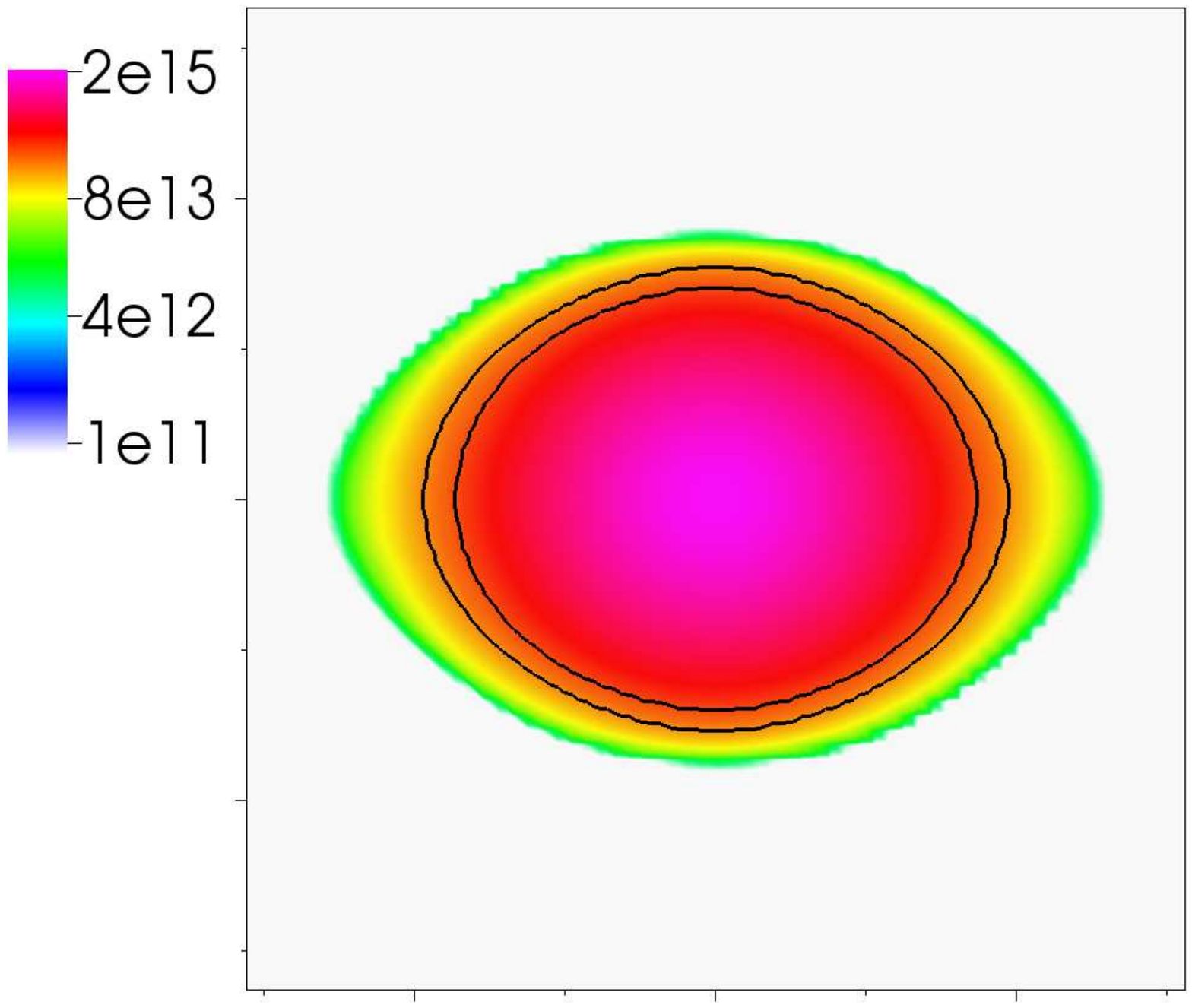,height=2.3in} 
 \epsfig{file=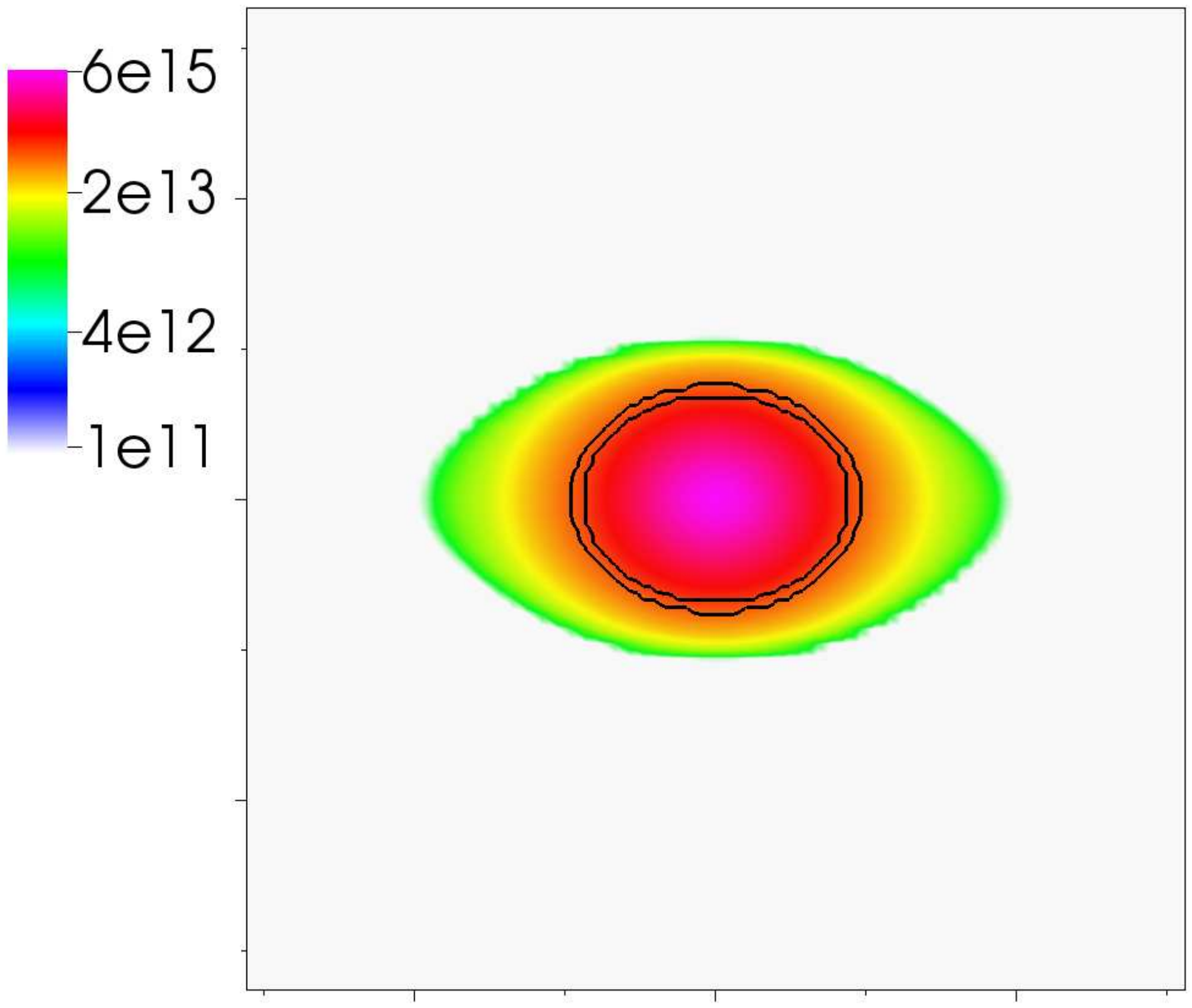,height=2.3in}
 \caption{ {\it Rotating, unstable star (force-free)}.
 Isosurfaces of the kernel function of Eq.~(\ref{eq:kernel}) at two different
times $t=-0.30$ms, left, and $t=-0.07$ms, right,  while the rotating star
collapses. Mapped in color is the density of the star in cgs units in a
 eridional plane. Black lines mark
two isosurfaces of the kernel function, corresponding to $F=(0.01,0.99)$.
}
 \label{fig:collapse_rhoft}
 \end{center}
 \end{figure}

Both the location of the stellar surface and the density in the transition layer
are dynamical. It is important that the choice of transition density $\rho^{\rm match}$
in Eq.~(\ref{eq:kernel}) respects the increasing density of the collapsing star.
In order to maintain a consistent depth of the transition layer, its position
is adjusted as the star collapses by scaling the matching density $\rho^{\rm match}$
in proportion to the peak density within the star: 
$\rho^{\rm match}(t) = \rho^{\rm match}(t=0) [\rho_{{\rm max}}(t)/\rho_{{\rm max}}(t=0)]$.
The transition layer is displayed at the beginning and near the end of the collapse in 
Fig.~\ref{fig:collapse_rhoft}.
Notice that with this conservative approach we are underestimating 
possible rotational effects.

\begin{figure}
 \begin{center}
 \epsfig{file=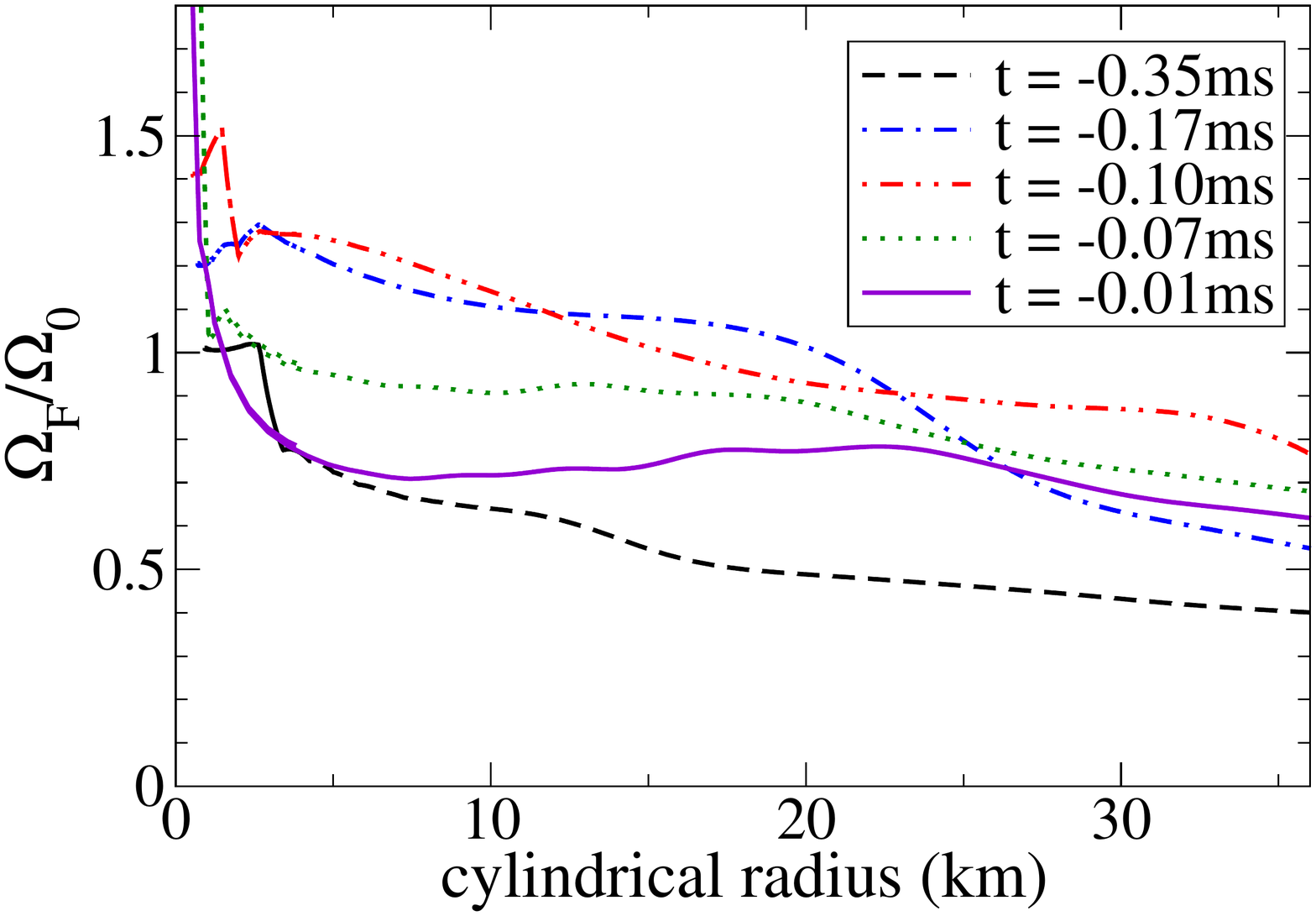,height=2.2in} 
 \epsfig{file=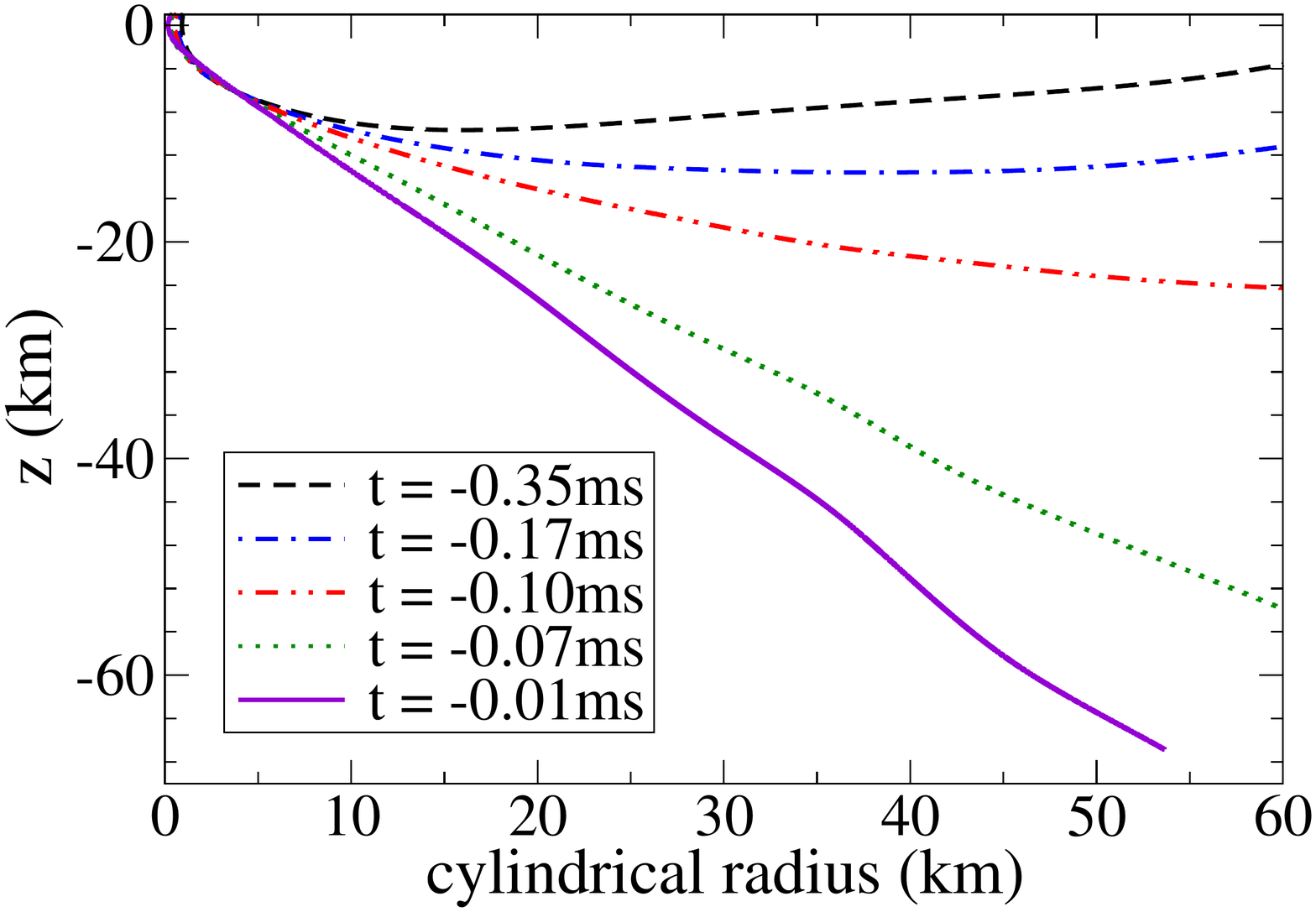,height=2.2in}
 \caption{ {\it Rotating, unstable star (force-free)}.
(Top panel) Angular velocity $\Omega_F$  
of the magnetic field line emanating from $\theta=122^{\circ},\phi=0^{\circ}$  
  as a function of cylindrical radius (normalized with respect to its initial value). 
(Bottom panel) Shape of 
such magnetic field line as time progresses. }
 \label{fig:collapse_omegaf_alongline}
 \end{center}
 \end{figure}

The angular frequency $\Omega_F$ of magnetic field lines anchored near the rotation axis of the star
is illustrated in Fig.~\ref{fig:collapse_omegaf_alongline}.  From the initial condition
(in which constancy of $\Omega_F$ is enforced), a negative gradient in $\Omega_F$ develops
in the transition layer during the earliest stages of the collapse.  This negative gradient has
then disappeared by the time that the stellar angular velocity has increased by $\sim 20\%$.
above the initial value $\Omega_0$ (see Fig.~\ref{fig:collapse_omegastar}).    
From then on, the qualitative behavior obtained is consistent with the expected one:
during the collapse, the star transfers angular momentum to the magnetic field
lines so that $\Omega_F = \Omega_{\rm star}$ near its surface, propagating
along the magnetic field lines with a speed $\sim c$.

The sharp negative gradient appearing in the last stages
of the collapse reflects the strong radial gradient in $\Omega_{\rm star}$ that
appears near the rotation axis, as well as the onset of strong general relativistic
effects. As the event horizon arises (prior to the formation of the apparent
horizon at $t = 0$), it disconnects the interior from the exterior solution
and causes $\Omega_F$ to decrease, tending to the value $\Omega_F = \Omega_H/2$
expected by the Blandford-Znajek solution of a spinning black hole.

%
%
\bibliographystyle{apsrev}

\end{document}